\documentclass[aps, onecolumn, notitlepage, floatfix, nofootinbib, amsmath, letterpaper,amssymb, showpacs,superscriptaddress]{revtex4-1} 
\usepackage{graphicx}
\usepackage{amssymb}
\usepackage{epstopdf}
\usepackage{slashed}
\usepackage{amsmath}
\usepackage{multirow}

\usepackage{url}
\usepackage{color}
\usepackage{listings}

\newcommand{\ignore}[1]{}
\newcommand{\be}{\begin{equation}} \newcommand{\ee}{\end{equation}}
\newcommand{\ba}{\begin{eqnarray}} \newcommand{\ea}{\end{eqnarray}}
\newcommand{\nn}{\nonumber}

\newcommand{\Pf}{\text{Pf}}
\newcommand{\tPf}{\text{tPf}}

\newcommand{\GeV}{{\rm\ GeV}}

\newcommand{\TeV}{{\rm\ TeV}}

\def\slashb#1{\setbox0=\hbox{$#1$}#1\hskip-\wd0\dimen0=5pt\advance
        \dimen0 by-\ht0\advance\dimen0 by\dp0\lower0.5\dimen0\hbox
          to\wd0{\hss\sl/\/\hss}}

\def\ket#1{\left| #1\right>}

\input{epsf}

\begin{document}

\pagenumbering{arabic}

{{\flushright{\tiny{\vspace*{-.4cm}CERN-PH-TH/2012-349} \vspace*{.4cm}}}
\title{Boosting the Standard Model Higgs Signal with the Template Overlap Method}

\author{Mihailo Backovi\'{c}} \email{mihailo.backovic@weizmann.ac.il} \affiliation{Department of Particle Physics and Astrophysics, \\ Weizmann Institute of Science, Rehovot 76100, Israel} 
\author{Jos\'e Juknevich} \email{jose.juknevich@weizmann.ac.il} \affiliation{Department of Particle Physics and Astrophysics, \\ Weizmann Institute of Science, Rehovot 76100, Israel}
\author{Gilad Perez}\email{gilad.perez@weizmann.ac.il} \affiliation{Department of Particle Physics and Astrophysics, \\ Weizmann Institute of Science, Rehovot 76100, Israel} \affiliation{CERN, Theory Division, CH1211 Geneva 23, Switzerland}

\begin{abstract}
We show that the Template Overlap Method can improve the signal to background ratio of boosted $H\to b \bar b$ events produced in association with a leptonically decaying $W$. We introduce several improvements on the previous formulations of the template method. Varying three-particle template subcones increases the rejection power against the backgrounds, while sequential template generation ensures an efficient coverage in template phase space.  We integrate $b$-tagging information into the template overlap framework and
introduce a new template based observable, the template stretch.   Our analysis takes into account the contamination from the charm daughters of top decays in $t\bar t$ events, and includes nearly-realistic effects of pileup and underlying events. We show that the Template Overlap Method displays very low sensitivity to pileup, hence providing a self-contained alternative to other methods of pile up subtraction. The developments described in this work are quite general, and may apply to other searches for massive boosted objects. 
\end{abstract}

\maketitle

\section{Introduction}
Recent results from ATLAS~\cite{:2012gk} and CMS~\cite{:2012gu} have confirmed the discovery of a new boson of mass roughly $125  \GeV$, decaying to $\gamma\gamma$, $ZZ^*$ and likely $W W^*$(for the Tevatron combination of the Higgs searches see Ref. ~\cite{Aaltonen:2012qt}). The results are so far consistent with the interpretation of the new particle as the Standard Model (SM) Higgs boson. 

While the data agrees with the SM Higgs boson, our understanding of the new particle's properties remains incomplete. A strong test of the SM Higgs theory consists of a detailed experimental study of its sharply predicted characteristics. This includes, among others, observing all the major decay modes of the SM Higgs, as well as establishing no deviations from the predicted SM Higgs production and decay rates. 
Global fits to the Higgs boson production and decay rates allow for extraction of its couplings to various other SM fields, as well as possible invisible channels (see Refs.~\cite{Carmi:2012in, Giardino:2012dp, Azatov:2012ga, Espinosa:2012im, Ellis:2012hz} for recent analyses).
 A SM Higgs boson of $125 \GeV$ predicts a dominant decay mode to a $b \bar{b}$ pair, which calls for a direct verification. An enormous QCD background, however,  makes it rather difficult to  observe this channel at the LHC. The hadronic Higgs decay mode has so far only been reported by the Tevatron experiments, at $2.8 \sigma$ in the CDF/D0 combinations. Despite an impressive progress by both CMS and ATLAS, the extraction of a statistically significant measurement of $h \to b \bar{b}$ rate from the LHC data at $8 \TeV$ remains challenging. 

One way to reduce the QCD background in $h \to b \bar{b}$ is by focusing on associated Higgs production with $W$ and $Z$ bosons.
However, the cross section of the background $pp\to V b\bar b$ process is  still much higher than the cross section of $pp\to Vh$. Even after a cut on the mass of the $b\bar b$ system around the known Higgs mass, the signal is swamped by the background. 
 Authors of Ref.~\cite{Butterworth:2008iy} showed that when considering moderately boosted Higgs events, traditional jet clustering algorithms with large cone sizes ($R \sim 1$) can be used to increase the signal to background ratio (S/B). The method of Ref.  \cite{Butterworth:2008iy} is based on the fact that decay products of a boosted Higgs  are collimated and can be captured within a single ``fat'' jet.
 This, in principle, reduces the  combinatorial background, contamination from soft and incoherent components and allows to better characterize the structure of the energy flow within the fat jet.
 
The last few years have seen a proliferation of new theoretical and experimental techniques to identify high-$p_T$ jets at the LHC  (see Refs. ~\cite{Ellis:2007ib,Abdesselam:2010pt,Salam:2009jx,Nath:2010zj,Almeida:2011ud,Plehn:2011tg, Soper:2011cr, Soper:2012pb}
for recent reviews and references therein).  
Two main classes of approaches have emerged: \textit{Filtering}~\cite{Butterworth:2008iy} (see also Refs. ~\cite{Krohn:2009th,Ellis:2009me}) and \textit{Template Overlap Method}~\cite{Almeida:2010pa}. 
Filtering algorithms act on the list of jet constituents by removing the soft components based on some measure which defines the ``hard'' part of the jet. The remaining constituents are then reclustered into the ``filtered'' jet. 
 The Template Overlap Method, discussed in detail below, does not manipulate the measured jet's list of constituents,neither does it require any special clustering algorithm. Instead the method compares the jet to a set of parton level states built according to
a fixed-order distribution of signal jets called \textit{templates}.  The comparison makes use of an ``overlap
function'' which evaluates the level of agreement between each measured jet and a set of templates.

Ref.~\cite{Almeida:2010pa} focused on building the templates according to the leading order decay modes, namely two body ($N=2$) for
the boosted Higgs and three body ($N=3$) for the boosted top.
In Ref.~\cite{:2012qa}, the ATLAS collaboration used the Template Overlap Method together with the the {\sc HEPTopTagger}~\cite{Plehn:2010st} to search for heavy $t\bar{t}$ resonances.
Authors of Ref.~\cite{Almeida:2011aa} showed how to extend the Template Overlap Method beyond leading order as well as how to construct templates
which describe the energy flow of say $h\to b\bar b g$ in an infrared (IR) safe manner.
The method allows one to gain access to ``partonic-like" observables which correspond to the template configurations with the maximal overlap score. The resulting information can further improve the ability to distinguish the signal from various background channels.

Furthermore, Ref.~\cite{Almeida:2011aa} introduced the concept of \textit{template jet shapes}, such as Template Planar Flow~\cite{Almeida:2008yp,Thaler:2008ju, Almeida:2011aa,Matan:2012vx} and template-angularity~\cite{Almeida:2008yp,Berger:2001ns}).
The results of Ref.~\cite{Almeida:2011aa} showed that template overlap is capable of delivering background rejection factors of $O(100)$ against the $Wj$'s background in the idealistic ultra-boosted Higgs  regime (\textit{i.e.} $p_T \sim 1$\,TeV), when combined with other jet substructure observables. 

In this paper, we examine the decay and radiation patterns of a boosted SM Higgs boson, with focus on a realistic $p_T$ kinematic regime (\textit{i.e,} $300 - 400$\,GeV). We argue that a boosted Higgs search using the Template Overlap Method is viable in the future LHC run.  We achieve the best signal sensitivity by combining templates in the full phase space for $N=3$ and $N=2$ overlaps in addition to other template based observables.  Moreover, we introduce a new variable, {\it Template Stretch}, which exploits the difference in plain distance of the two leading $b$-tagged subjets relative to the signal expectations. 

Our treatment of Template Overlap Method improves on the previous formulations in several ways. First, we define templates in terms of longitudinally boost-invariant variables. Second, and more importantly, we entirely revamp the method of template generation. In Ref. \cite{Almeida:2011aa}, the minimum number of templates required to adequately describe  the jet energy flow in the medium $p_T$ range, was roughly two orders of magnitude larger than in this paper. The reason is that in Ref. \cite{Almeida:2011aa}, templates were generated in the Higgs rest frame
 (with a MonteCarlo-like method) and then boosted to the lab frame on an event by event basis. Generating templates in the lab  frame and ``tiling'' them according to the event kinematics leads to better coverage of phase space at lower $p_T$ and to a great improvement in the overall performance of the analysis. Third, we introduce $b$-tagging into the Template Overlap framework. Information about $b$-jets combined with peak templates serves to improve the rejection power, defined as the signal efficiency divided by the efficiency for the background. Finally, the optimal radius of the template subcones is not necessarily the same for every parton in a template, as low momentum subjets tend to have wider angular profiles. We allow the three-body template subcones to vary with $p_T$ providing a more adequate description of the showering patterns within a fat jet. Varying cones improve the tagging performance of three-body overlap as well as most of the other template-derived observables.

Our analysis includes nearly-realistic effects of pileup and underlying event (UE). 
In high luminosity environments, the large jet cone radius allows for severe effects of pileup on the spectrum of both jet and substructure observables. We show that the template jet shapes give us an additional handle on pileup, as they are based on best matched templates and not jet constituents.  The ``spikiness'' of the jet energy distribution naturally avoids the complication of soft un-correlated backgrounds. 
The Template Overlap Method is thus less susceptible to pileup compared to other kinematic observables such as jet invariant mass and $p_T$. This feature invites us to re-consider several jet-shape observables in terms of the ``partonic'' distribution of the peak templates. For instance, jet Planar Flow (Pf) is known to exhibit high susceptibility to pileup~\cite{Krohn:2009wm,Soyez:2012hv}, however it is a useful background discriminant when the hard and coherent part of the massive jet is considered~\cite{Almeida:2008yp,Thaler:2008ju, Almeida:2008tp}. We thus introduce a pileup-insensitive alternative to Planar Flow constructed from the template states alone. 

We use parton-shower simulations to illustrate the non-susceptibility of the Template Overlap Method to a high-pileup environment. Our results agree with the 7\,TeV ATLAS data analysis in Ref.~\cite{:2012qa}, which showed that the overlap method is indeed fairly robust to presence of moderate pileup contamination. 

In Section~\ref{Sec:TemplateOverlapMethod}, we give an overview of the Template Overlap Method and introduce several template based observables sensitive to the QCD radiation patterns. Section~\ref{sec:data} describes our Monte Carlo (MC) data generation, and shows
the results for boosted Higgs searches with a mass of $125\GeV$ at $\sqrt{s} = 8 \TeV$ and $\sqrt{s} = \, 13 \TeV$. Section \ref{sec:data} also contains a detailed discussion of pileup effects. We give a detailed review of template Planar Flow in Appendix \ref{app:PF}, while Appendix \ref{sec:Templates} describes the new method of template states generation.

\section{Template Overlap Method} \label{Sec:TemplateOverlapMethod}

Template Overlap Method is based on the quantitative comparison between the energy flow inside physical jets and the energy carried by partons modeled after the boosted signal events (templates).
 We define libraries of templates as sets of $N$ four-momenta $\ket{f} = |p_1, p_2, \cdots ,p_N\rangle$ representing the decay products of a SM Higgs boson at a fixed momenta $P$ and Higgs mass $m_h$:
\be
	\sum_{a=1}^{N} p_a = P,\,\,\,\,\,\,\,\,\, P^2 = m_h^2.
\ee
We require that the $N$ quanta of energy be captured within an anti-$k_T$ jet of varying size, scaled according to the $p_T$ of the Higgs (or according to the $p_T$ of the associated vector boson).
The number of partons in the templates is not necessarily fixed, but is calculated in fixed order perturbation theory\footnote{In principle, one can re-sum the soft radiation from each of the partons.  Here we simply stick to a fixed order perturbation theory description.}, and ``next-to-leading-order" templates
with more than the minimum number of partons are possible. Below we focus  on combining the information from templates in the full phase space for $N=3$ 
and $N=2$ partonic configurations, and show that they provide some additional rejection power against the dominant $Wb\bar b$ and $t\bar t$ backgrounds.

Next, a functional measure quantifies the agreement in energy flow between a given Higgs decay hypothesis (a template) $f$ and an observed jet $j$. A scan over a large set of templates that cover the $N$-body phase space of a Higgs decay results in $f[j]$, the template which maximizes the functional measure.  Our primary jet substructure observable is the $N$-body overlap $Ov_N$,  the value of the functional measure for the best matched template.

For each jet candidate, we define the (maximum) overlap as
\begin{equation}
 Ov_N(j,f[j]) = \max_{\{f\}}\,\left[ \exp\left[ -\sum_{a=1}^N \frac{1}{\sigma_a^2}\left( \epsilon \,  p_{T,a} -\sum_{i\in j} p_{T,i} \,F(\hat n_i,\hat n_a) \right)^2 \right] \right],
\end{equation}
where $\{f\}$ collectively denotes a template library for the given jet $p_T$, $p_{T,a}$ is the transverse momentum of the $a^{th}$ template parton and $p_{T,i}$ is the transverse momentum of the $i^{th}$ jet constituent (or calorimeter tower, topocluster, etc.). 
The first sum is over the $N$ partons in the template and the sum inside the parentheses is over jet constituents. The kernel functions $F(\hat n, \hat n_a)$ restrict the angular sums to (nonintersecting) regions surrounding each of the template momenta.  We refer to the template state which maximizes the functional measure as the ``peak template'' $f[j]$ of the jet $j$. The parameter $\epsilon$ allows us to correct for the energy not captured by the template overlap. In the case of $N=3$, we use $\epsilon = 0.8$. For more details see Section \ref{sec:Scale}.

In this analysis, we take the kernel function to be
 a normalized step function that is nonzero only in definite angular regions around the directions of the template momenta $p_a$:
\begin{equation}
 F(\hat n_i,\hat n_a) = \left\{ \begin{array}{rl}
 1 &\mbox{ if $ \Delta R < r_a$} \\
  0 &\mbox{otherwise}
       \end{array} \right. ,
\end{equation}
 where $\Delta R$ is the plain distance between the template parton and a jet constituent in the ($\eta$,$\phi$) plane.
The parameters $r_a$ determine the angular scale of the template subjet. Together with the energy resolutions $\sigma_a$, these are the only tunable parameters of the model. 

A few ideas for possible strategies to determine the values of $R_a$ and $\sigma_a$ are listed below:
\begin{itemize}
 \item Choose the single best parameter according to some optimization criterion ({\it e.g.,} optimize the tagging efficiency and background rejection), and use the same values for all the partons within a set of templates. 
 \item Choose the parameters separately for each template, {\it e.g.} using a $p_T$-dependent scale for template matching. 
\end{itemize}

Based on a combination of these two criteria, we fix $\sigma_a$ (for the $a$th parton) by that parton's transverse momentum,
\begin{equation}
 \sigma_a = p_{T,a}/3.
\end{equation}
We use a subcone of radius $r_2 = 0.3$ for the two-body template analysis, while the three-body subcone is dynamically determined on a template-by-template basis. Section \ref{sec:Scale} contains a detailed discussion on the optimal scaling of subcone radius. We should nevertheless emphasize that the overall performance of the method can be maximized for a wide range of $r$ by rescaling other parameters. 

Next, we generate libraries of $O(10^4)$ 2-body templates and $O(10^6)$ 3-body templates in steps of Higgs $p_T$ of $30\GeV$ starting from $315\GeV$. For each event, we achieve the best signal sensitivity by dynamically selecting a set of templates based on the transverse momentum of the $W$ boson,
\begin{equation}
 p_{T}^{W} \in \left[p^{\rm bin}_{T \, {\rm min}},\,p^{\rm bin}_{T \, {\rm max}}\right),
\end{equation}
where $p^{\rm bin}_{T \, {\rm min}},p^{\rm bin}_{T \, {\rm max}}$ are the limits of the bin corresponding to the template set with  $p_T^{\rm temp} = (p^{\rm bin}_{T \, \rm max} - p^{\rm bin}_{T\, \rm min} ) / 2$. For instance, an event with $p_T^W = 320\GeV$ would be analyzed by a template set with $p_T = 315 \GeV$ etc. We give more details on template properties and generation in the Appendix.

Replacing the criteria for the template set selection from jet $p_T$ to $p_T^W$ has a enormous advantage when considering effects of pileup, however it is not the only choice. As an alternative, one could analyze each jet with all template sets, but at a huge expense in computation time. 

We use the {\sc TemplateTagger}~\cite{templatetagger} numerical package for the template matching analysis \footnote{Publicly available at: \url{tom.hepforge.org}}.
The package is a {\sc C++} code which provides basic implementation 
of the Template Overlap Method for jet substructure, as well as several other jet analysis tools. 

 \subsection{Other Peak Template Observables} 
 Template overlap provides a mapping of final states $j$ to partonic configurations $f[j]$ at any given order in perturbation theory. Once the best matched template $f[j]$ is found, it can be used to characterize the energy flow of the state, giving information on the likelihood that the event is signal or background.
 The scope of template overlap does not stop with $Ov_2$ and $Ov_3$. Peak templates contain additional information about correlations within the fat jet, some of which we explore in the following sections.
 
\subsubsection{Angular Correlations}
Of particular value are angular correlations between template momenta which can otherwise be concealed in the numerical values of the peak overlap. For instance, the angular distribution of jet radiation can be measured with the variable $\bar \theta$~\cite{Almeida:2011aa}, defined as
\be
	\overline{\theta} = \sum_i \sin \Delta R_{iJ},
\ee
where $\Delta R_{iJ}$ is the distance in the $(\eta,\phi)$ plane between the $i^{th}$ template partonic momentum and the jet axis. 
When measured using three-body templates, the $\bar \theta$ variable exploits the fact that signal events tend to have smaller emission angles. Notice that for highly boosted jets, the 2-body version of $\overline{\theta}$ simply reduces to the angle between the two partons~\cite{Almeida:2008yp}.

\subsubsection{Template Planar Flow }
Another useful background discriminant is Planar Flow  ~\cite{Thaler:2008ju, Almeida:2008yp} (see Ref. \cite{Matan:2012vx} for a recent study of the Pf distribution of QCD massive boosted jets).
To define the Pf variable we first introduce the ``jet inertia tensor'' as
\be
	I ^{kl} \equiv \frac{1}{m_J} \, \sum_i^N \frac{p_i^k p_i^l}{ p^T_i}.
\ee
Here $m_J$ is the jet mass, and $i$ runs over the jet constituents. We choose to use the boost-invariant definition  for $I ^{kl}, $ whereby we define the two dimensional vectors $p_i^k$ as
\be
 p_i^k \equiv p^T_i (\eta_i, \phi_i),
  \ee   
  with $\eta_i, \, \phi_i$ are measured relative to the jet axis. 
The Pf jet shape is given by
\be
	\Pf \equiv \frac{4 \,\det(I)}{{\rm tr}(I)^2}\,.
\ee
Notice that because of the fact that the trace of $\big(I ^{kl}\big)$ is proportional to the jet mass~\cite{GurAri:2011vx}, Planar Flow is only well defined for massive jets~\cite{Soyez:2012hv}.
Planar flow is particularly helpful in distinguishing energy flow distributions which lie on a line ($\Pf \rightarrow 0$) from uniformly distributed energy flow ($\Pf \rightarrow 1$). For instance, Planar Flow of a boosted Higgs will tend to be smaller than that of a massive QCD jet that in turn will be smaller than that of a boosted top or gluino~\cite{Eshel:2011vs}. 

Planar flow of a \textit{jet} is useful when considering only the hard and coherent part of the jet. Because of high sensitivity of jet Planar Flow to pileup and UE, here we consider Template Planar Flow ($\tPf$) as an alternative. The advantage of $\tPf$ is that it is constructed purely out of peak template states and thus less susceptible to pileup. We demonstrate this point using a Monte Carlo simulation in Section \ref{sec:pileup}. The non zero mass of the peak template states guarantees infra-red safety of $\tPf$. We define $\tPf$ using peak template momenta as well as the template subcones to include physical effects of energy smearing.  Appendix \ref{app:PF} gives a more detailed discussion.

\subsubsection{Template $b$ Identification }

A simple but useful way to reject backgrounds and accept signal events is to incorporate the information related to $b$-tagging into the template overlap framework.  The identification of $b$-tagged jets relies on information beyond what is provided by the calorimeters (say from the presence of a displaced vertex and/or a hard lepton). This information ({\it e.g.} direction of the $b$-tagged jets in $\eta$ and $\phi$) is fairly uncorrelated with the information about the direction of the peak template partons. 
We integrate the $b$-tagging information into the template overlap framework by assigning  a $b$-quark tag, $t_b^{(f)}$ to each peak template $f[j]$. A two body template parton is assigned a $b$-tag if an anti-$k_T$ ($r = 0.4$) $b$-jet lies within a template cone of radius $r_2$  around the template parton axis. For simplicity, we take a jet to be $b$-tagged if it has $p_T > p_T^{\rm tag}$ (default: $p_T^{\rm tag} = 20\GeV$) and contains a $b$ or $\bar b$ quark. 

Information about template $b$-tags can be of particular use in discriminating the large $t \overline{t}$ background. For instance, consider a typical doubly-$b$-tagged jet coming from a $t \overline{t}$ event. A fragment of another light (or $c$) jet is likely to fall in the cone of $R \sim 1.$ If only the criterion of a doubly $b$-tagged jet is used, there is no guarantee that the peak two body template will select the two $b$ subjets, making the $t \overline{t}$ event more likely to pass the kinematic constraints of the template states. On the contrary, if we require that the $b$-tagged jets 
coincide with the template momenta, we discriminate against jets in which only one $b$-jet is tagged by the template. Notice that the effect of $b$-tagging on the templates should not have a large effect on the signal.

\subsubsection{Template {\it Stretch}}

In order to increase the signal efficiency of the Tempate Overlap Method we chose a working point where the energy resolution of each of the template partons is rather loose ({\it i.e.} $\sigma_a = p_{T,a} / 3$). This implies that even after both $Ov_2$ and $Ov_3$ cuts the mass distribution for the background events is still broad and certainly more spread than that of the signal. This, as well as the fact that for  template $b$-tagging we require a large $r = 0.4$ anti-$k_T$ jet parameter (motivated by the current experimental defaults), implies that the angular distance between the partonic, ``$b$" candidates with a high $Ov_2$ score would still have some smearing with respect to the actual distance between the two anti-$k_T$ $b$-tagged jets. 
To capture this effect, we define a new observable, {\it Template Stretch}, as 
\be 
S_{b\bar{b}}^{(t)}= \frac{\Delta R_{b\bar{b}} }{ \Delta R_{t}}.  
\ee
where $\Delta R_{t}$ is the distance between the peak two-body template momenta and $\Delta R_{b\bar{b}}$ is the distance between the two $b$-tagged subjets.
We expect that the background events will have a broader distribution of $S_{b\bar{b}} ^{(t)}$ compared to the signal events.
The functionality of the template stretch is correlated with the mass of the jet, but with an important advantage. The mass of a fat jet is subject to a large jet cone radius $R \sim 1,$ making it highly susceptible to effects of pileup and underlying event. Since $S_{b\bar{b}}^{(t)}$ is constructed out of subjets with $r = 0.4$ and template states, it is bound to be less sensitive to pileup. We will illustrate this point further in the following sections.

\subsection{Varying Subcone Templates and Showering Correction} \label{sec:Scale}

Fixed template cones are limited by the fact that different $p_T$ subjets yield a different energy profile in $\eta, \phi$. This fact is important for three body template analysis, where we expect the $p_T$ of the three \textit{peak} template partons to be non-uniform.  For instance, one would find that a template subcone of radius $r = 0.05$ is adequate to capture the radiation pattern of a $500 \GeV$ quark. Yet, the same subcone would completely fail to adequately describe the radiation pattern of a quark with energy of $ 100 \GeV$, resulting in a poor overlap score.  We thus introduce the concept of scaled three body subcones into the template overlap framework. Varying template subcones allow us to correct for energy deposition outside the template subcone radii. This in turn leads to an an improved template-level energy resolution while keeping systematic uncertanties well under control \footnote{At low $p_T$, the leakage of QCD radiation outside the template subcones can be especially large when using fixed 
subcones. To account for this, one has to include energy correction factors into the $p_T$ of the templates, which are largely affected by systematic uncertanties. }. 
Jets become narrower as $p_T$ increases, meaning that a smaller jet area is needed to collect some fixed fraction of the jet energy at higher transverse momentum. The corresponding distribution of jet areas (at a fixed energy fraction) is generically non calculable but is measured by experiments, and is commonly denoted as the ``jet shape" variable. 
 Refs.~\cite{Aad:2011kq} and~\cite{Chatrchyan:2012mec} present ATLAS and CMS studies of the jet shape variable respectively.
\begin{figure}[htb]
\begin{center}
\includegraphics[width=4in]{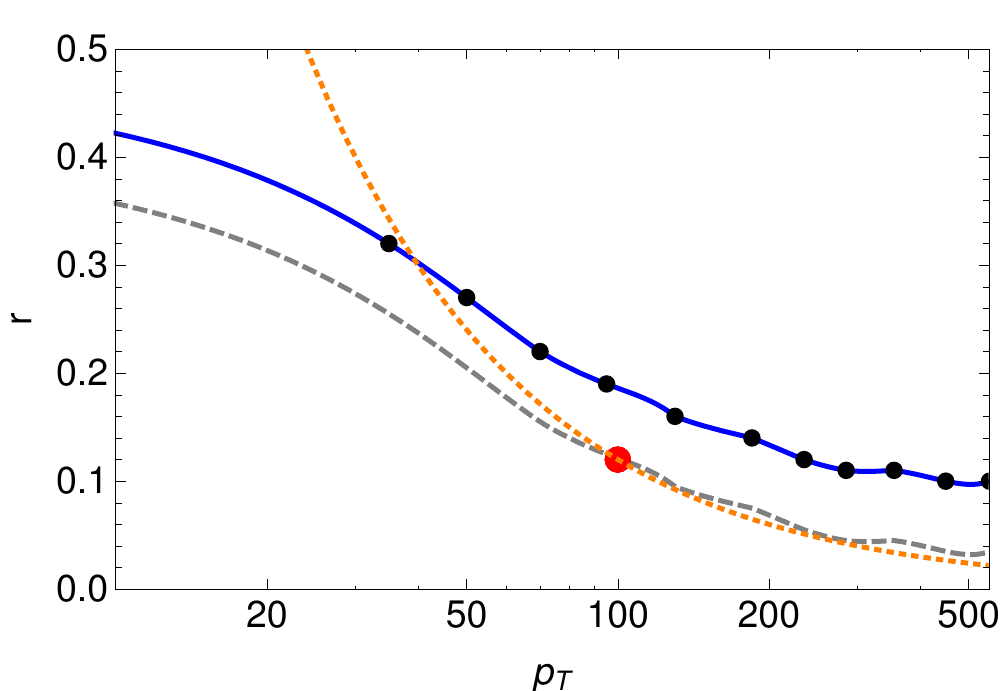} 
\caption{Jet shape results on optimal template subcone scaling. The blue, solid curve shows the minimal anti-$k_T$ radius necessary to capture $80 \%$ of jet energy, obtained from ATLAS data. The grey, dashed curve is the ATLAS result shifted down by $\delta r = 0.063,$ in order to match the optimized value of $r_3(100 \GeV) = 0.12$  (large, red dot). The orange, dotted line is the naive scaling of Eq.~\eqref{eq:r3scale}.  }
\label{fig:ConeScaling}
\end{center}
\end{figure}

 We used the ATLAS jet shape study in Ref. ~\cite{Aad:2011kq}  to establish a scaling rule for the template subcones. The differential jet shape drops rapidly as $r$ increases: at low $p_T$, more than 80\% of the transverse momentum is contained within a cone of radius $r=0.3$ around the jet direction. This fraction increases up to 95\% at very high $p_T$. We fit the numerical values of the integrated jet shape in different $p_T$ regions to obtain the minimum cone radius required to capture $80\%$ of the jet transverse momentum.  In the overlap analysis, we correct for the $80 \%$ efficiency by scaling the template $p_T$ accordingly.
Fig.~\ref{fig:ConeScaling} shows the result of our dynamical scaling. The points represent the minimum  radius necessary to capture $80 \%$ of  jet's energy as a function of jet $p_T.$  The resulting curve gives a shape to the optimal scaling rule for template subcones. The error bars on the data points are small enough that they can be omitted for the purpose of our analysis. To obtain the subcone values in the $p_T < 30 \GeV$ region, we extrapolate the data. 
An overall shift in the jet-shape curve remains a free parameter. To calibrate it we choose a benchmark point of $r_3(100\GeV)$. We demonstrate in the following section that other calibrations are possible and may perform in a similar manner, within a reasonable range. The dotted curve shows that a shift of $ \delta r = 0.063 $ units provides an excellent fit.
 Notice that the naive scaling
\ba
	r_{3}(p^t_T) &=& r_{3}(100 \GeV)\frac{100 \GeV}{p^t_T}  \label{eq:r3scale},
\ea
is in excellent agreement in the $p_T > 60 \GeV$ region,  while the discrepancy with data becomes large at lower jet $p_T.$

\begin{figure}[htb]
\begin{center}
\begin{tabular}{cc}
\includegraphics[width=3in]{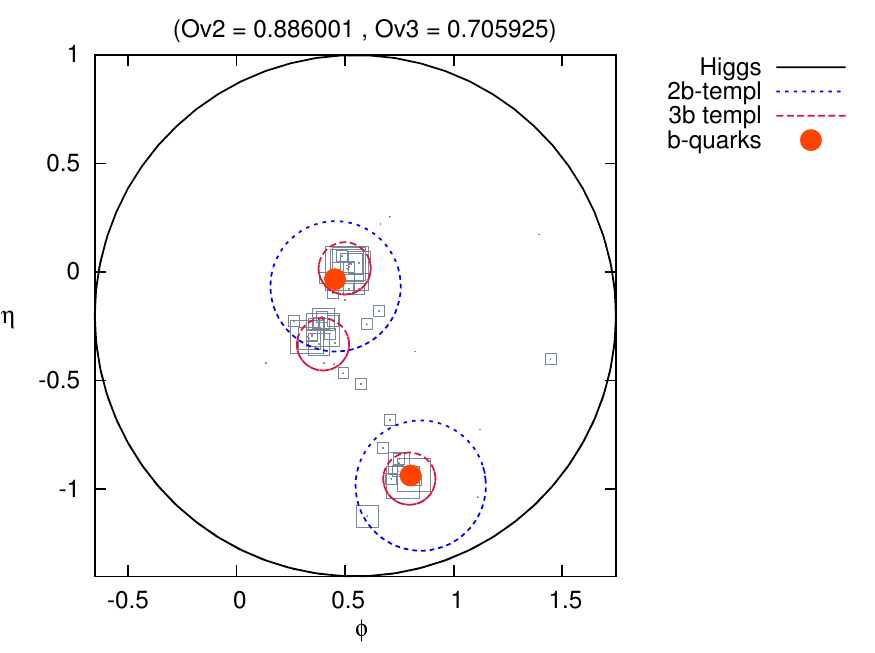} & \includegraphics[width=3in]{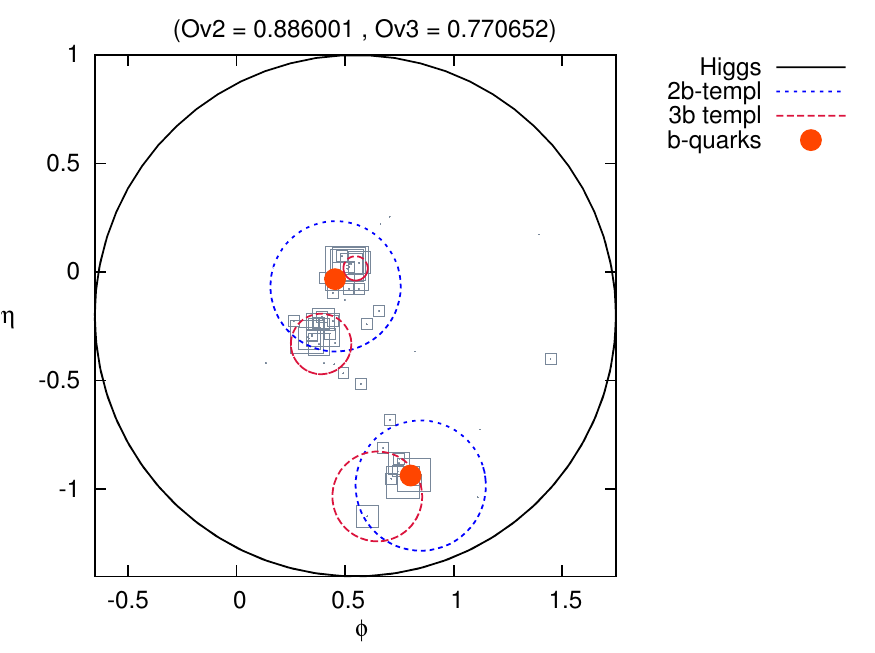} 
\end{tabular}
\caption{Template analysis of a boosted Higgs jet. Left panel shows a Higgs jet analyzed with fixed three body subcones of radius $r_3= 0.12$. The right panel corresponds to a repeated the analysis using the scaled subcones shown in Fig.~\ref{fig:ConeScaling}. Grey squares represent the jet constituents, the $p_T$ of which is proportional to the size of the square. The solid circles are positions of $b$-quarks in the hard process.}
\label{fig:EventView0}
\end{center}
\end{figure}
Fig.~\ref{fig:EventView0} shows an example of the effect of varying subcones on peak templates compared to fixed ones. The blue, dotted circles represent the peak two body template with radius $r_2 = 0.3$. The red, dashed circles are peak three body templates with a fixed $r_3 = 0.12$ (left panel) and varying $r_3$ (right panel). Notice that a higher percentage of the lowest $p_T$ subject is ``encompassed'' by the varying cone resulting in an overall increase in the $Ov_3$ score.
\begin{figure}[htb]
\begin{center}
\begin{tabular}{cc}
 \includegraphics[width=3in]{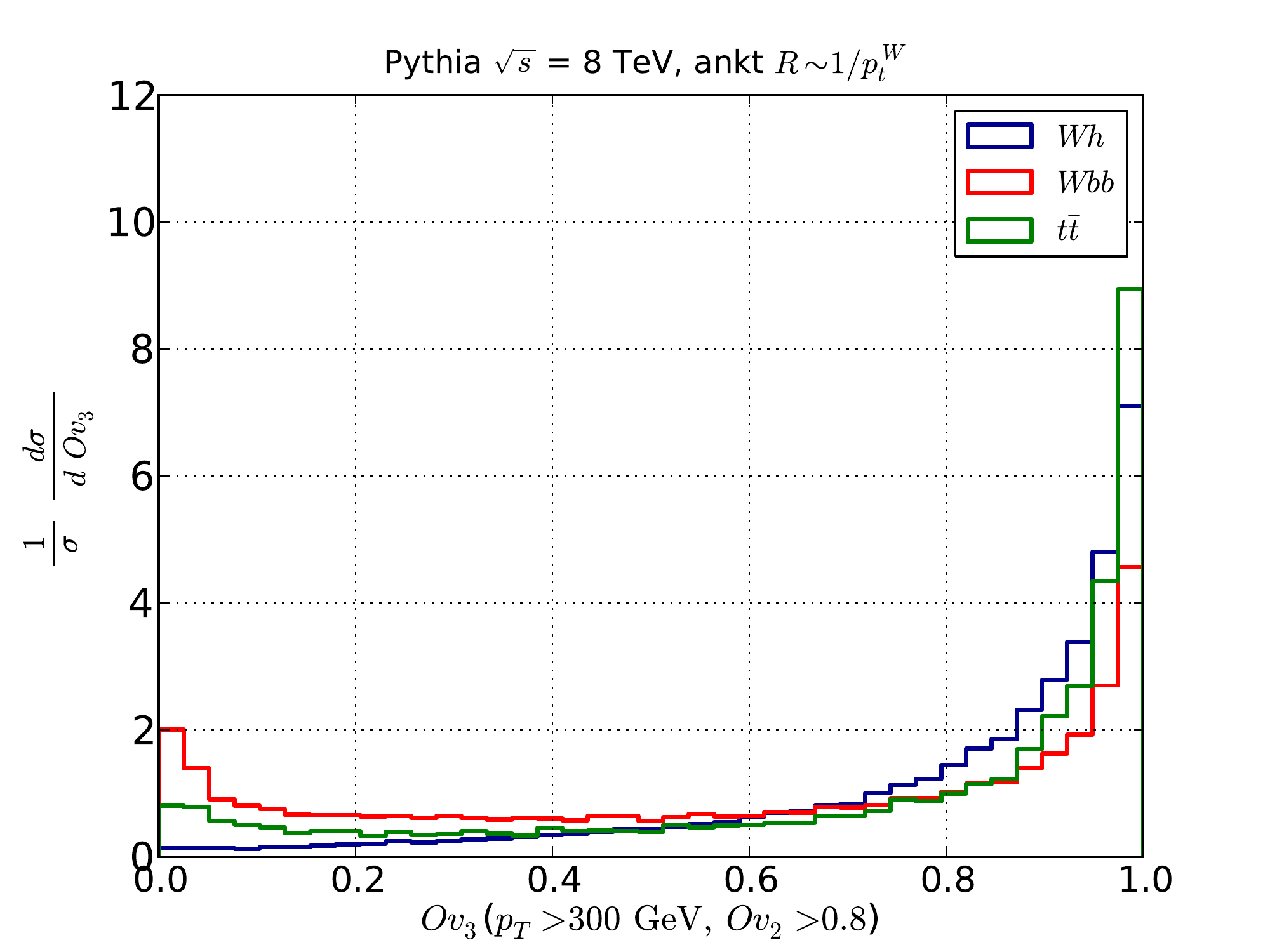}  & \includegraphics[width=3in]{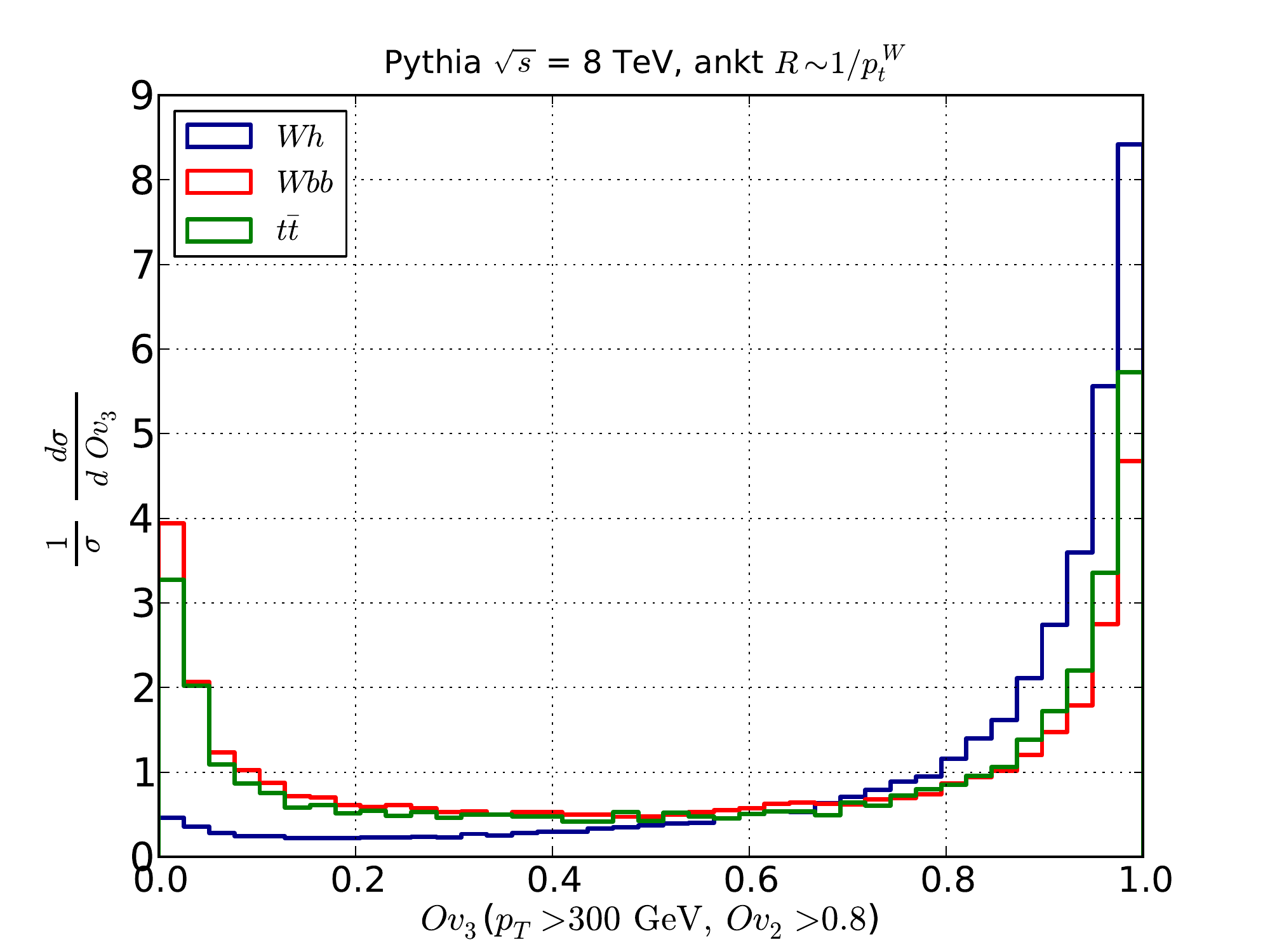} 
\end{tabular}
\caption{Template observable distributions for doubly-$b$-tagged events with $Ov_2 > 0.8$  and $b$-tagged 2-body templates ($r_2 = 0.3$) for $p_T^W > 300 \GeV$ and \textit{no mass cut}. Jets are reconstructed using {\sc FastJet}, and the anti-$k_T$ algorithm with a varying cone size $R$ (here denoted  ``ankt''. See more details on $R$ scaling below). The left panel represents distributions with $r_3 = 0.12$. The right panel represents the same distributions using the scaled subcones of Eq.~\eqref{eq:r3scale}. }
\label{fig:TempDists}
\end{center}
\end{figure}

Varying subcones improve the performance of templates on the distribution level as well.
 Fig.~\ref{fig:TempDists} shows a comparative example for $Ov_3$. The panel on the left was obtained using a fixed $r_3 = 0.12$ while $r_3$ was allowed to vary according to the scaling rule of Fig.~\ref{fig:ConeScaling} in the right panel. Notice that the varying subcones result in background overlap distributions which are significantly more peaked in the region of low overlap values. On the other hand the signal remains mildly affected, resulting in improved performance.

\subsection{Stability of the Template Overlap Method}

How sensitive is the Template Overlap Method rejection power to the choice of subcone radius $r_{\rm shape} \equiv r_3(100 GeV)$ and the overall normalization of the working curve of Fig.~\ref{fig:ConeScaling}?  
Significantly modifying the overall scale for subcone radius $r_{\rm shape}$ does alter the distribution of our kinematic variables; in particular, reducing the size of $r_{\rm shape}$ tends to shift $Ov_2$ and $Ov_3$ to lower values and vice versa.
The sensitivity of the template method to the specific choice of $r_{\rm shape}$ calls for a dedicated experimental study on a control sample to fix the corresponding value. One can use either boosted top analysis in the signal area (in fact the $Ov_3$ distribution of boosted tops and their backgrounds was already studied experimentally by ATLAS in Ref.~\cite{:2012qa}) or looking at hadronic $W$ inside a boosted top jet as a way to experimentally analyze the above dependence. The sensitivity to $r_{\rm shape}$ is of course not unique to our proposal, as even if fixed cones are used,  all jet substructure methods will depend on the choice of the corresponding  subcone parameters.
Varying the subcone size to keep the enclosed energy fraction fixed is in fact a more covariant way to proceed with the substructure analysis.
We further wish to point out that it is in general quite possible to choose values of cuts on $Ov_2$ and $Ov_3$ such that the overall signal and background efficiencies are essentially the same for different subcone radii. To illustrate this point, in Fig.~\ref{fig:stability} we show curves of rejection power for several choices of the subcone parameter $r_{\rm shape}$. The meaning of $r_{\rm shape}$ in Fig.~\ref{fig:stability} is defined as the choice of $r_3(100 \GeV)$, while the actual value of $r_3$ is allowed to vary according to the scaling rule in Fig.~\ref{fig:ConeScaling}. The blue squares represent rejection power against the $Wb\bar{b}$ background as a function of the three-body subcone radius $r_3 (100GeV)$, and a fixed $Ov_3 > 0.6$ cut, while the $Ov_3$ cut is allowed to scale with $r_{\rm shape}$ to provide a five percent efficiency for each $r_{\rm shape}$ . Signal and background efficiency of a fixed $Ov_3$ cut shift with the choice of $r_{\rm shape},$ but disproportionately.  Rigid $Ov_3$ 
cuts thus show high sensitivity to the choice of $r_{\rm shape}$ as shown by the blue squares in Fig.~\ref{fig:stability}. Alternatively, fixing signal efficiency also fixes the background efficiency for a wide range of $r_{\rm shape},$ thus preserving the rejection power over a wide range of $r_{\rm shape}$ ({\it i.e.} $r_{\rm shape } < 0.12$).
\begin{figure}[!h]
\begin{center}
 \includegraphics[width=4in]{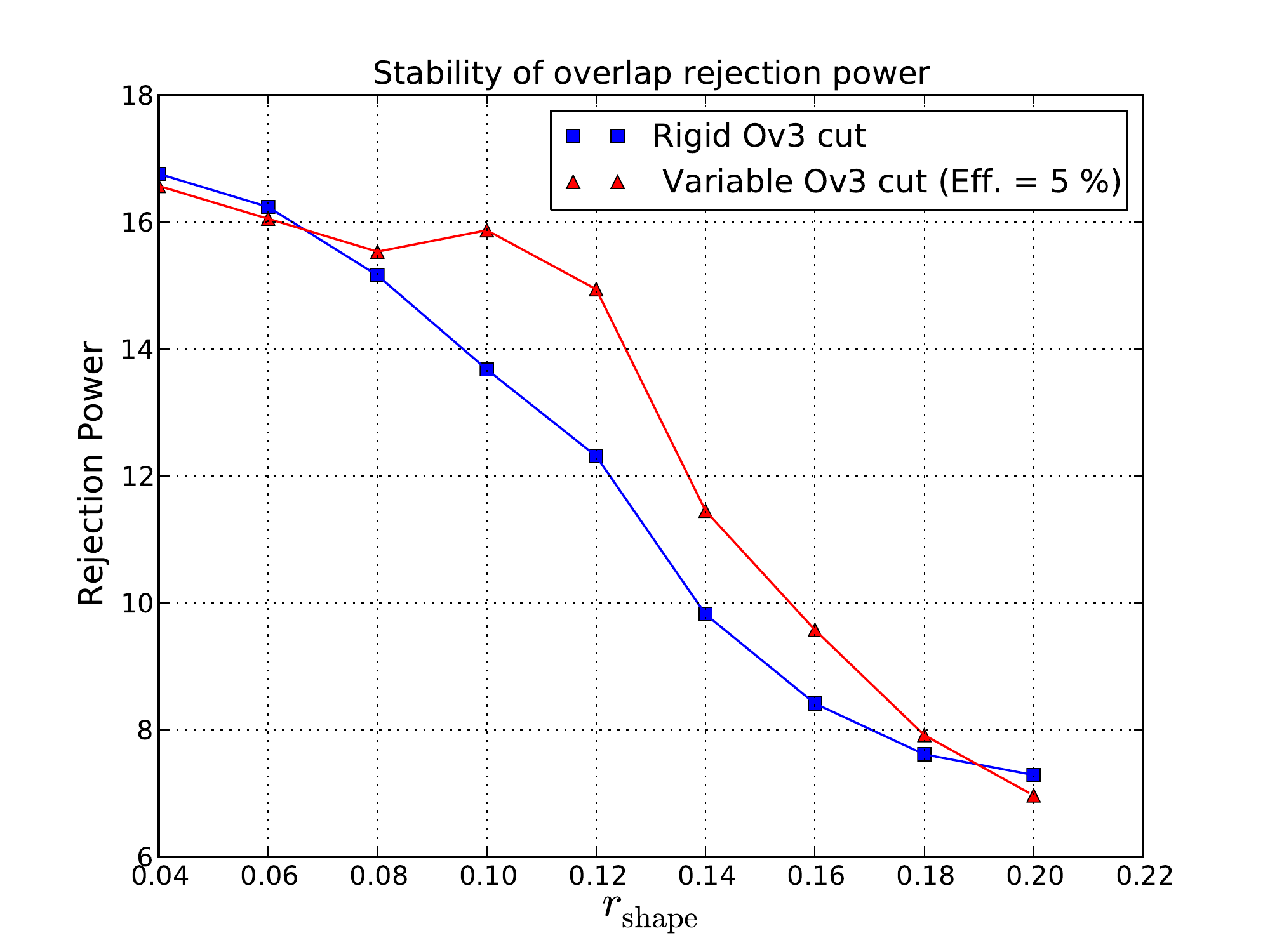}
\caption{ Sensitivity of $Wb \bar{b}$ rejection power to the choice of radius $r_{\rm shape} = r_3(100 \GeV).$ The blue line shows rejection power with a rigid $Ov_3 > 0.6$ cut. 
 The green line is the rejection power with a cut on $Ov_3$ varied with $r_{\rm shape}$, while keeping a fixed efficiency. The rejection power is relative to cross sections with no Basic Cuts (see next section).
}
\label{fig:stability}
\end{center}
\end{figure}

\section{Data Simulation and Analysis} \label{sec:data}
In this section, we investigate the tagging efficiencies for Higgs jets and the mistag rates for QCD backgrounds using template overlap. 
We consider events in which a boosted $H \rightarrow b \overline{b}$ jet is produced in association with a leptonically decaying $W$ boson (only first two generations of leptons). The most dominant backgrounds for this process come from $ t\overline{t}$ and $W b \overline{b}$, while other channels do not significantly contribute after $b$-tagging requirement.
\begin{figure}[!]
\begin{center}
\begin{tabular}{cc}
\includegraphics[width=3in]{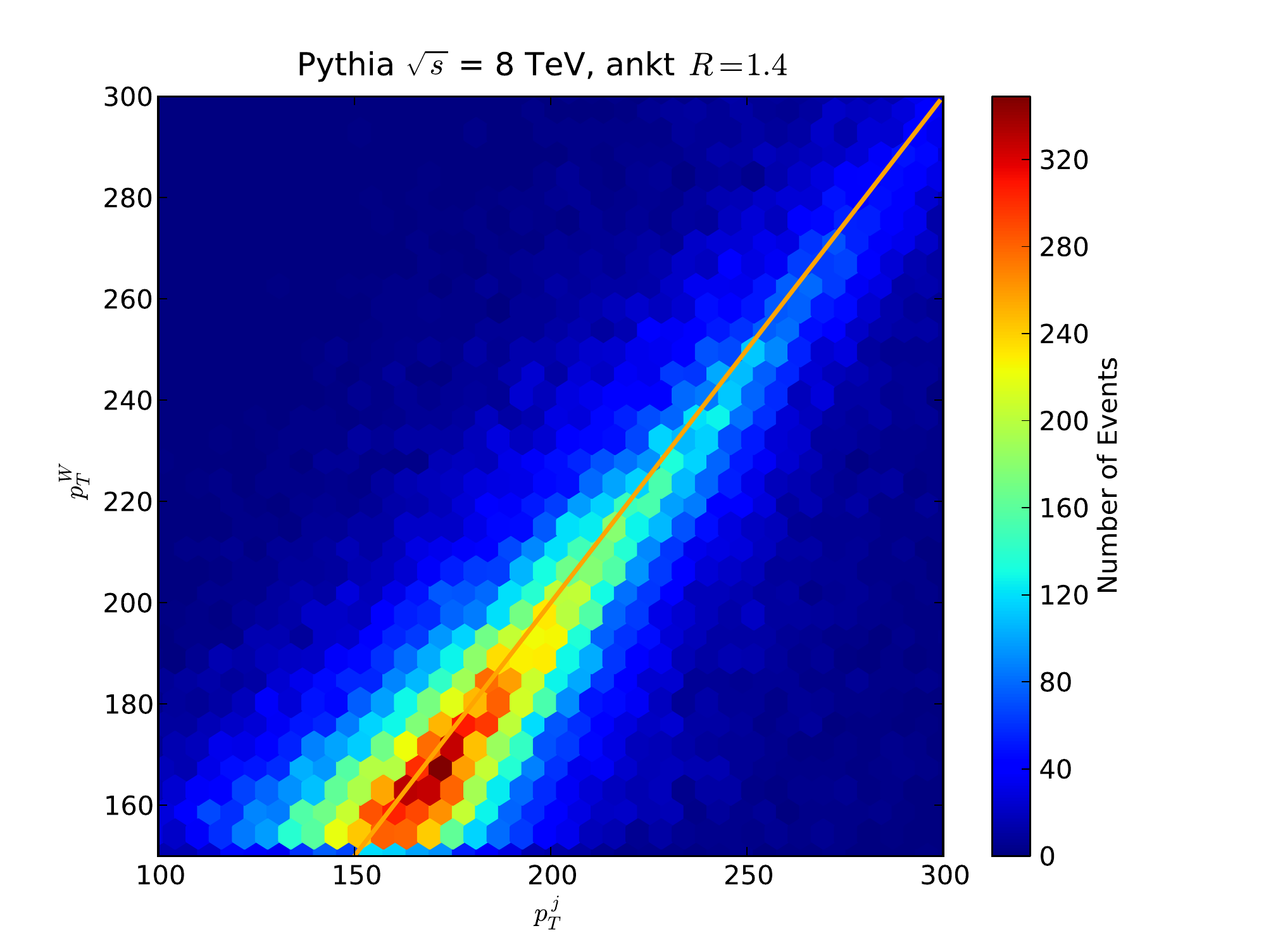} & \includegraphics[width=3in]{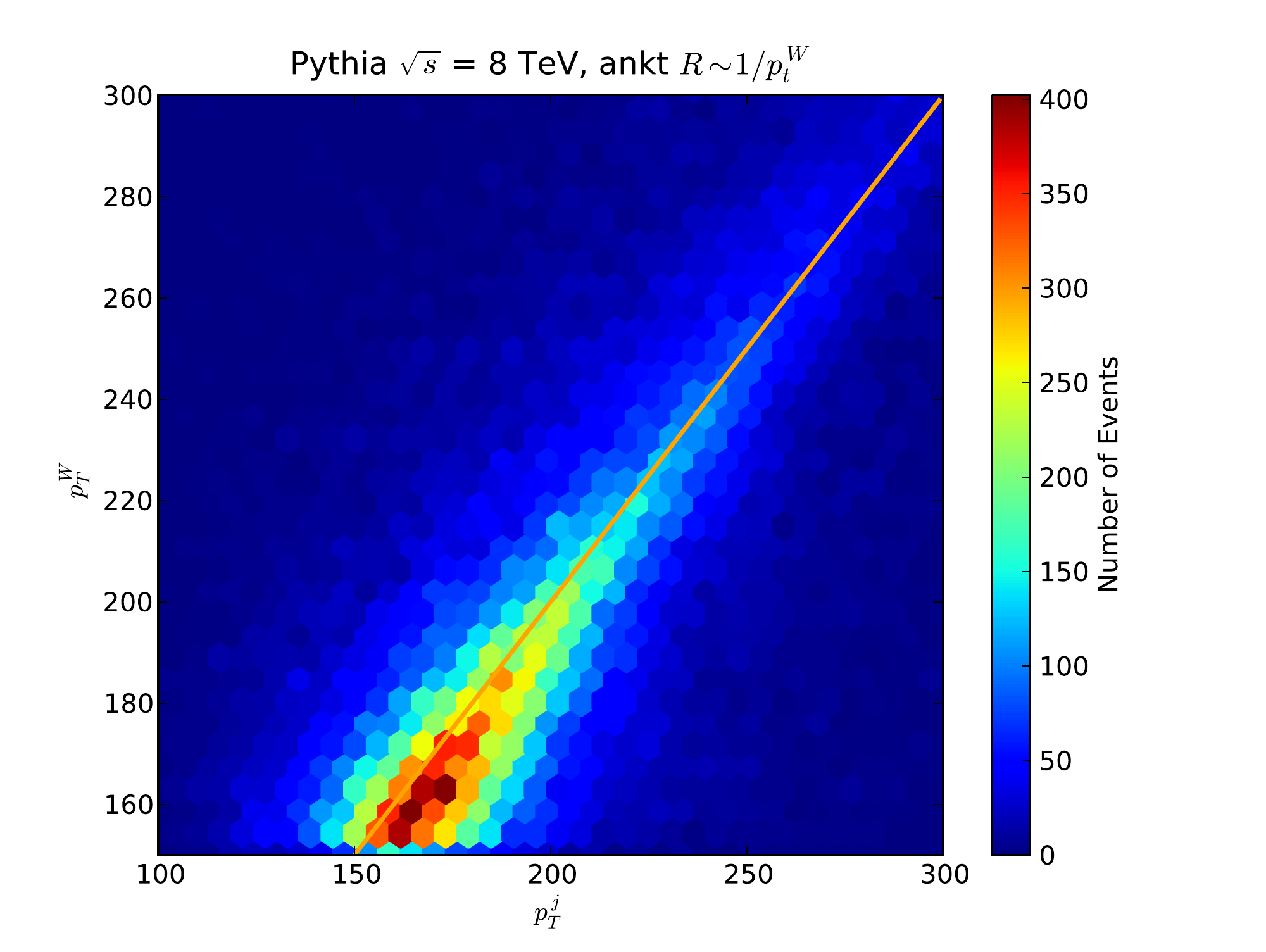}
\end{tabular}
\caption{The left panel shows the correlation between the Higgs-jet $p_T$ and the $p_T$ of the associated $W$ for a cone of fixed size $R=1.4$. The right panel shows the same correlation with the jet cone rescaled according to $p_T^W$ as in Eq.~\eqref{eq:scale}.  }
\label{fig:jetPtwPt}
\end{center}
\end{figure}

The scope of our analysis includes data at both $\sqrt{s} = 8\TeV$ and $\sqrt{s} = 13\,\TeV$  with and without pileup. We use {\sc MG/ME v5.1.33}~\cite{Maltoni:2002qb} interfaced to {\sc Pythia v6}~\cite{Sjostrand:2006za} (with MLM matching~\cite{Mangano:2006rw})  as well as {\sc Sherpa v1.4.0}~\cite{Gleisberg:2008ta} (with CKKW matching~\cite{Hoeche:2009rj})  for data generation and the CTEQ6L1 ~\cite{Pumplin:2002vw} parton distribution functions.  We perform jet clustering with a \!{ \sc Fastjet}~\cite{Cacciari:2008gp} implementation of the anti-$k_T$ jet clustering algorithm. 
To simplify the notation from here on we solely refer to {\sc Pythia} and {\sc Sherpa} results and leave the fact that all our data samples are matched implicit.
Our {\sc Sherpa} simulations serve to illustrate the effects of various showering algorithms. We only include the $Wh$ and $Wb \bar{b} $ data in the comparison, as the $t\bar t$ distributions are characterized by hard scales and therefore less sensitive to detail of the showering. 
We scale the fat anti-$k_T$ jet radius according to the $W$ momenta \be
	R = \max \left(1.4\, \frac{200 \GeV}{p_T^W}, 0.8\right) \label{eq:scale}, 
\ee
where $p_T^W$ is the transverse momentum of $p^l + p_T\!\!\! \!\!\!\slash \ \, \,. \, $ 
Note that the momentum of the $W$ is highly correlated with the momentum of the Higgs as they recoil against each other. However, scaling the cone according to $p_T^W$  has an advantage in that it is not susceptible to pileup.

Continuing, we limit the value of $R$ from below to be higher than 0.8 as to be able to accommodate two 0.4 anti-$k_T$ jets used for $b$-tagging. The scaled fat-jet cone fulfills three tasks; it is designed to capture the $b\bar b$ at a fixed efficiency rate of $\sim 80$\%\,; it reduces the amount of contamination from soft radiation and pileup for events with high $p_T$ Higgs jets; and it lowers the overall $t \overline{t}$ background.  Note that the scaling rule of Eq.~\eqref{eq:scale} has only a minor effect on the correlation between the Higgs fat jet momenta and that of the $W$ as is shown in Fig.~\ref{fig:jetPtwPt}. 

Next,we normalize our data to next-to-leading-order (NLO) cross sections obtained from {\sc MCFM 6.3} \cite{Campbell:2010ff}. The cross sections assume a fixed renormalization/factorization scale of $\mu = p_T^{min} = 300 \GeV$ at $8 \TeV$ and $\mu = p_T^{min} = 400 \GeV$ at $13 \TeV$ and {\sc CTEQ6.6M} \cite{Nadolsky:2008zw} parton distribution functions. For each event, we
find the jet with the highest transverse momentum $j$ and impose the following \textit{Basic Cuts}:
\ba
	p_T^{j}  > p_T^{min}\,, & \,\,\,\,\,\,\,\,\,\,\,\,\,\,p_T^W >  p_T^{min}\,, \nn\\
	\eta_j,\eta_l <  2.5\,,  & p_T \!\!\! \!\!   \!\! \slash \ \, > 40 \, \GeV\,, \nn\\
	N_{b} = 2\,,  & \Delta R_{bb} \ge 0.4\,, \nn\\
	N_{k}(p_T > 20\, \GeV)  < 2\,, &\,\,\,\,\,\,\,\,\,\,\,\,\,\,\,\,\,\,\,\,\,\,\,\,\,\,\,\,  N_{l}(p_T > 20\, \GeV)  =  1\,,
	\label{eq:BasicCuts}
\ea
where $N_{k, l}$ is the number of jets (anti-$k_T$, $r=0.4$) and leptons outside the highest $p_T$ fat jet (of radius $R$) and $N_b$ is the required number of $b$-tagged (anti-$k_T$ $r=0.4$) subjets. For a fat anti-$k_T$ jet $j$ (of radius $R$),  and an anti-$k_T$, $r =0.4$ jet $k$, a jet is considered to be outside the fat jet if the plain distance $\Delta R(j, k ) > R + r$. Similarly, a lepton $l$ is considered outside if $ \Delta R (j , l) > R$. Table \ref{sigma_table} summarizes the cross section results with and without Basic Cuts.

We consider $p^{min}_T = 300, 350 \GeV$ respectively at $8 \TeV$ and $13 \TeV$.  For $b$-tagging we assume an efficiency of $75\%$ and fake rate of $1\%$ for light jets~\cite{Aad:2009wy}. The current studies suggest a charm fake rate
of $18\%$~\cite{Aad:2009wy}, which is likely a conservative estimate. Charms are extremely important when considering boosted Higgs decays, as the largest part of the $t\bar t$ background comes from events in which one top decays leptonically, while the hadronic $W$ from the other top decays to a charm. 
We emphasize that omitting the charms as a source of background (as is done in some of the boosted Higgs analyses) will result in an improved performance for our tagger. Yet, at present, it is not clear whether this is possible, and the burden of proof is thus placed on the experimental collaborations.

We analyze the cases with and without pileup separately in order to illustrate the sensitivity of the Template Overlap Method to a pileup environment. This allows us to determine the range of background rejection power as a function of the efficiency of pileup subtraction. In addition, it also allows for a comparative study of the various jet substructure observables in a pileup environment. 
\begin{center}
\begin{table}[htb]
\begin{tabular}{|c|c|c|c|c|}
	\hline
 fb	 & $t\bar{t}$ &$Wb\bar{b}$ & $Wh$ & $S/B$ \nn \\
	 \hline
$\sigma( \, \sqrt{s} = 8\TeV ,\,  p_T^W > 300 \GeV )$ & 565.0 & 56.0 & 1.6 & \nn \\
$\sigma( \, \sqrt{s} = 8\TeV ,\,  {\rm Basic \, Cuts} )$ & 2.0 & 2.5 & 0.2 &  0.05 \nn \\ \hline
$\sigma( \, \sqrt{s} = 13\TeV ,\,  p_T^W > 350 \GeV )$ & 956.0 & 47.0 & 1.2& \nn \\ 
$\sigma( \, \sqrt{s} = 13\TeV ,\, {\rm Basic \, Cuts })$ & 3.0 & 1.7&  0.3 & 0.06 \nn\\   
\hline
\end{tabular}

\caption{NLO signal and background cross sections at $\sqrt{s} = 8 \,\TeV$ and $\sqrt{s} = 13 \,\TeV$. The listed numbers assume a leptonically decaying $W^{\pm}$ with first 2 generations of  leptons included. Basic Cuts include Eq.~\eqref{eq:BasicCuts} as well as the $b$-tagging efficiencies.} 
\label{sigma_table}
\end{table}
\end{center}

\subsection{Higgs Tagging with Template Overlap - No Pileup} 

We proceed to discuss the ability of the Template Overlap Method to discriminate between different sources of coherent QCD radiation.

In terms of pileup filtering, analysis without pileup is equivalent to stating that the efficiency of pileup subtraction is $100 \%.$ In this section we present only the results on jets with $p_T > 300 \GeV,$ simulated at $\sqrt{s} = 8 \,\TeV,$ while we postpone the discussion of the future $13 \,\TeV$ LHC run until upcoming sections. 
The first important feature of the Template Overlap Method is that it is designed to identify a particular kinematic jet substructure configuration, including the jet $p_T$ and mass. High peak overlap score implies that the kinematics of a fat jet matches the kinematics of the peak template state.
In Fig.~\ref{fig:JetMassCuts} we plot the jet mass distribution without (left panel) and with (right panel) template overlap cuts of $Ov_2 > 0.9$ and $Ov_3 > 0.8$, after the Basic Cuts of Eq.~\eqref{eq:BasicCuts} have been applied. It is evident from Fig.~\ref{fig:JetMassCuts} that sizable chunk of the background is removed as a result of the overlap cuts though
the resolution of the fat jet Higgs mass is only moderately improved.

\begin{figure}[htb]
\begin{center}
\begin{tabular}{cc}
\includegraphics[width=3in]{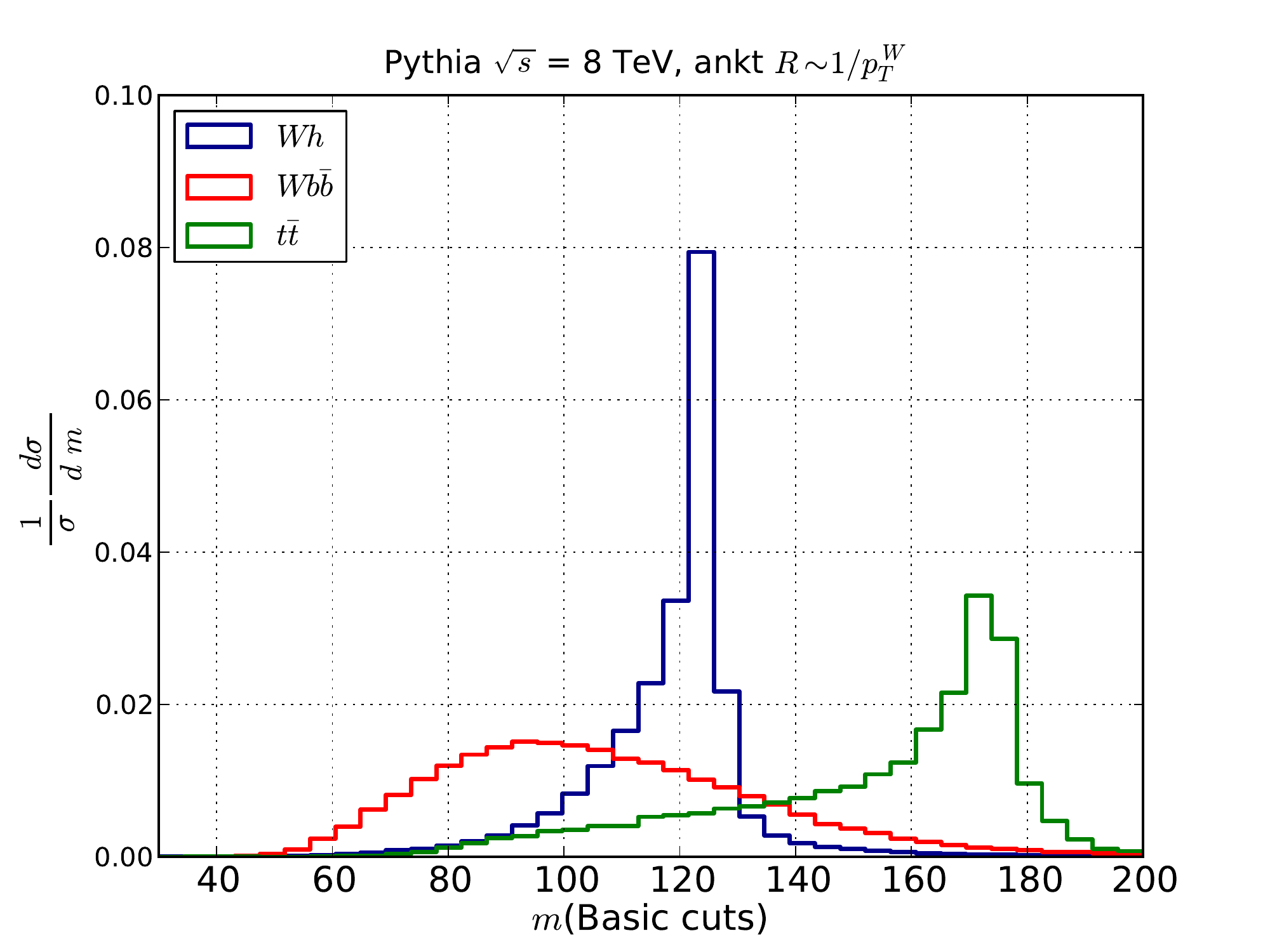}  & \includegraphics[width=3in]{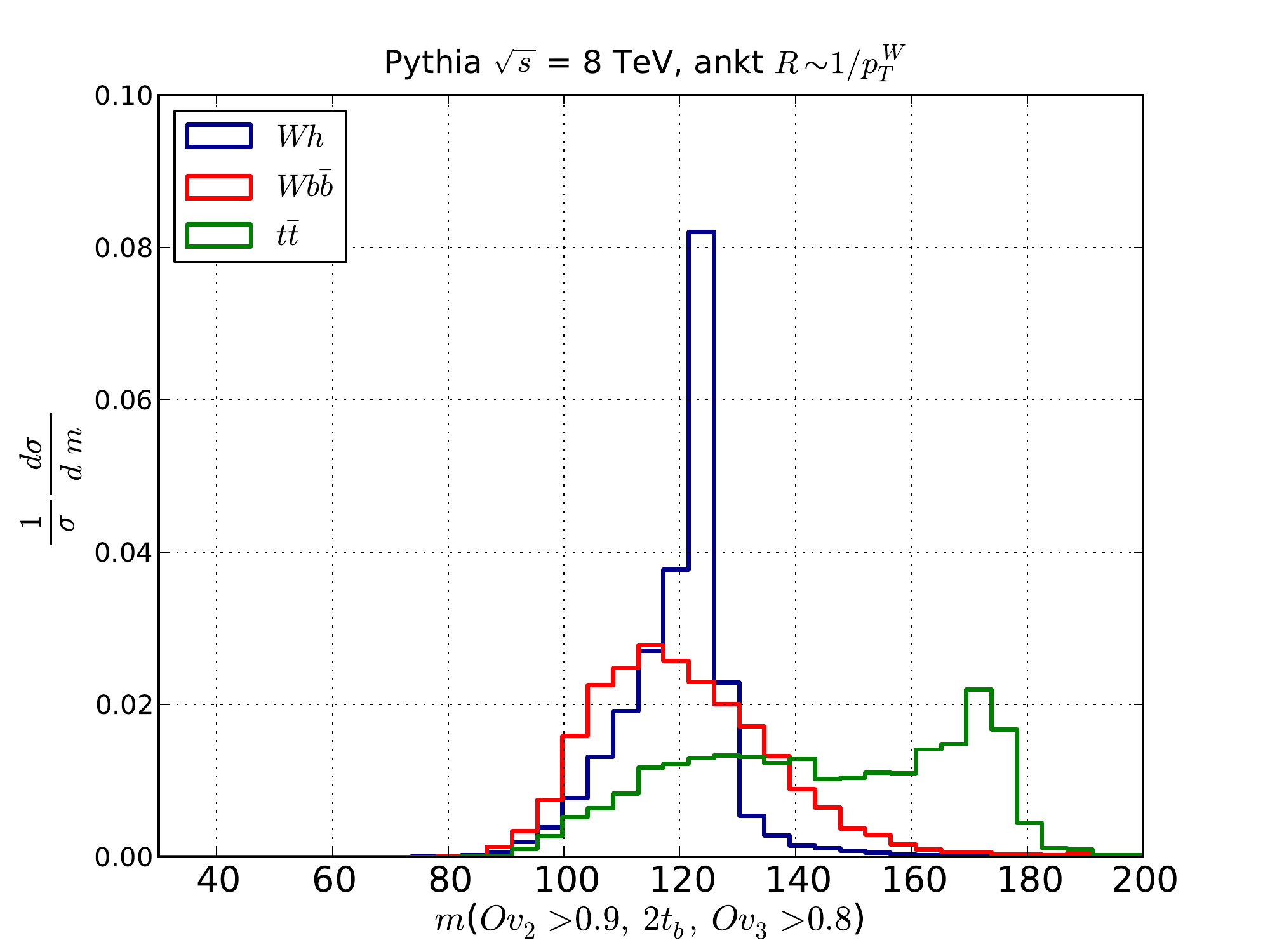}  
\end{tabular}
\caption{Invariant-mass peak searches with templates. The left panel shows the mass distributions with no cuts on template observables. The right shows the same distributions when overlap cuts are applied to the data: $Ov_2 > 0.9$ and $Ov_3 > 0.8$.}
\label{fig:JetMassCuts}
\end{center}
\end{figure}

The mass resolution of the peak templates depends largely on the chosen parameters of the method, namely the template cone radii $r_a$ and their energy resolution $\sigma_a$ which we choose rather loosely as to keep the signal efficiency at a reasonable level.  Recently, authors of Ref.~\cite{:2012qa} presented a jet substructure analysis of boosted $t\bar{t}$ pairs at ATLAS. Their results showed that even with a high peak overlap cut, an additional mass window improved the background rejection power by a factor of two. Our result in Fig.~\ref{fig:JetMassCuts} agrees with the ATLAS result. It appears that even after the overlap cuts, a mass window of (say) $110 \,\GeV < m_j < 130 \, \GeV$ would improve the background rejection power (this would, however,  require an additional procedure of pileup removal). In the following sections we remain agnostic about this issue and show results with an without a mass cut.

Fig.~\ref{fig:TemplateDists} shows distributions of several template-inspired observables obtained from both the  {\sc Pythia} and {\sc Sherpa} data.
Since our focus is on the difference in the shapes of various observables, all of the kinematic distributions are shown after cuts slightly different from the ones of Eq.~\eqref{eq:cuts}.  
 Ref.~\cite{Almeida:2011aa} showed that at very high $p_T$ (say above the TeV scale), $\bar{\theta}$ displays a sizable rejection power. Our result shows that at lower $p_T$, in the region where small $R$ approximation does not hold, the background discriminating power of $\bar{\theta}$ is highly diminished. In addition to $Ov_3$,  $\tPf$ and especially $\Delta R_{bb} / \Delta R_{t} $ appear to be promising variables.  We discuss $\tPf$ in more detail in Appendix \ref{app:PF}.

\begin{figure}[htb]
\begin{center}
\begin{tabular}{cc} 
\includegraphics[width=2.3in]{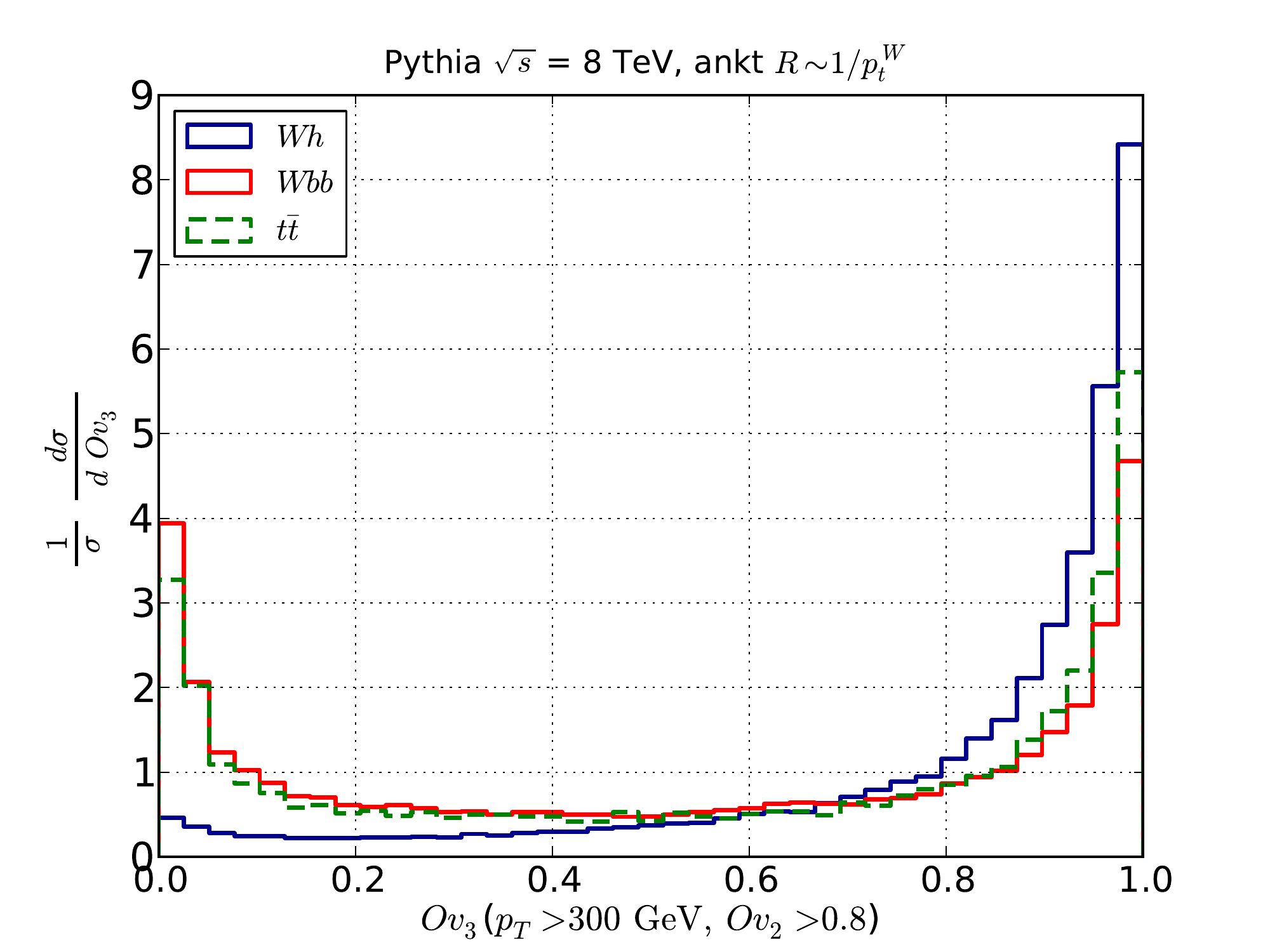}  & \includegraphics[width=2.3in]{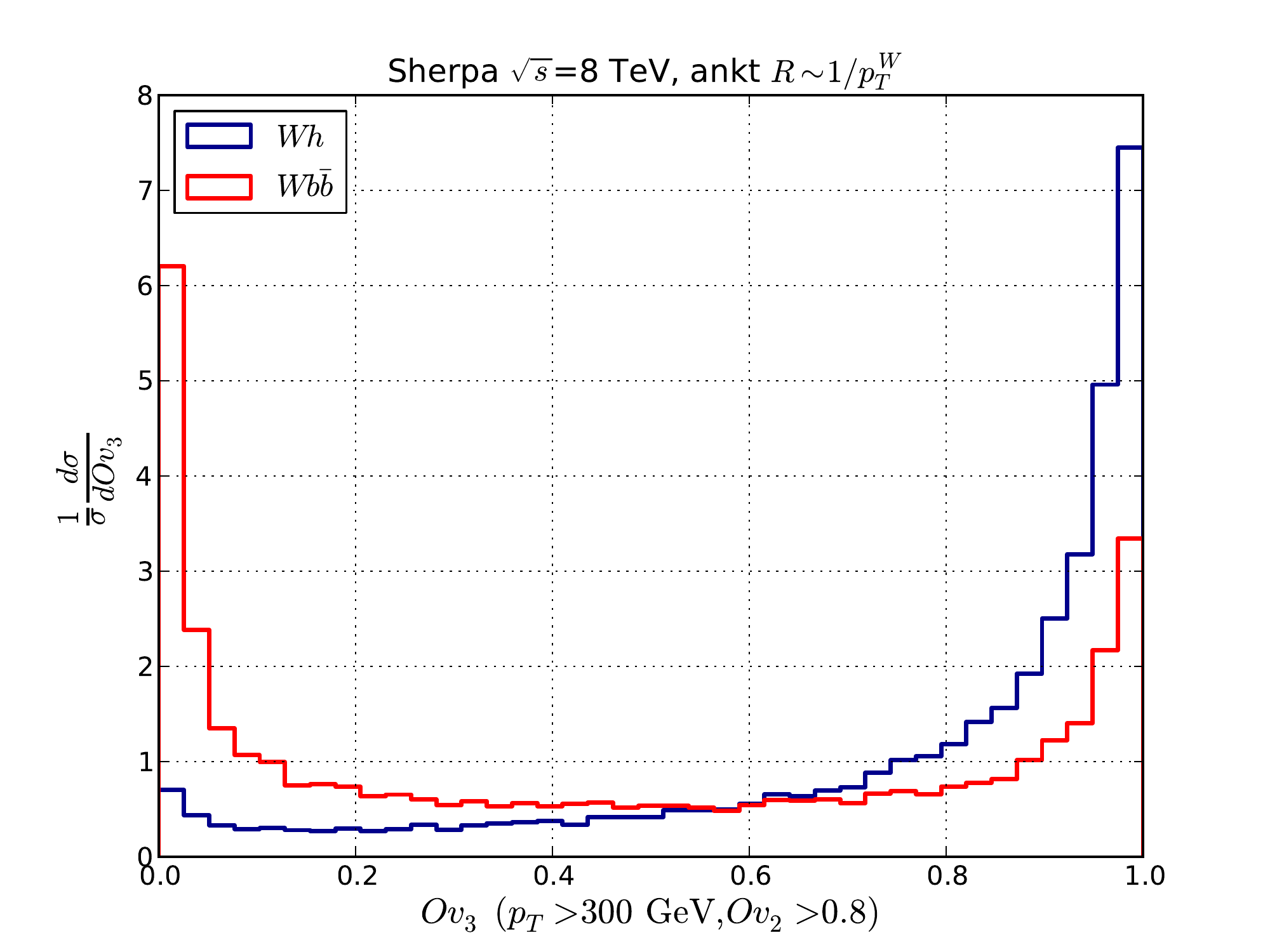} \\    
\includegraphics[width=2.3in]{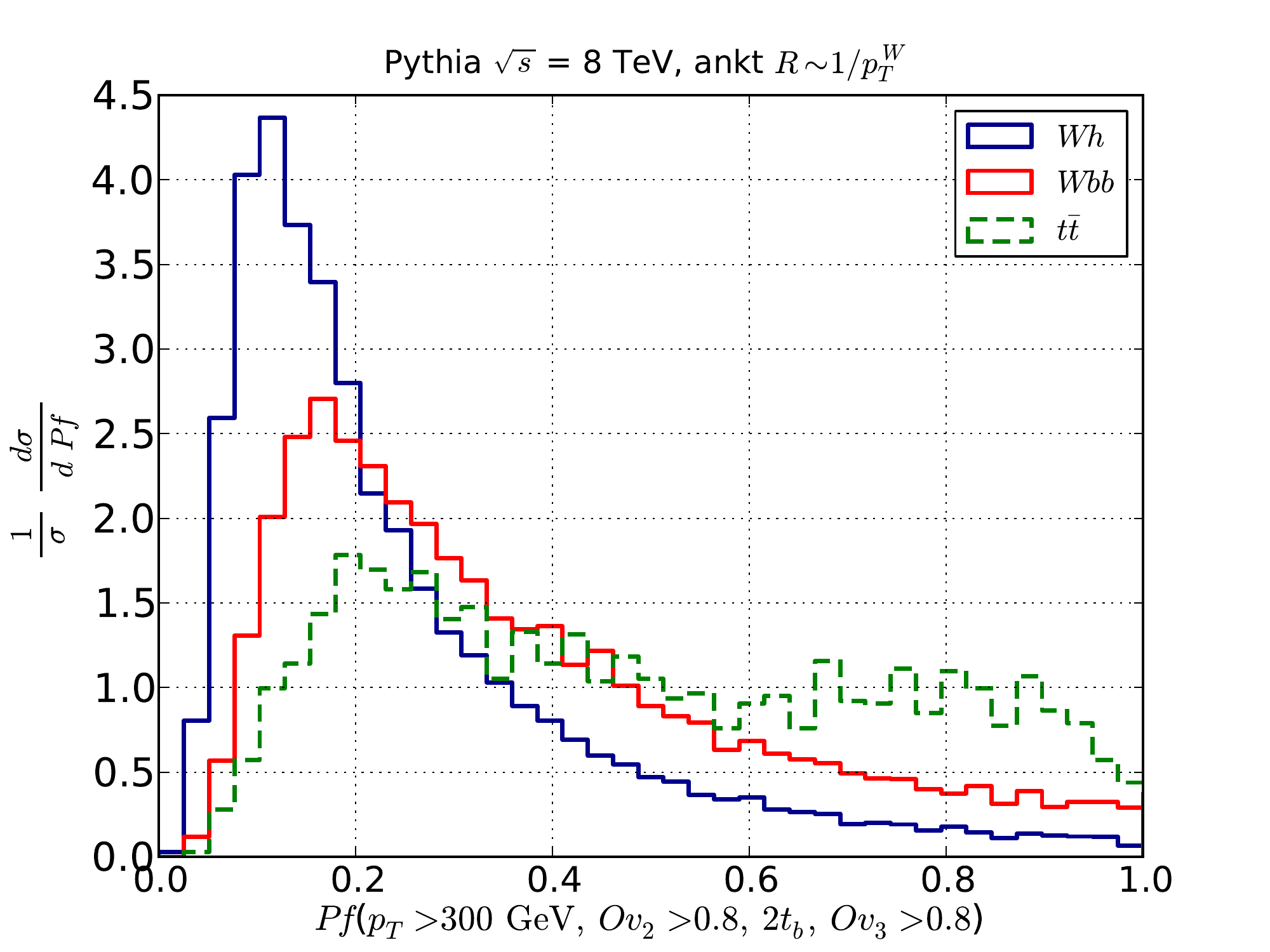}  &\includegraphics[width=2.3in]{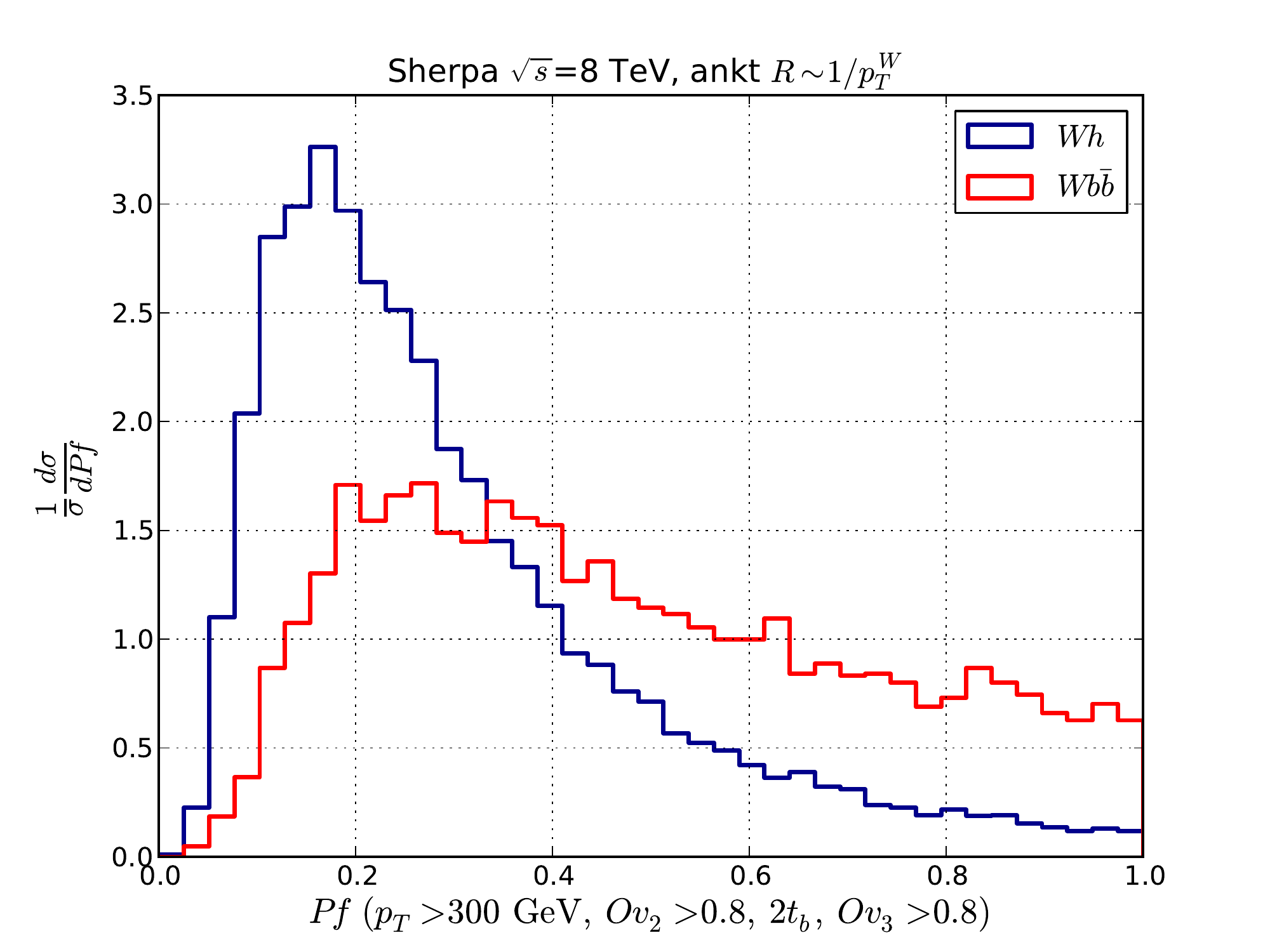}\\
\includegraphics[width=2.3in]{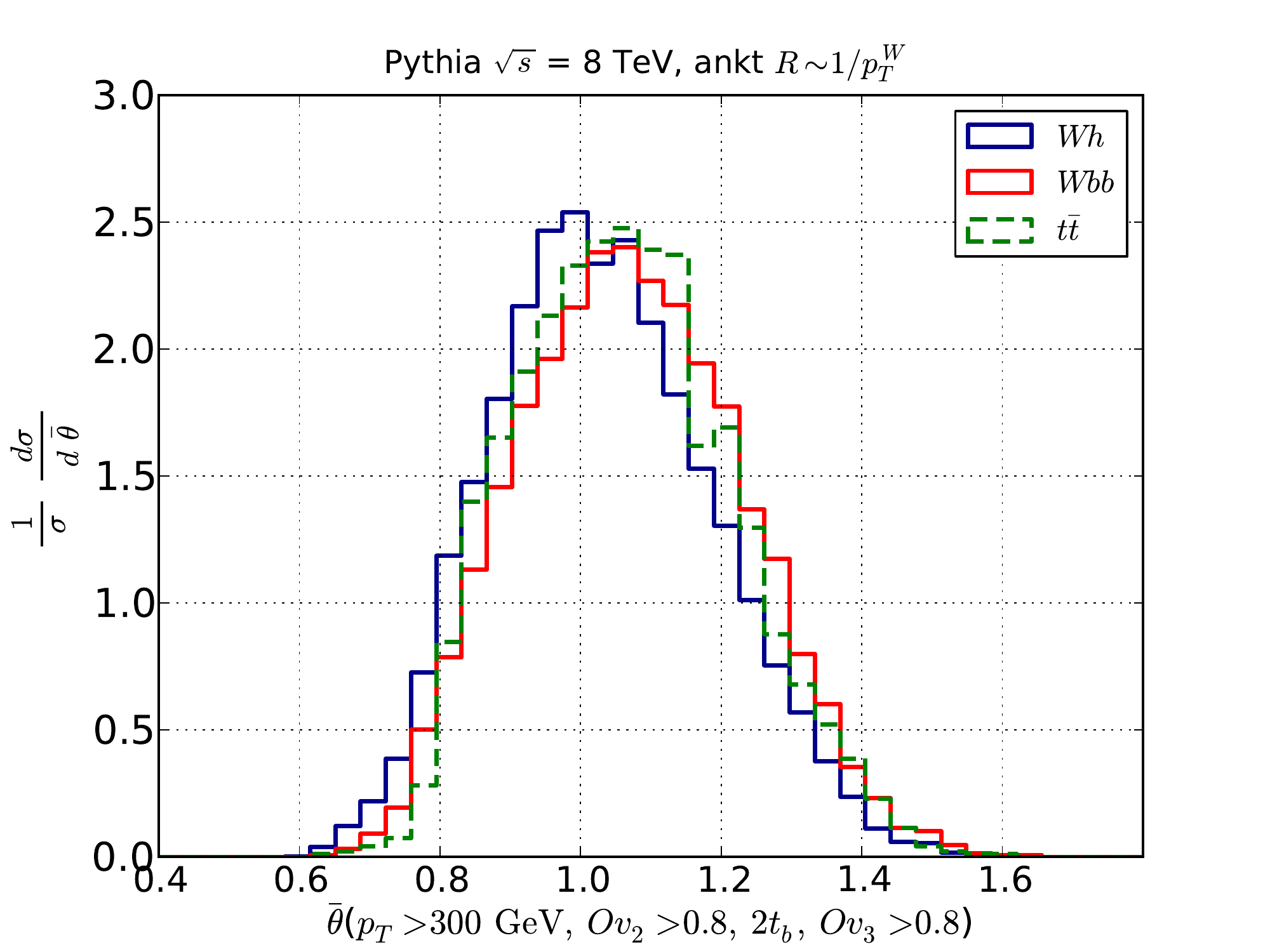}  &\includegraphics[width=2.3in]{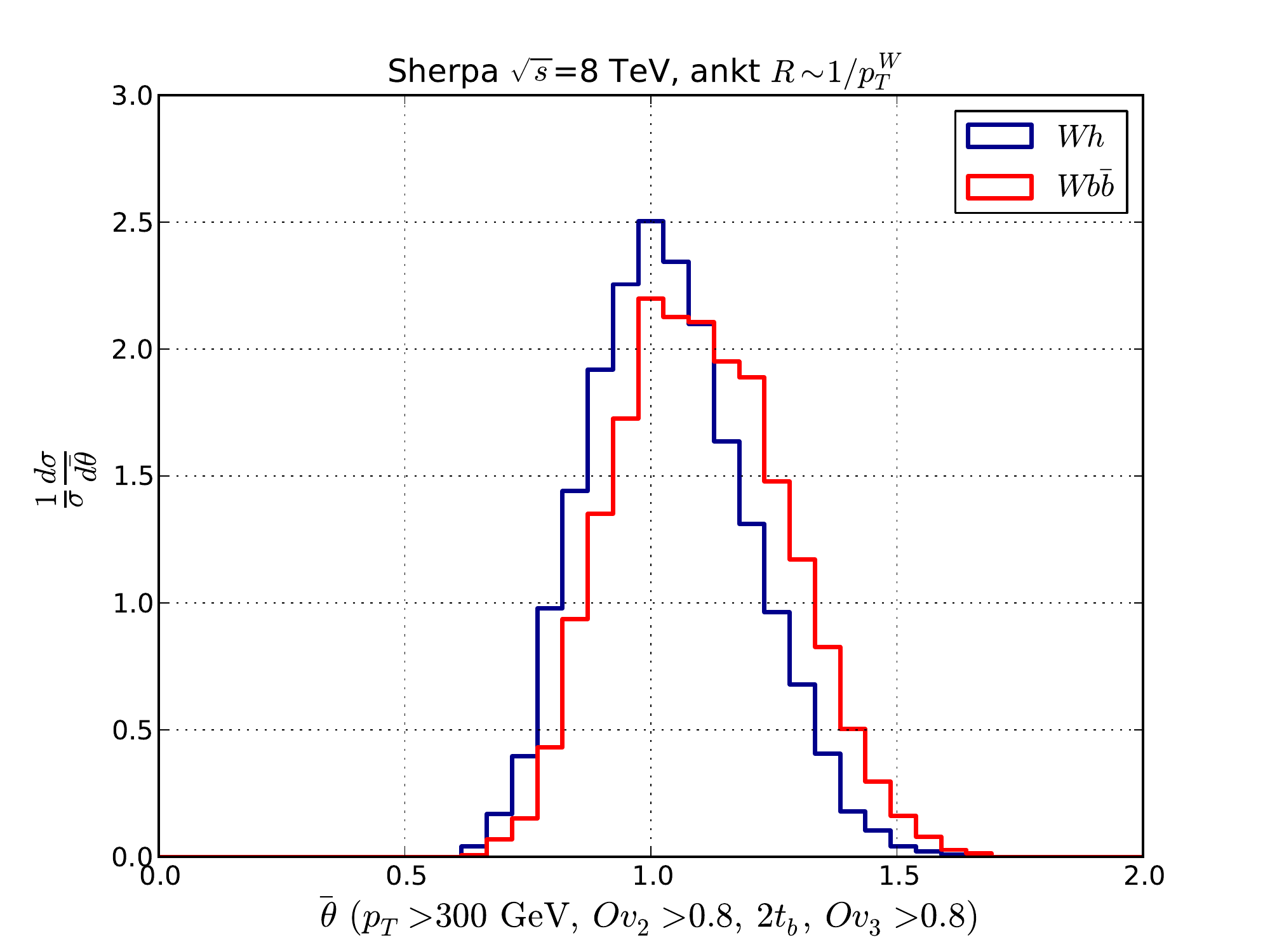} \\
\includegraphics[width=2.3in]{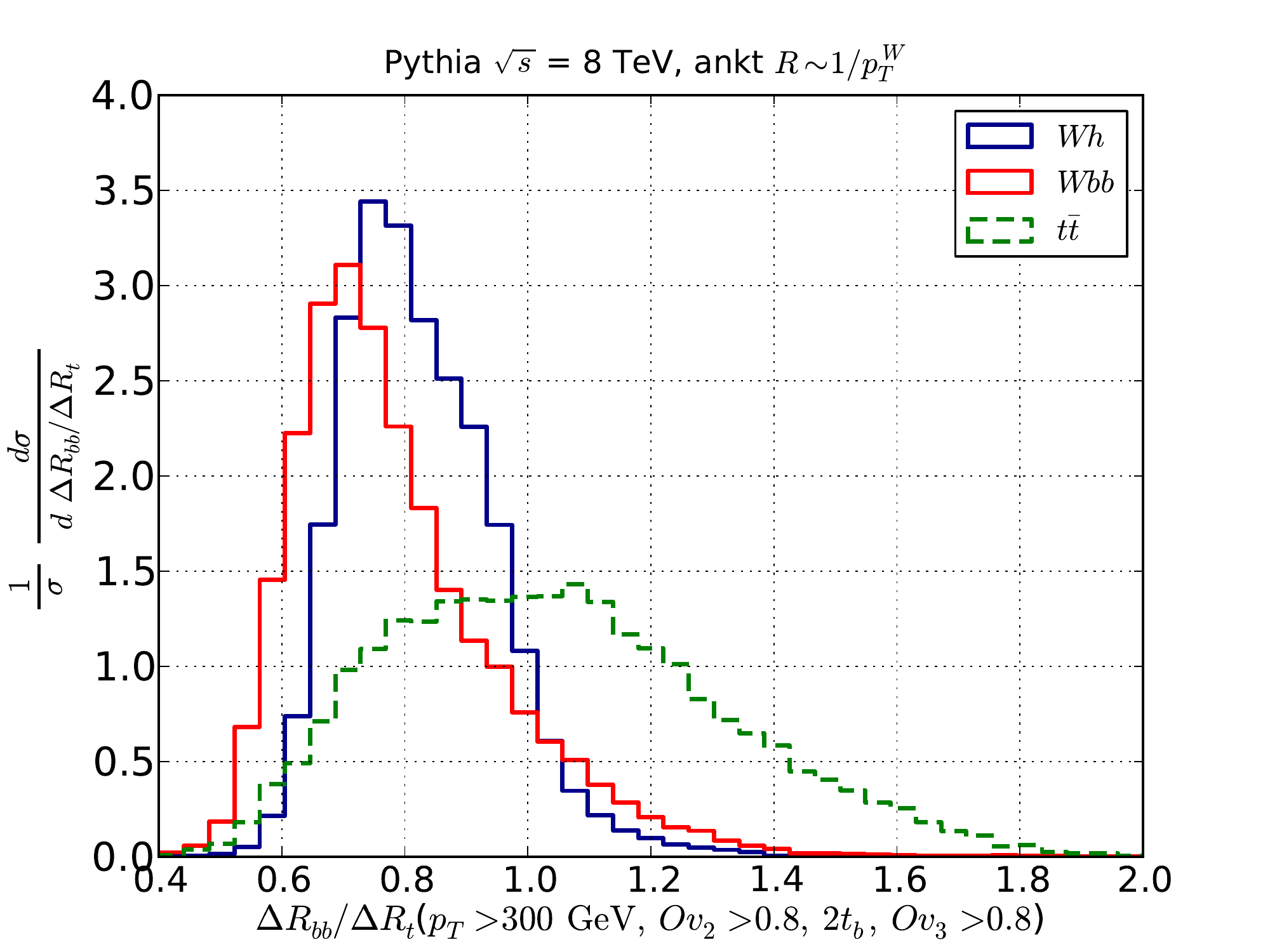}  &\includegraphics[width=2.3in]{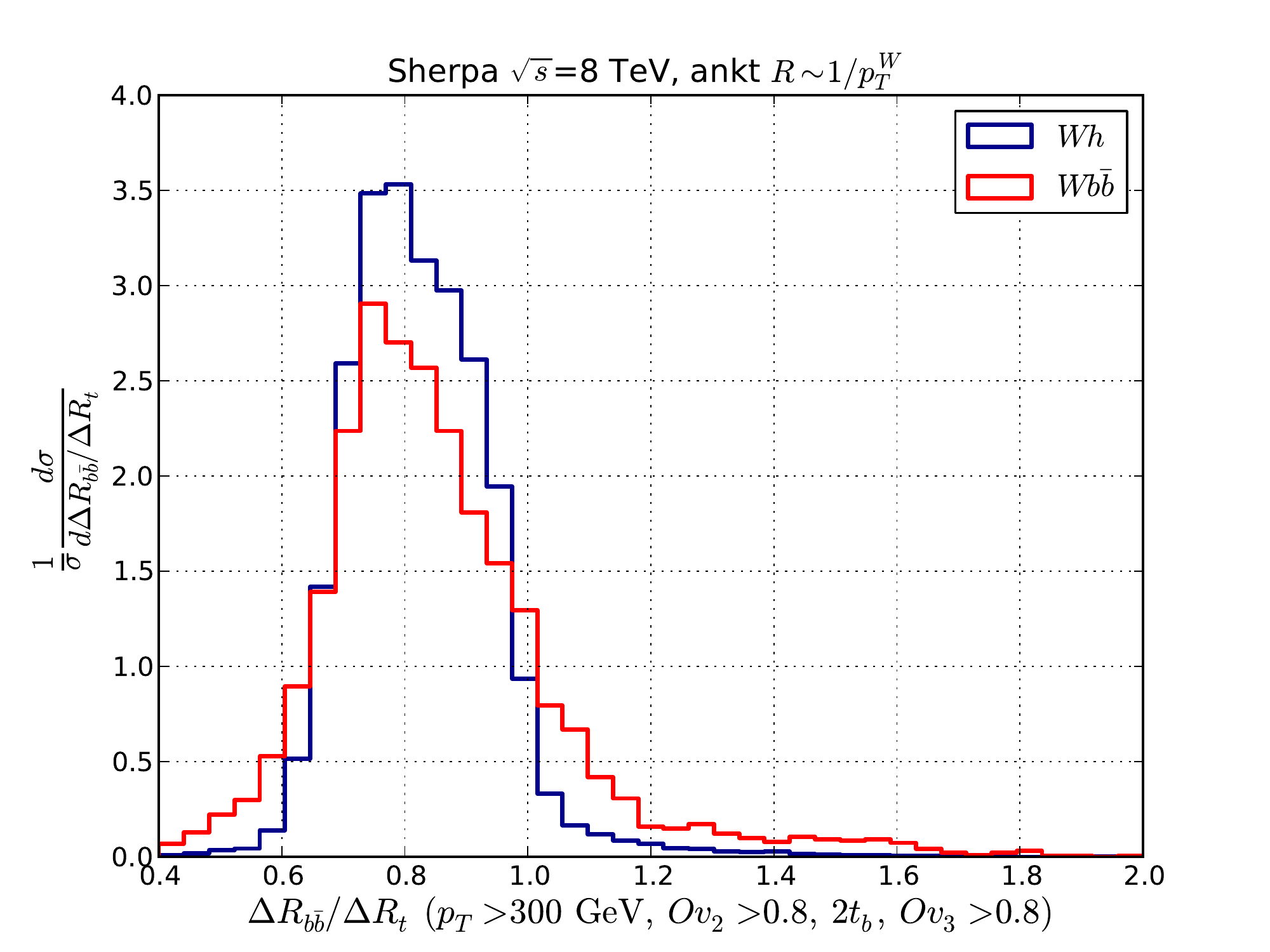}\\
\includegraphics[width=2.3in]{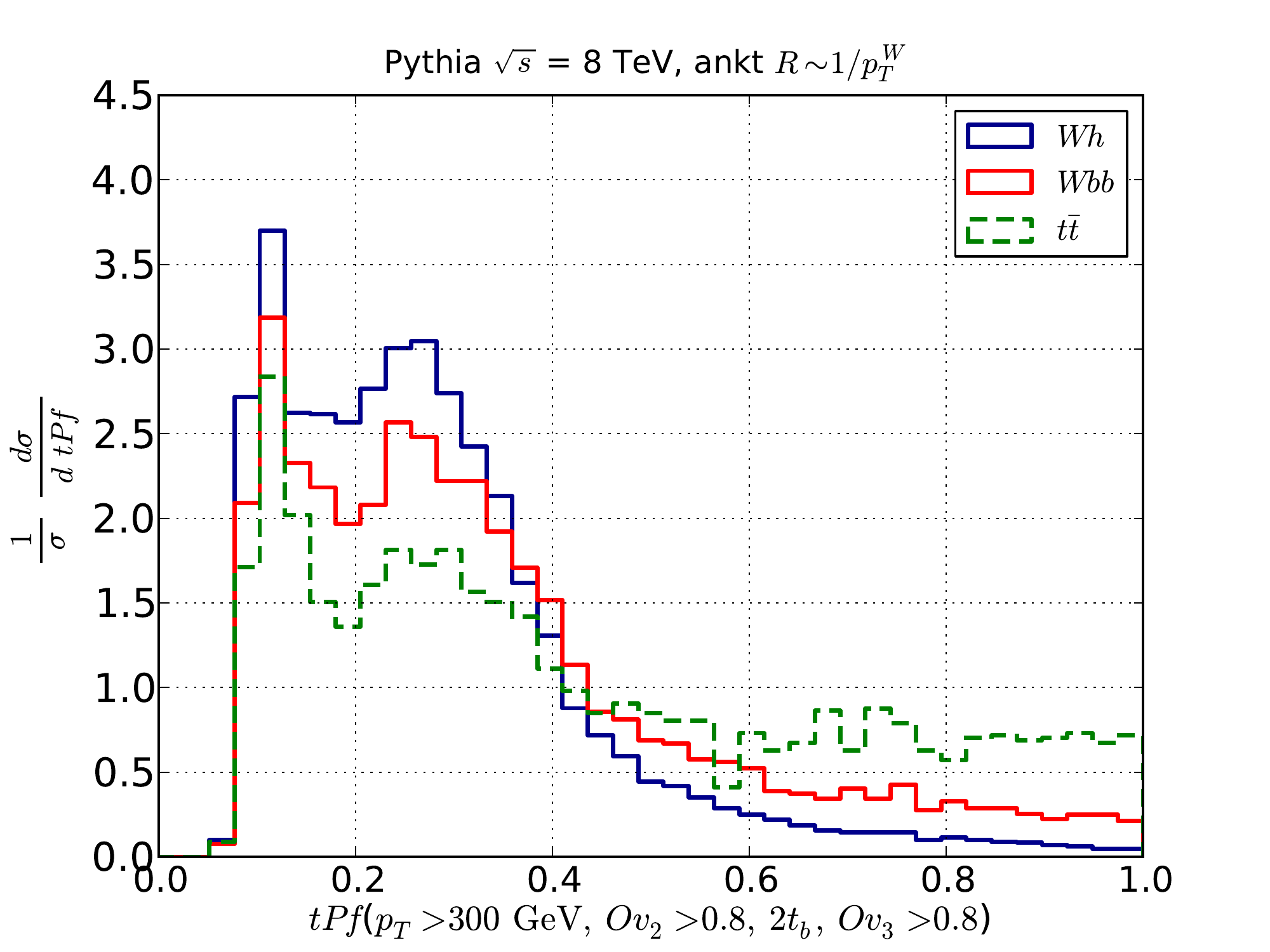}  &\includegraphics[width=2.3in]{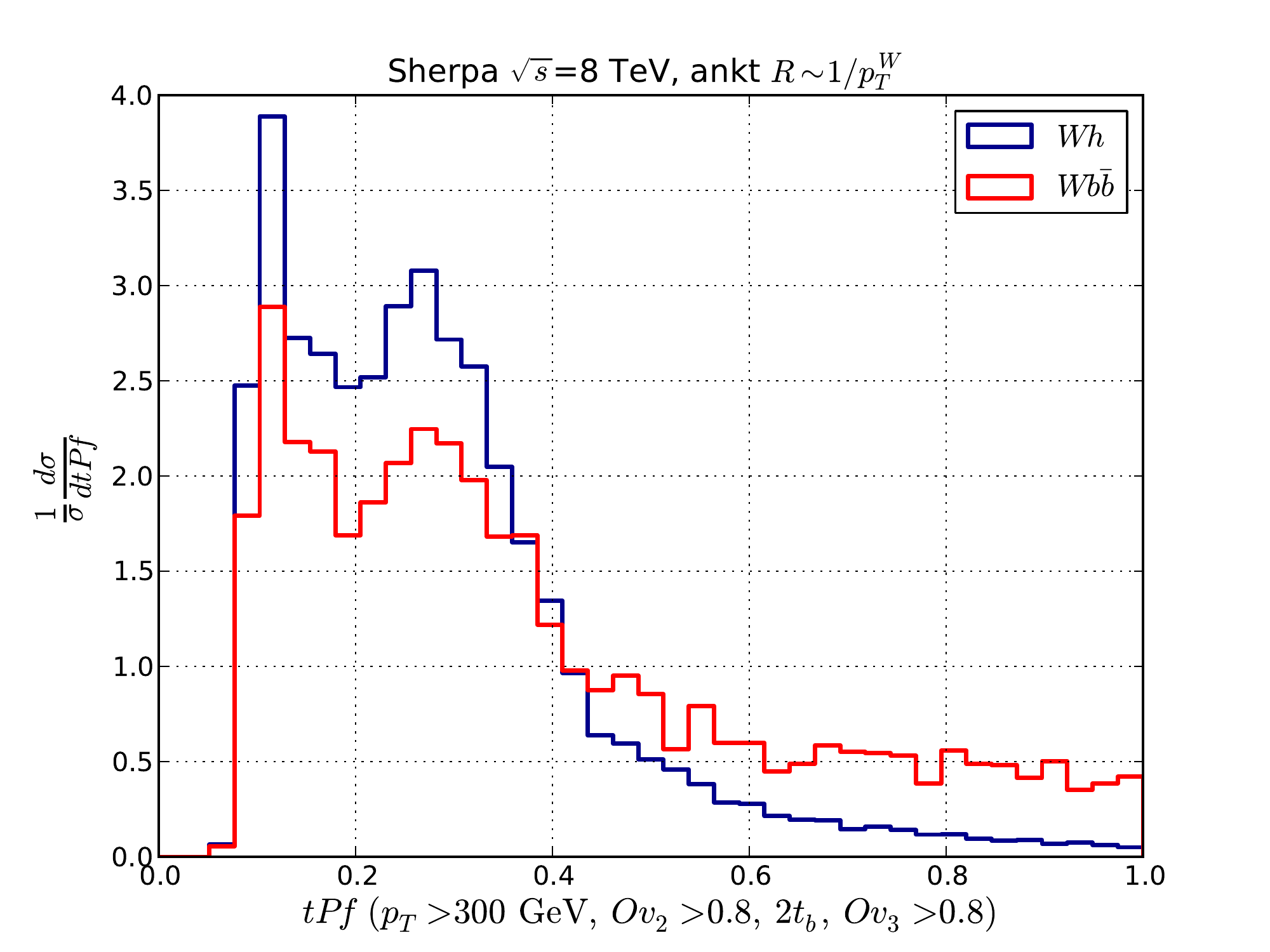} \\

\end{tabular}
\caption{Distributions of various substructure and template-based observables obtained from {\sc Pythia} (left) and {\sc Sherpa} (right). Basic Cuts are applied to all panels in addition to the cuts specified in the label.}
\label{fig:TemplateDists}
\end{center}
\end{figure}

\subsubsection{Background Rejection Power at $\sqrt{s} = 8 \TeV$}

We proceed to discuss the rejection power of the method for jets with $p_T > 300 \GeV$ at $\sqrt{s} = 8 \TeV$. For the purpose of illustration, we consider several combinations of cuts on both template and jet observables, while we leave $Ov_3^{min}$ a free parameter. We label the cuts as following:
\ba
	\mathbf{Cuts \, 1:}\,\,\,\,\,\,\, & Ov_2 > 0.9. \nn\\
	\mathbf{Cuts \, 2:}\,\,\,\,\,\,\, & Ov_2 > 0.9, \,2 t^b. \nn \\
	\mathbf{Cuts \, 3:}\,\,\,\,\,\,\, & Ov_2 > 0.9, \, 2 t^b, \,  \Delta R_{bb} / \Delta R_t < 1.0. \nn \\
	\mathbf{Cuts \, 4:}\,\,\,\,\,\,\, & Ov_2 > 0.9, \, 2 t^b, \, \tPf < 0.3. \nn \\
	\mathbf{Cuts \, 5:}\,\,\,\,\,\,\, & Ov_2 > 0.9, \, 2 t^b, \, \Delta R_{bb} / \Delta R_t < 1.0.\nn \\
				&  110 \GeV < m < 130 \GeV ,
\label{eq:cuts}
\ea
where $2 t^b$ denotes that both two-particle peak template momenta are $b$-tagged.

\begin{figure}[htb]
\begin{center}
\begin{tabular}{cc}
\includegraphics[width=3.5in]{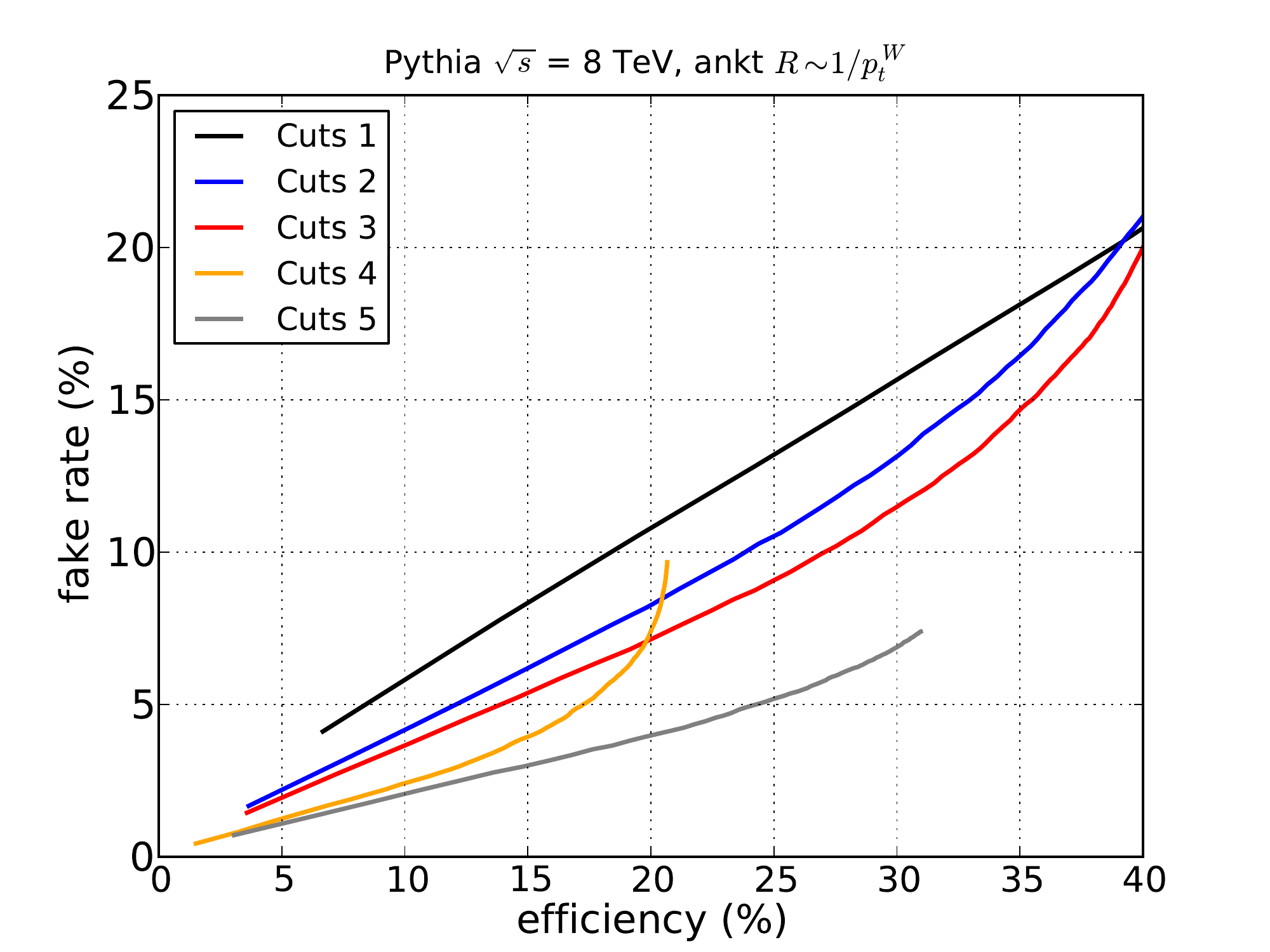}  
\end{tabular}
\caption{ Background rejection power of the template analysis. The left panel shows the overall efficiency and fake rate with fixed cuts of Eq.~\eqref{eq:cuts}. A cut on $Ov_3^{min}$ runs along the curves. The right panel shows rejection power for the $t \bar{t}$ and $W b \bar{b}$ separately for Cuts 1. All efficiencies are relative to Basic Cuts of Eq.~\eqref{eq:BasicCuts}.}
\label{fig:rejrate1}
\end{center}
\end{figure}

Fig.~\ref{fig:rejrate1} summarizes the results. The left panel shows rejection power obtained from our analysis, with and without a mass window cut. The signal efficiency and fake rates are measured relative to the cross sections with Basic Cuts from Table~\ref{sigma_table}. The curves also include $b$-tagging efficiencies we discussed the previous section. Each curve represents fake rate as a function of signal efficiency with a set of fixed cuts on template observables, while a cut  on $Ov_3$ in the range of $(0, 1)$ runs along the curve. Our results show that template observables can significantly improve the background rejection power relative to Basic Cuts of Eq.~\eqref{eq:BasicCuts}.  Fig.~\ref{fig:rejrate2} illustrates the rejection power over individual background channels.  Template Overlap method alone performs significantly better in rejecting $Wb\bar{b}$ events for most signal efficiencies, as shows in the left panel of Fig.  Fig.~\ref{fig:rejrate2}. This is reasonable since $t\bar{t}$ events typically consist of two $b$-tagged jets and an additional 
fragment of a 
hadronically decaying $W$ boson. Such a configuration is more likely to be tagged with a higher $Ov_3$ score than the typical two body substructure of a light QCD jet. However, cuts on additional kinematic observables such as  the Template Stretch or $tPf$, as well heavy flavor tagging requirements, can result in $t\bar{t}$ events being rejected at a rate higher than $Wb\bar{b}$.
\begin{figure}[htb]
\begin{center}
\begin{tabular}{cc}
\includegraphics[width=3.5in]{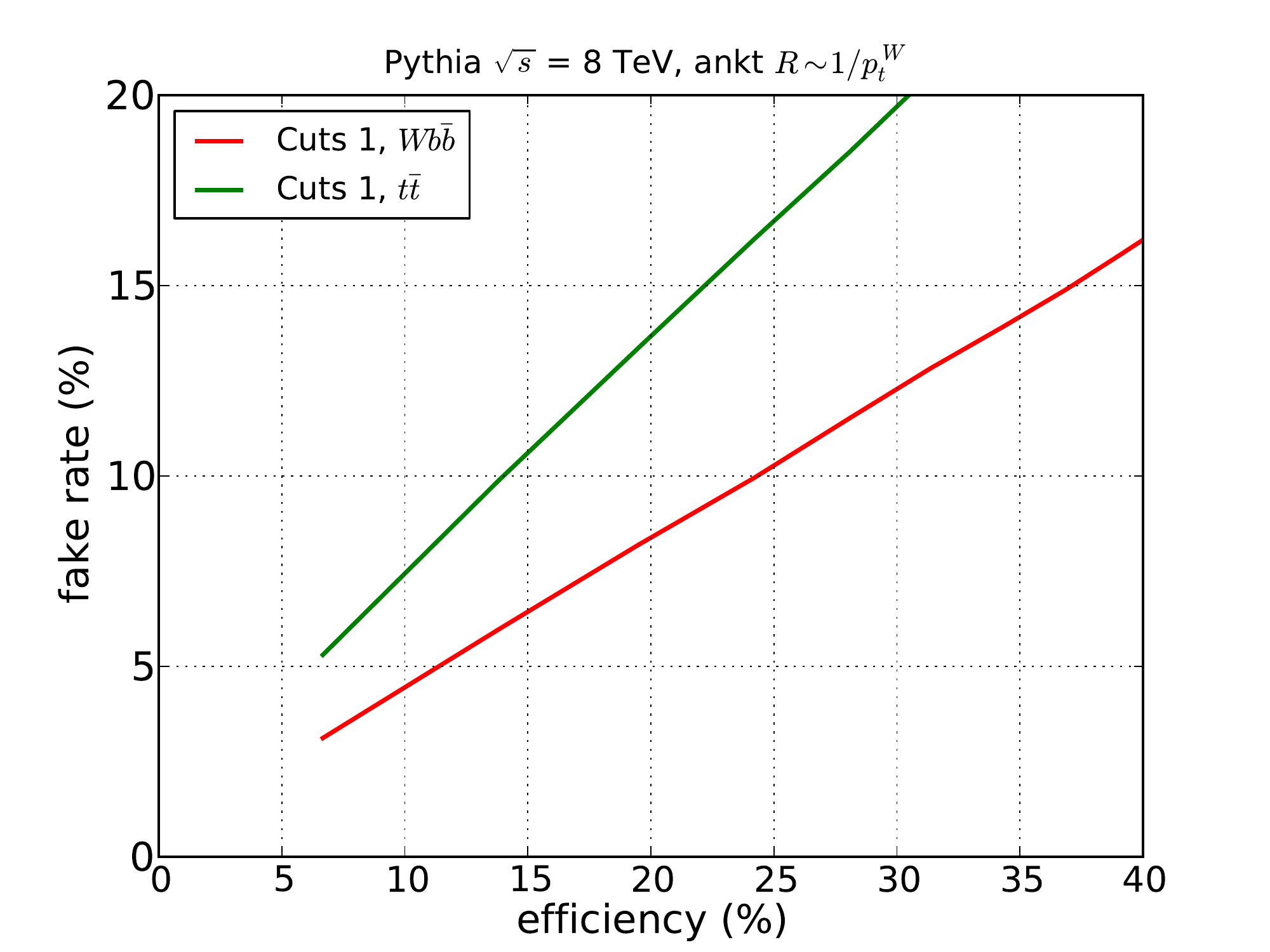} & \includegraphics[width=3.5in]{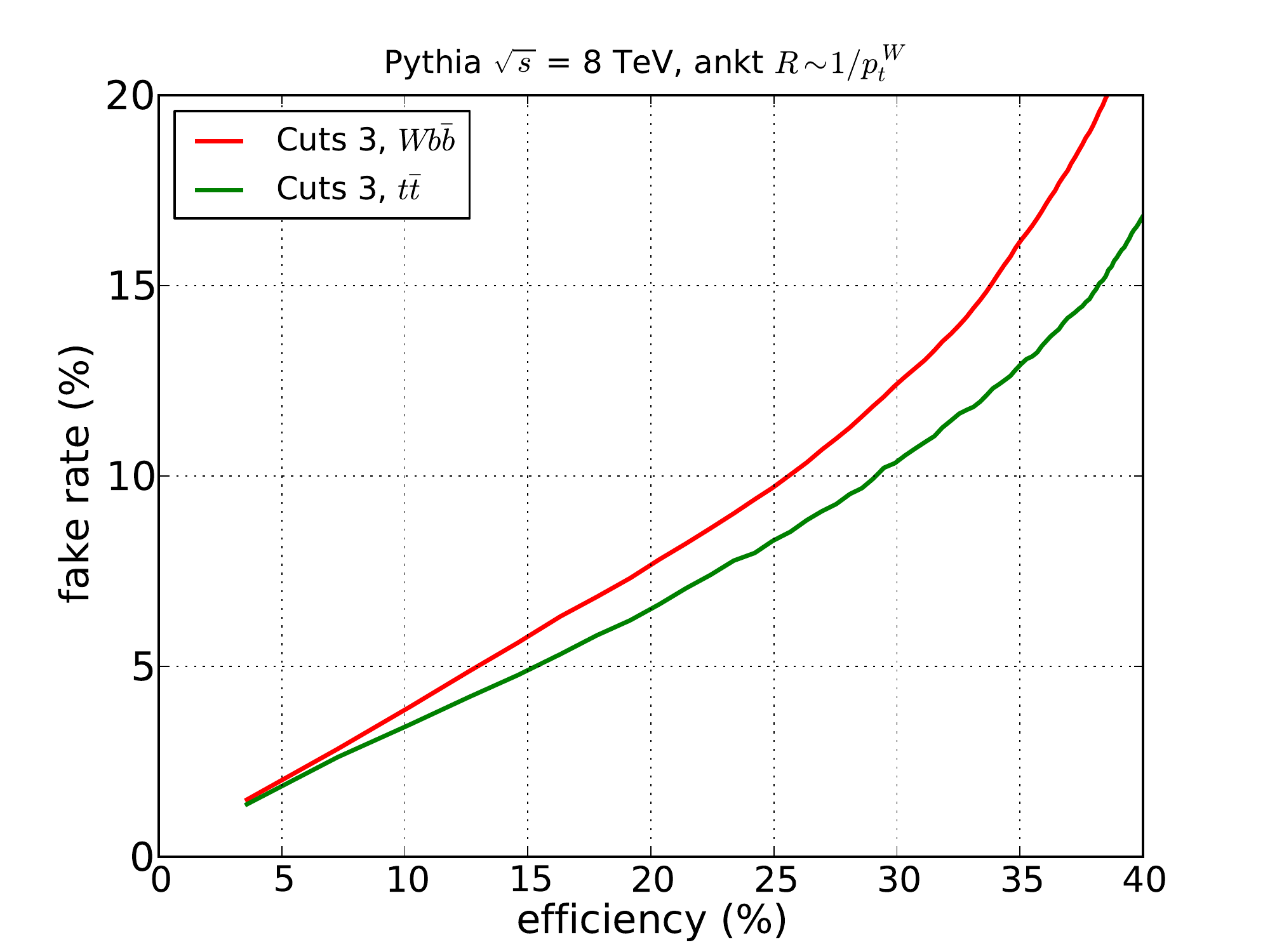} 
\end{tabular}
\caption{ Background rejection power of the template analysis for individual background channels. The left panel shows the overall efficiency and fake rate for $t \bar{t}$ and $W b \bar{b}$ separately with Cuts 1 of Eq.~\eqref{eq:cuts}. A cut on $Ov_3^{min}$ runs along the curves. The right panel shows the signal efficiency and fake rate for Cuts 3 of Eq.~\eqref{eq:cuts} . All efficiencies are relative to Basic Cuts of Eq.~\eqref{eq:BasicCuts}.}
\label{fig:rejrate2}
\end{center}
\end{figure}

Table \ref{tab:Rej0} shows an example of benchmark signal efficiency points. At $14 \%$ efficiency, a rejection factor of $\approx 4$ is achievable with template based observables only.  Additional mass window boosts the rejection power by $\approx 25 \%$, leading to a $S/B \approx 0.3$, with roughly 20\% efficiency. 

Template Overlap Method can achieve enough rejection power at $8\TeV$ to overcome the backgrounds at the cost of signal efficiency. Yet, the current estimates for integrated luminosity of the LHC $8\TeV$ run are not enough to yield practical results, even if we could combine the CMS and ATLAS data. We thus turn to projections for the future $13 \TeV$ run.

\begin{table}[!]
\begin{center}
\begin {tabular}{cc}

{\sc PYTHIA} &
\begin{tabular}{|c|c|c|c|c|c|}
\hline
Cut Set & $Ov_3^{min}$ & $Wh$ efficiency (\%) & Ê$Wb\bar{b}$ fake rate (\%) & $t\bar{t}$ fake rate (\%) & overall rejection power \\
\hline
Cuts 3 & 0.3 & 39.0 & 20.0 & 16.0 & 2.1 \\
Cuts 3 & 0.8 & 27.0 & 10.0 & 10.0 &2.7 Ê\\
\hline 
 Cuts 4 & 0.3 & 20.0 & 9.0 & 4.0 & 3.0 \\
Cuts 4 & 0.8 & 14.0 & 5.0 & 2.0 & 4.0 \\
\hline
Cuts 5 & 0.8 & 23.0 & 6.0 & 3.0 & 5.0 \\
\hline
\end{tabular}\\
{\sc SHERPA} &
\begin{tabular}{|c|c|c|c|c|c|}
\hline
Cut Set & $Ov_3^{min}$ & $Wh$ efficiency (\%) & Ê$Wb\bar{b}$ fake rate (\%) & $t\bar{t}$ fake rate (\%) & overall rejection power \\
\hline
Cuts 3 & 0.3 & 39.0 & 17.0 & -- & 2.3\\ 
Cuts 3 & 0.8 & 26.0 & 8.0 & Ê-- & 3.1\\
\hline
Cuts 4 & 0.3 & 23.0 & 11.0 & Ê-- & Ê2.1\\ 
Cuts 4 & 0.8 & 15.0 Ê& 5.0 & -- & 2.9\\
\hline
Cuts 5 & 0.8 & 17.0 & 4.0 & -- & 4.3\\
\hline
\end{tabular}
\end{tabular}

 \caption{Background Rejection Rates at $\sqrt{s} = 8 \TeV$. The values in the table show the signal efficiencies and fake rates relative to the cross sections with Basic Cuts of Eq.~\eqref{eq:BasicCuts}. The overlap rejection power includes both the $Wb\bar{b}$ and $t \bar{t}$. \label{tab:Rej0}}
\end{center}
\end{table}

\subsubsection{Background Rejection Power at $\sqrt{s} = 13 \TeV$}

The composition of background channels at $\sqrt{s} = 13 \TeV$ changes relative to $8 \TeV$, with $t \bar{t}$  amounting to $60 \%$ of the total. Higher center of mass energy also allows us to push the minimum $p_T$ to higher values; we opt for $p_T \ge 350 \GeV$. Fig.~\ref{fig:rejrate2} shows the result. We again find that template overlap can significantly improve the rejection rate over traditional jet observables. The overall performance of templates improves relative to $\sqrt{s} = 8 \TeV$ and $p_T \ge 300 \GeV, $ as expected. 

\begin{figure}[htb]
\begin{center}

\includegraphics[width=3.5in]{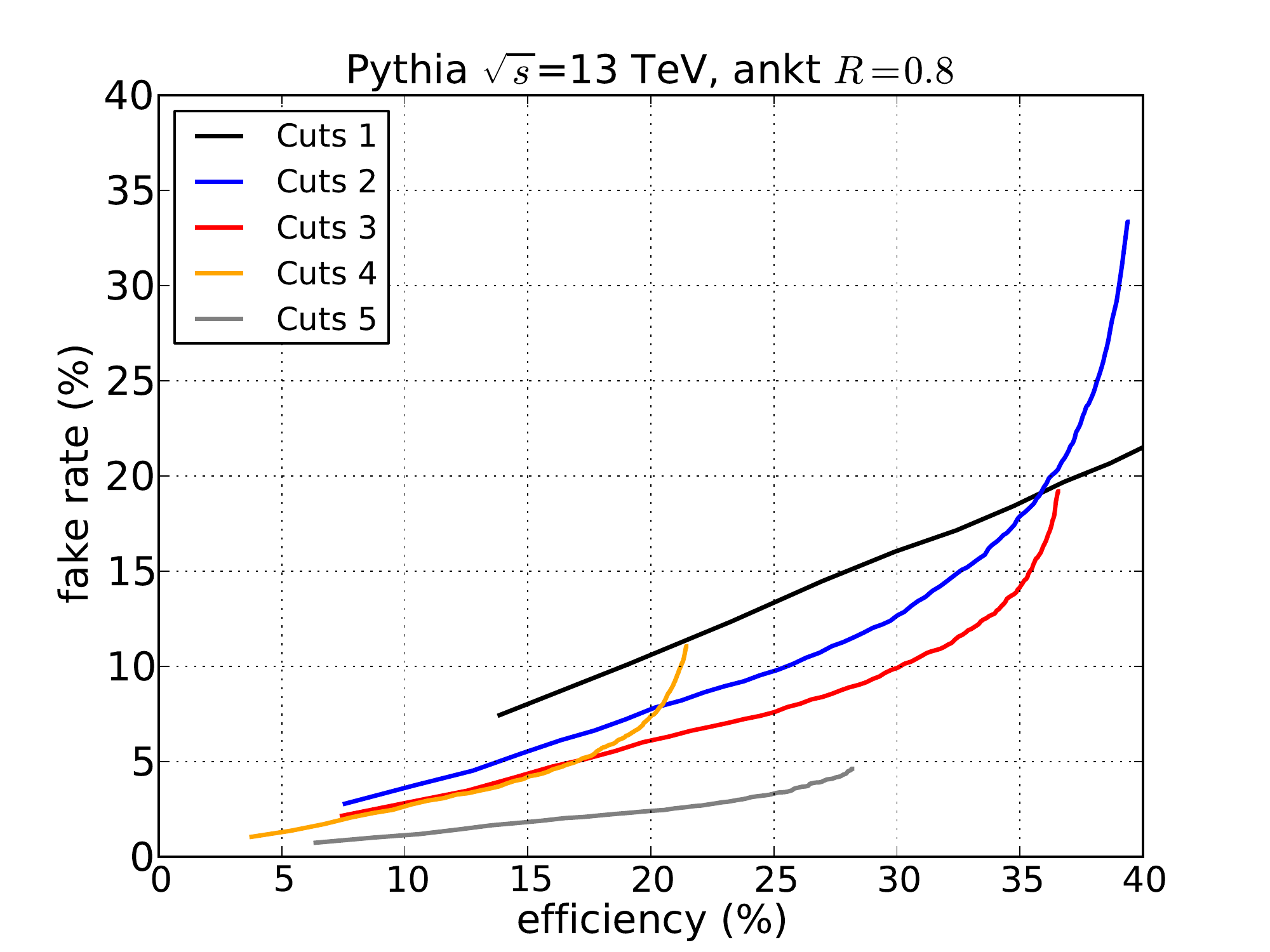}

\caption{ Background rejection power of the Template Overlap Method at $\sqrt{s} = 13 \TeV$. The left panel shows the overall efficiency and fake rate with fixed cuts of Eq. \eqref{eq:cuts}. A cut on $Ov_3^{min}$ runs along the curves. All efficiencies are relative to Basic Cuts of Eq.~\eqref{eq:BasicCuts}.}
\label{fig:rejrate2}
\end{center}
\end{figure}

 Fig.~\ref{fig:rejrate2} shows that an overall rejection power of $\approx 10 $ (Eff: 20\%) is achievable at $\sqrt{s}=13 \TeV$, leading to an overall $S/B \approx 0.5$. This constitutes an improvement over the $8\TeV$ result where the maximum rejection power  was $\approx 5$. 

\begin{table}[htb]
\begin{center}
\begin{tabular}{cc}

{\sc Pythia} &
\begin{tabular}{|c|c|c|c|c|c|}
\hline
Cut Set & $Ov_3^{min}$ & $Wh$ efficiency (\%) & Ê$Wb\bar{b}$ fake rate (\%) & $t\bar{t}$ fake rate (\%) & Overall Rejection Power \\
\hline
Cuts 3 & 0.3 & 35.2 & 19.1 & 11.8 & 2.4\\ 
Cuts 3 & 0.8 & 27.0 & 10.5 & 10.0 & 3.2\\ 
\hline
Cuts 4 & 0.3 & 20.4 & 8.5 & Ê7.6 & Ê2.6\\ 
Cuts 4 & 0.8 & 15.3 Ê& 4.5 & 4.2 & 3.6\\ 
\hline
 Cuts 5 & 0.8 & 22.1 & 5.1 & 1.3 & 8.2\\
\hline
\end{tabular}
\end{tabular}
 \caption{Background Rejection Rates at $\sqrt{s} = 13 \TeV$. The values in the table show the signal efficiencies and fake rates relative to the cross sections with Basic Cuts. The Overall Rejection Power includes both the $Wb\bar{b}$ and $t \bar{t}$. \label{tab:Rej}}
\end{center}
\end{table}

\subsection{Higgs Tagging with Template Overlap - Effects of Pileup} \label{sec:pileup}

A foe to most jet substructure observables, pileup has become an LHC fact of life. The strive for high luminosity resulted in pileup levels of whopping 20 average interactions per bunch crossing during the current $8 \TeV$ run.  Pileup events contribute both to the fat jet constituent multiplicity and the energy distribution within a jet, resulting in possibly dramatic effects on  \textit{any} jet substructure observable constructed out of jet constituents.

\begin{figure}[htb]
\begin{center}
\begin{tabular}{cc}
\includegraphics[width=3in]{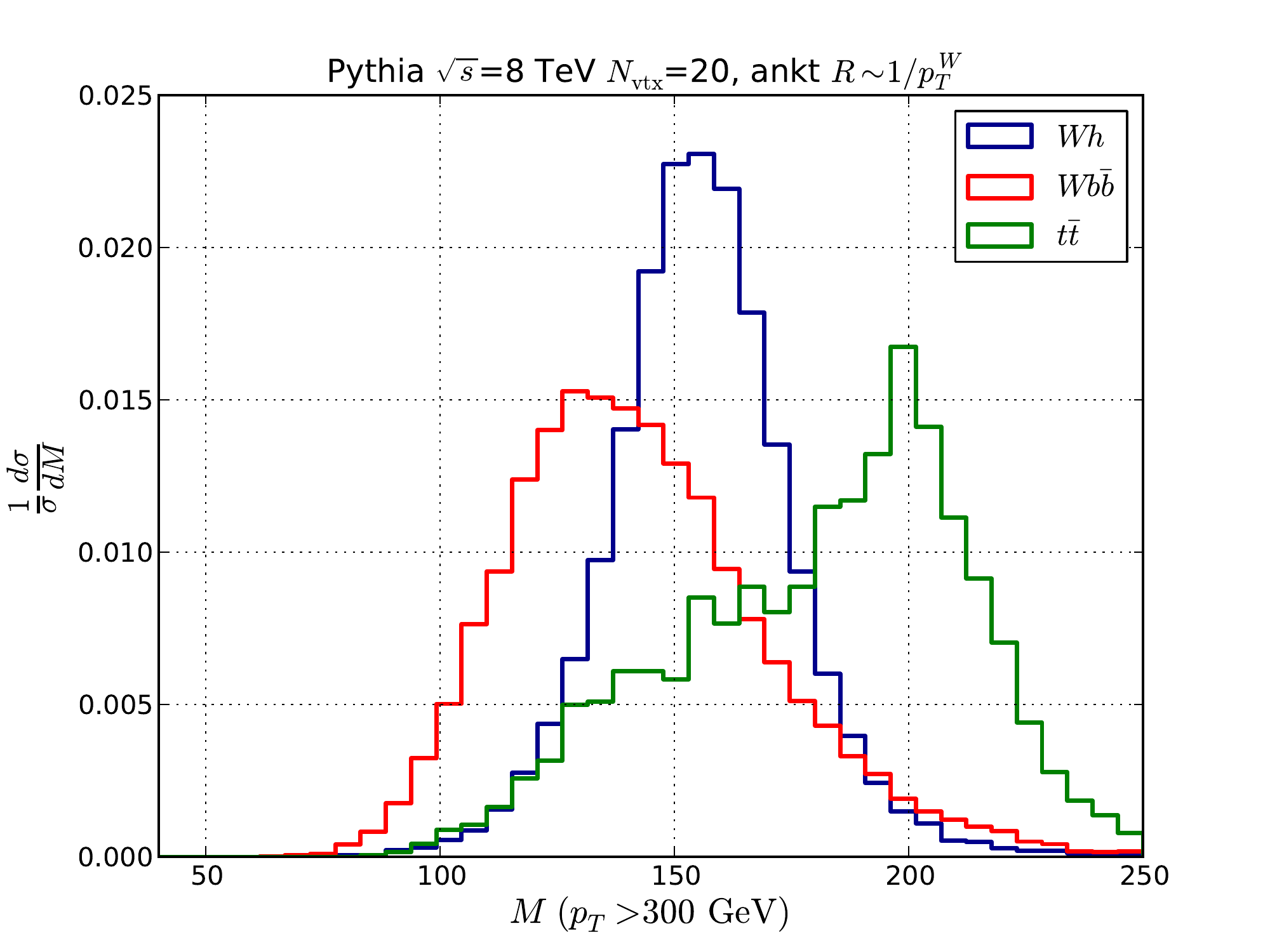} & \includegraphics[width=3in]{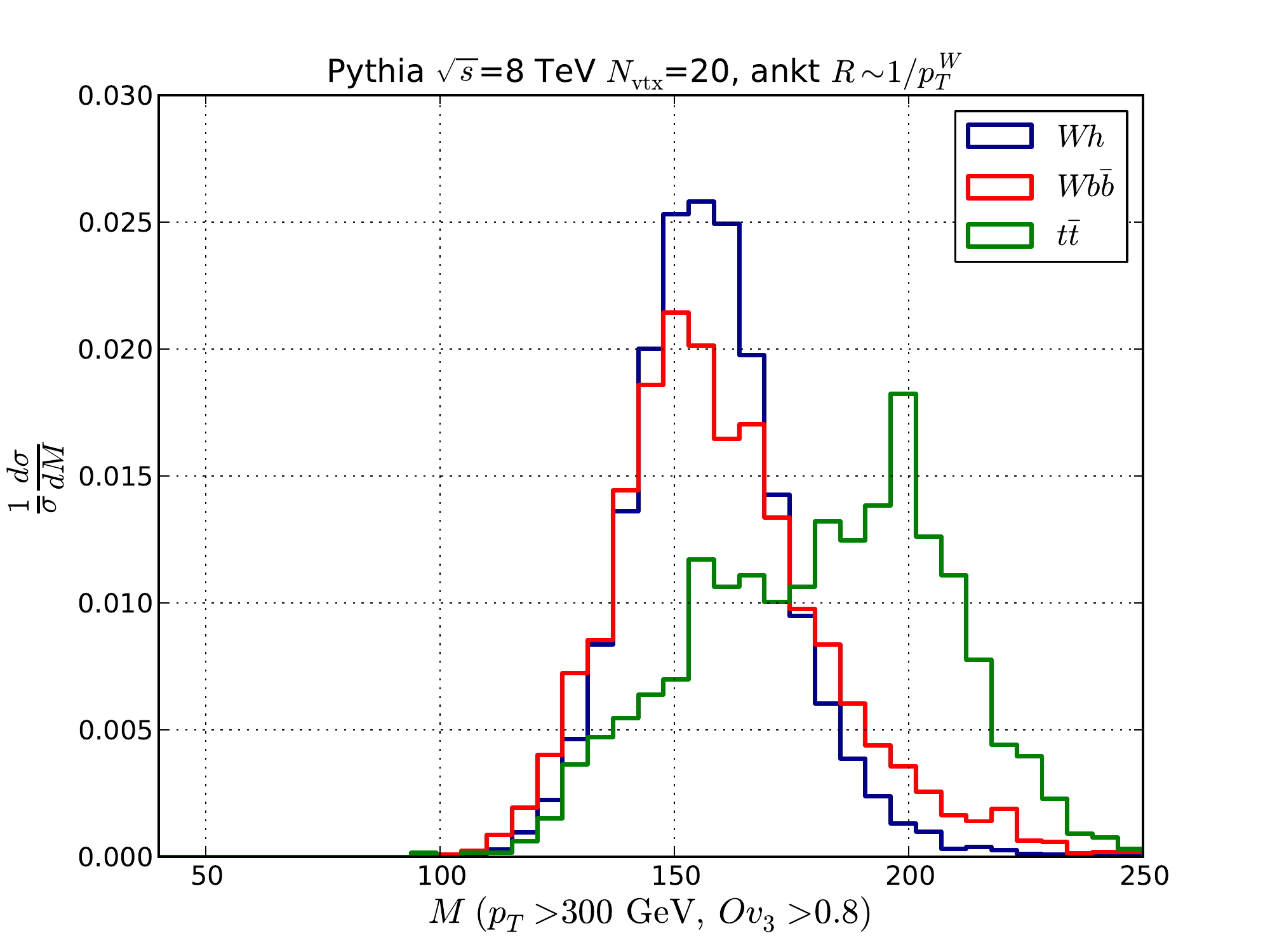} 
\end{tabular}
\caption{
Jet mass distributions for signal and dominant backgrounds with pileup, $N_{\rm vtx}=20$. 
The left panels show distributions after the Basic Cuts of Eq.~\eqref{eq:BasicCuts}, while the right panel shows the same distributions with $Ov_2>0.9$ $Ov_3>0.8$. 
}
\label{fig:PileupJetMass}
\end{center}
\end{figure}

Jet mass is perhaps the best illustration of this point. Fig. \ref{fig:PileupJetMass} shows an example. The left panel shows mass distributions in the presence of average 20 interactions per bunch crossing. In addition to shifting the mass peaks by as much as $30 \GeV$ (relative to the plots in Fig.  \ref{fig:JetMassCuts}) pileup reduces the mass resolution. In a recent study of Ref. \cite{:2012qa}, the ATLAS collabration tested this feature with the LHC $7 \TeV$ data. The mass resolution does not improve significantly even after cuts on the overlap are applied, again due to a low energy scale resolution. Note that reducing the value of $\sigma_a$ and thus increasing the mass resolution of the Template Overlap Method could be used to study effects of pileup on jet observables.

Many ``post processing'' pileup subtraction techniques exist in the literature, such as trimming \cite{Krohn:2009th},  pruning \cite{Ellis:2009me} and jet area techniques \cite{Cacciari:2007fd,Soyez:2012hv}.
Alternatively, Ref.~\cite{Alon:2011xb}  presents a data driven method of correcting for pileup effects for jet shape variable of massive narrow jets.
Finally, particle tracking information can be used to subtract pileup events, a method already used by the CMS collaboration  \cite {2011JInst...611002C}. In this section we do not consider any pileup subtraction. Instead, we show that the Template Overlap Method is largely unaffected by pileup.

\begin{figure}[!]
\begin{center}
\begin{tabular}{cc}
\includegraphics[width=3.5in]{event_view_nopileup_variable_05-eps-converted-to.pdf} & \includegraphics[width=3.5in]{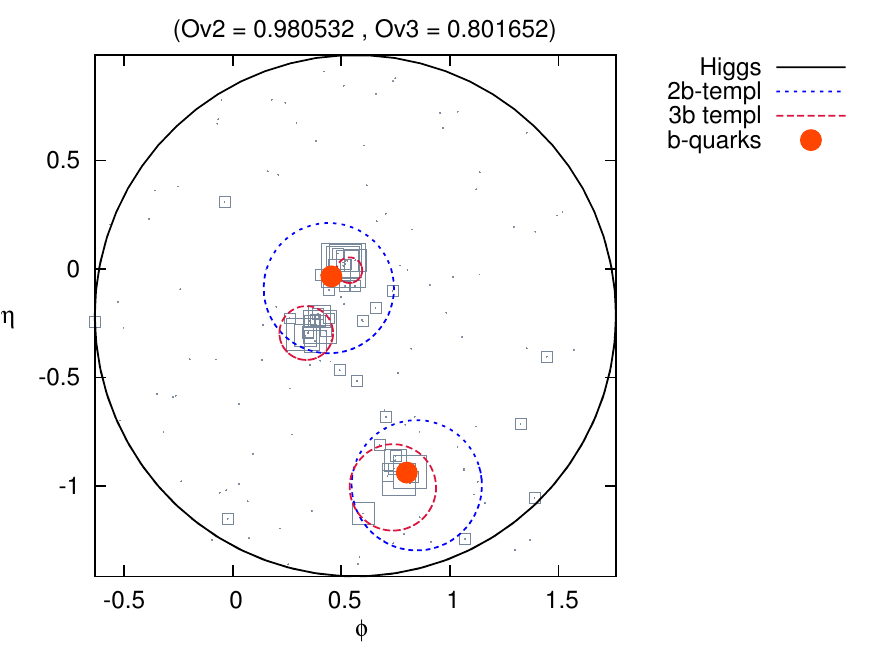} 
\end{tabular}
\caption{Comparison of boosted Higgs template results without and with pileup ($20$ average interactions per bunch crossing). Left panel shows a Higgs jet analyzed with no pileup. The right panel is the same jet with pileup added.  Grey squares represent the jet constituents, the $p_T$ of which is proportional to the size of the square. The solid circles are positions of $b$-quarks in the hard process. Particles with $p_T< 1\GeV$ are not shown on the plot, but are included in the analysis.}
\label{fig:EventView1}
\end{center}
\end{figure}

Robustness of the Template Overlap Method against pileup comes from the definition of template overlap. Consider for instance a single template momentum $p^t$. The core of the overlap measure is the difference
\be
	\delta p_T = p^t_T - \sum_j p^j_T \times \theta (r_3 - \Delta R_{tj}),
\ee 
where $p^j$ are momenta of jet constituents and $\theta$ selects the ones which fall into a cone of radius $r$ around $p^t$. The size of the template subcone $r_3$ thus limits the effects of pileup.  

\begin{figure}[htb]
\begin{center}
\begin{tabular}{ccc}
\includegraphics[width=2in]{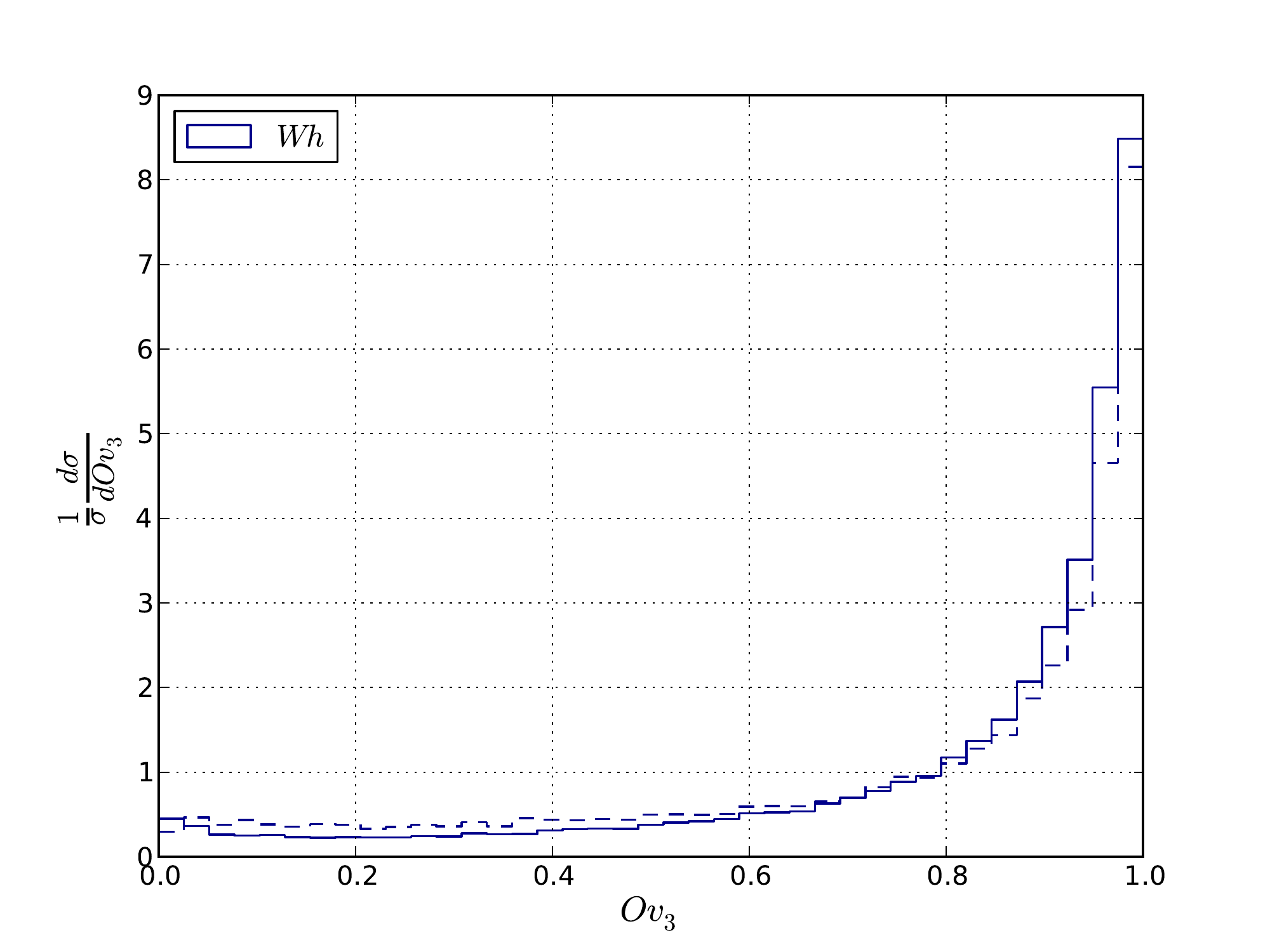} & \includegraphics[width=2in]{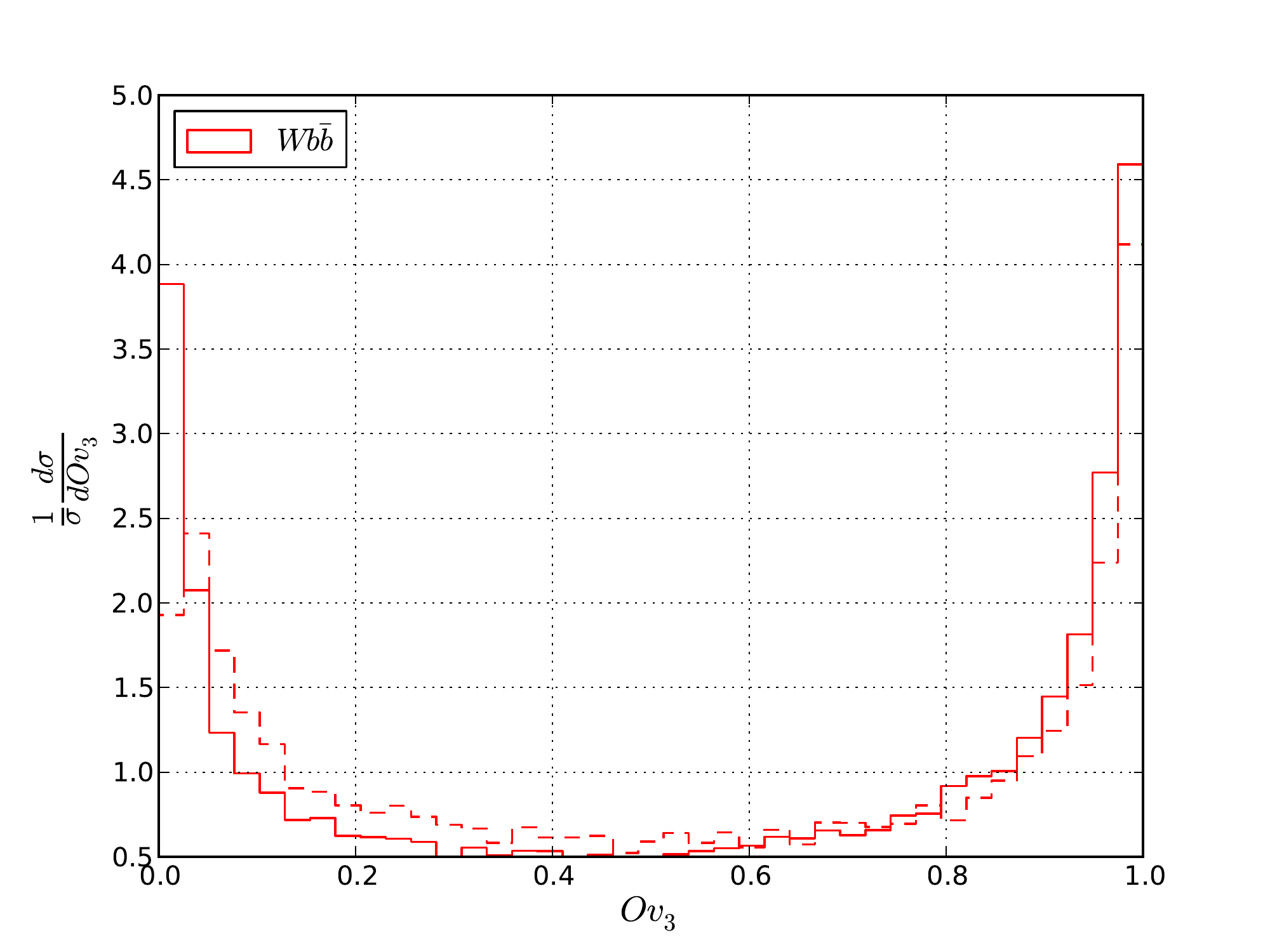} & \includegraphics[width=2in]{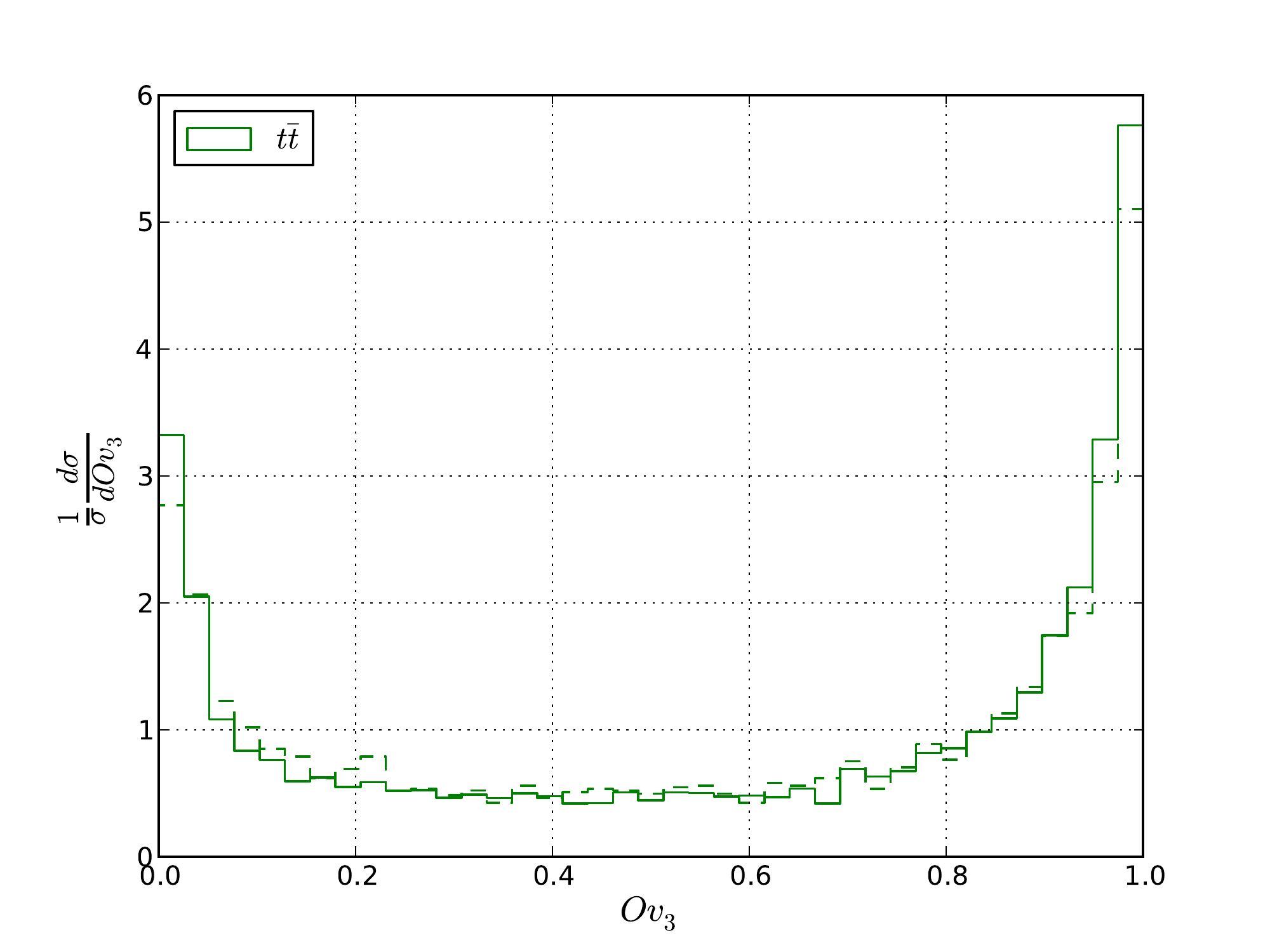}\\
\includegraphics[width=2in]{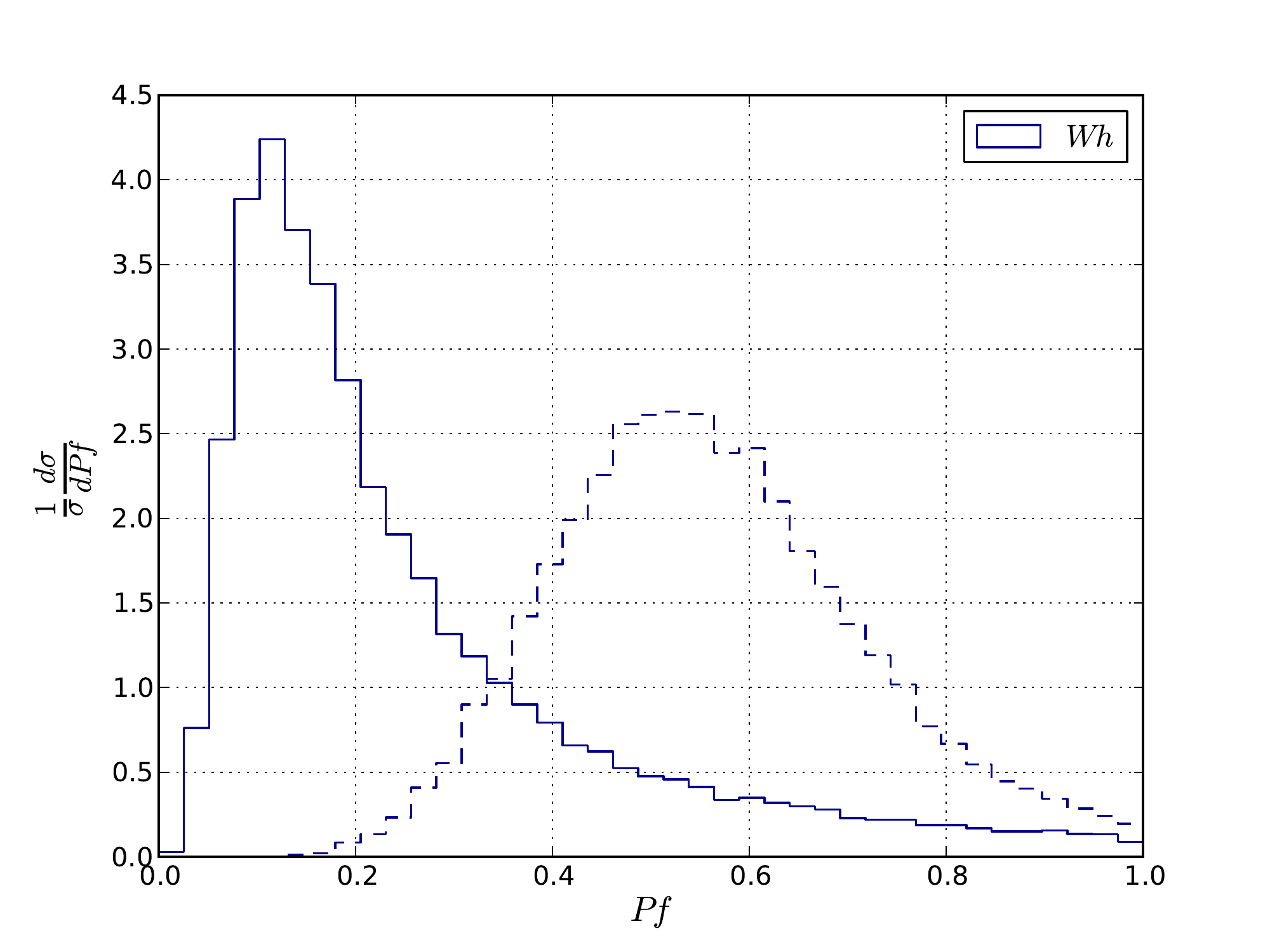} & \includegraphics[width=2in]{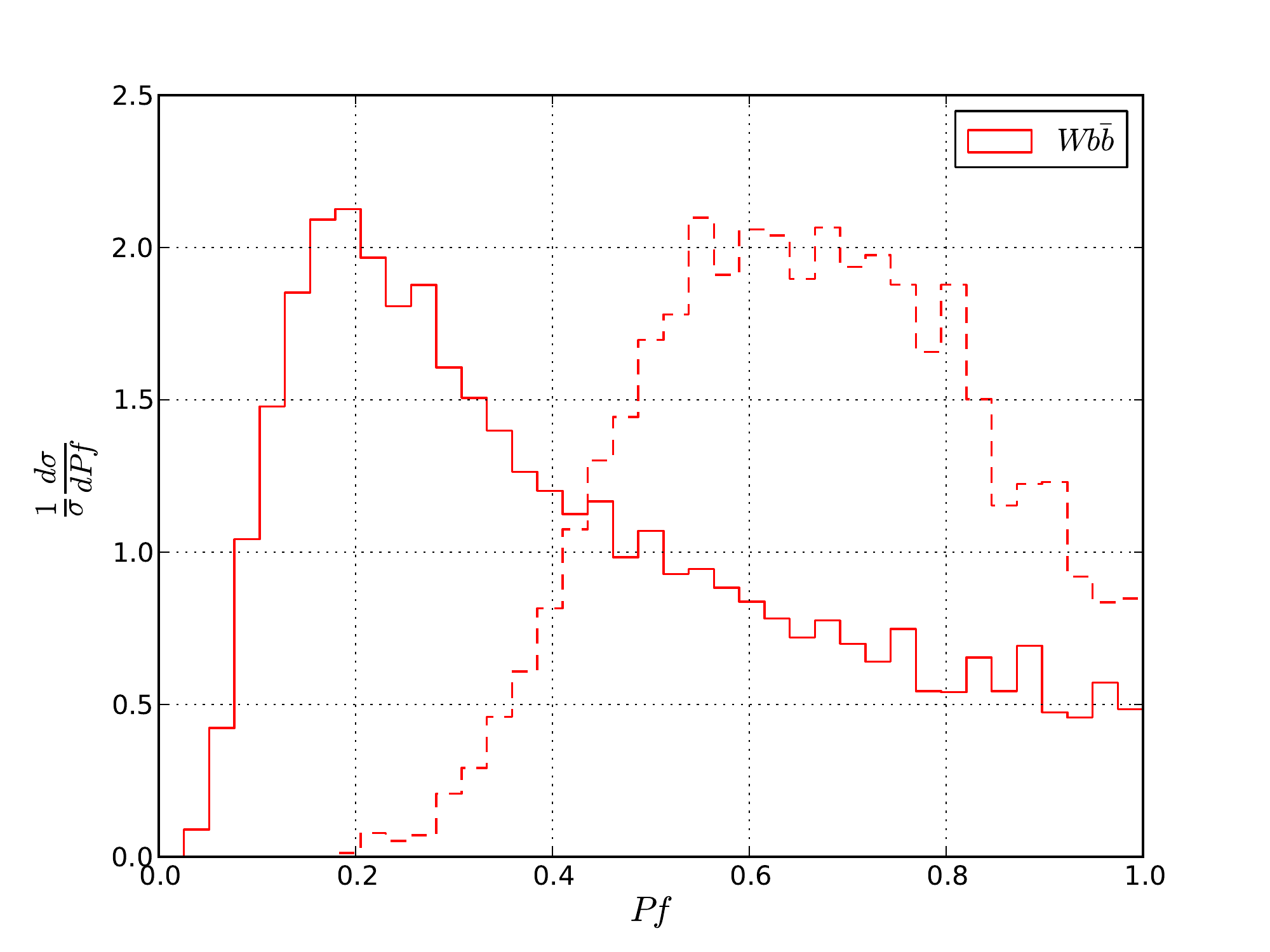} & \includegraphics[width=2in]{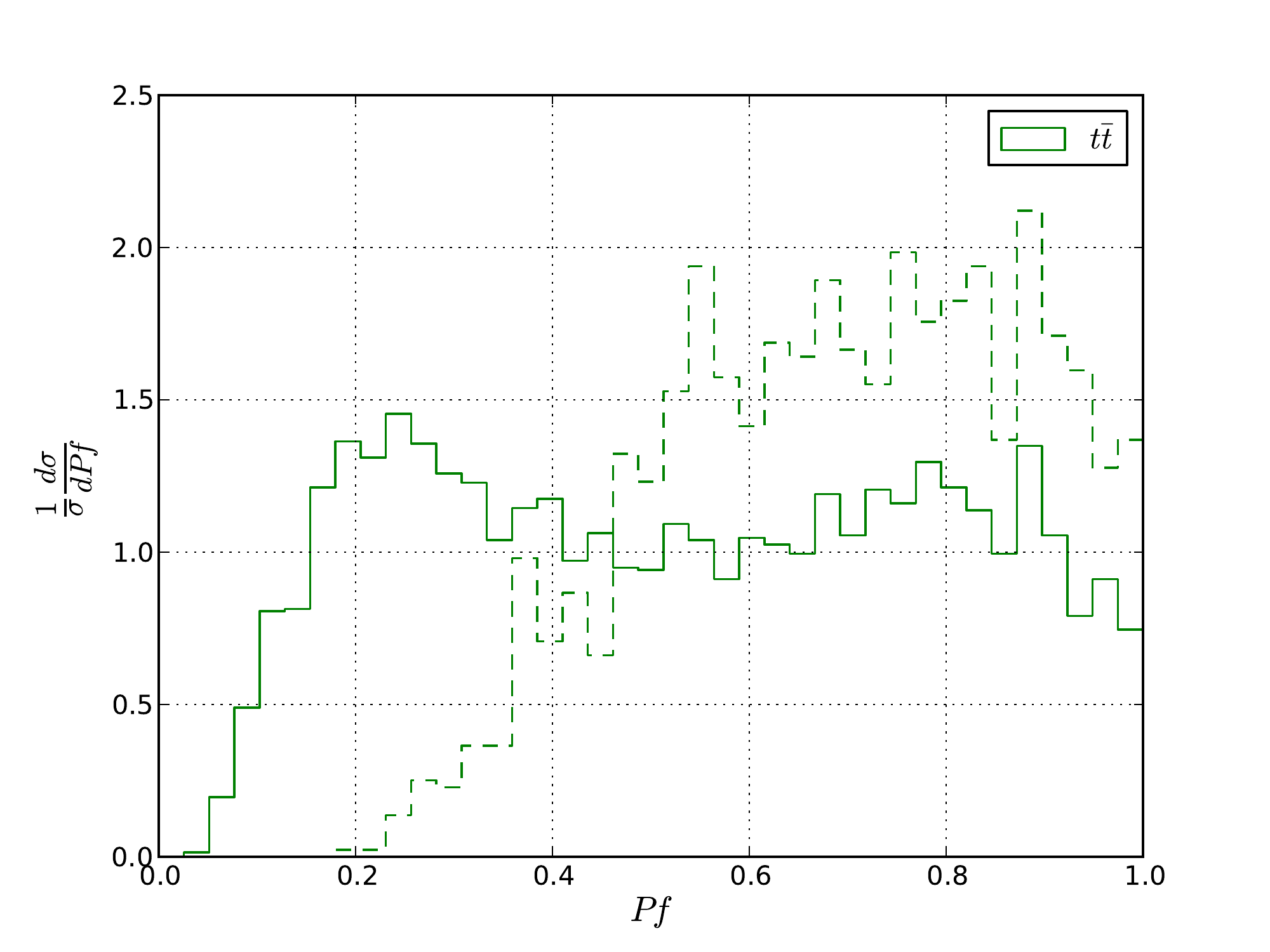}\\
\includegraphics[width=2in]{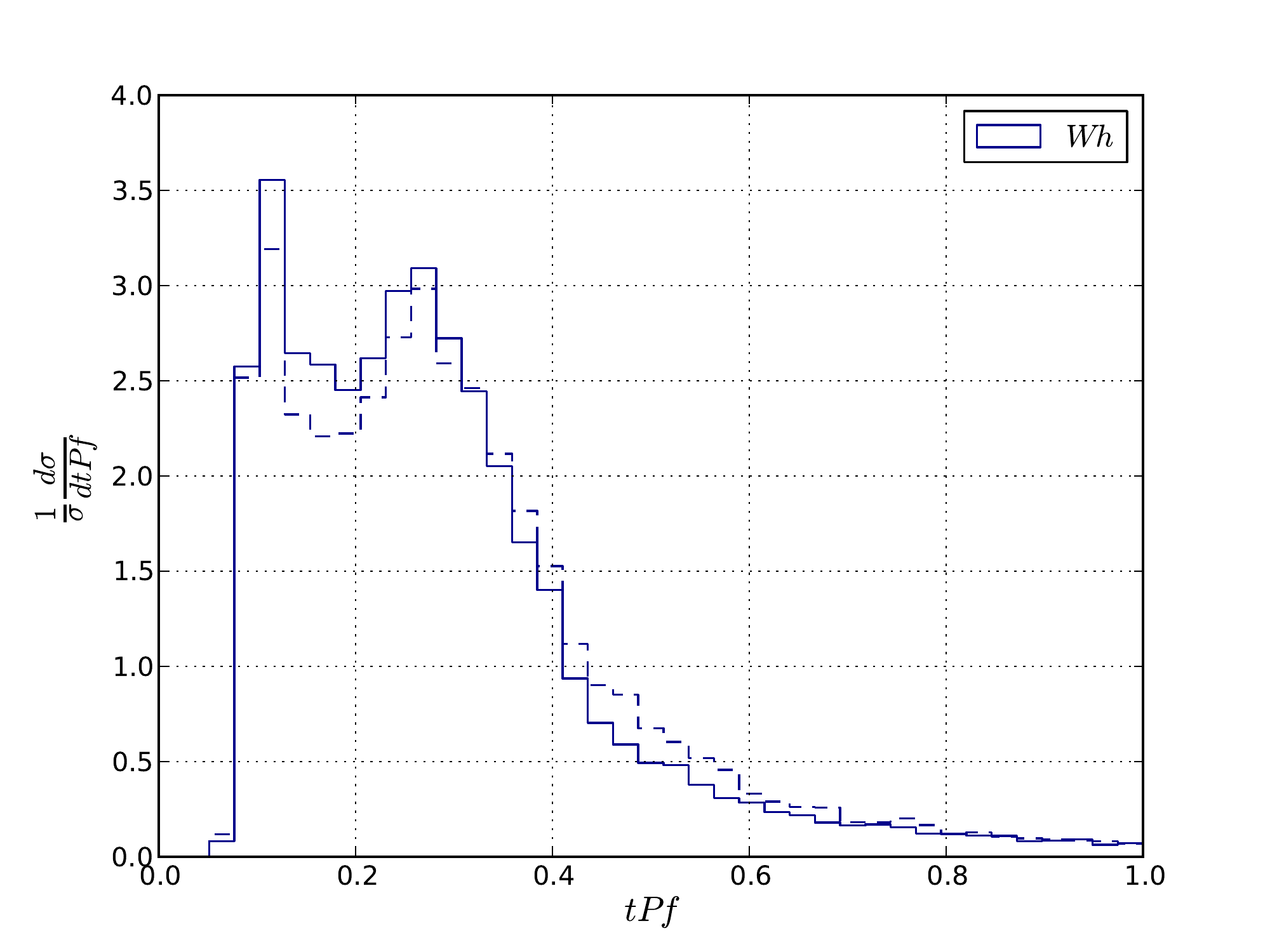} & \includegraphics[width=2in]{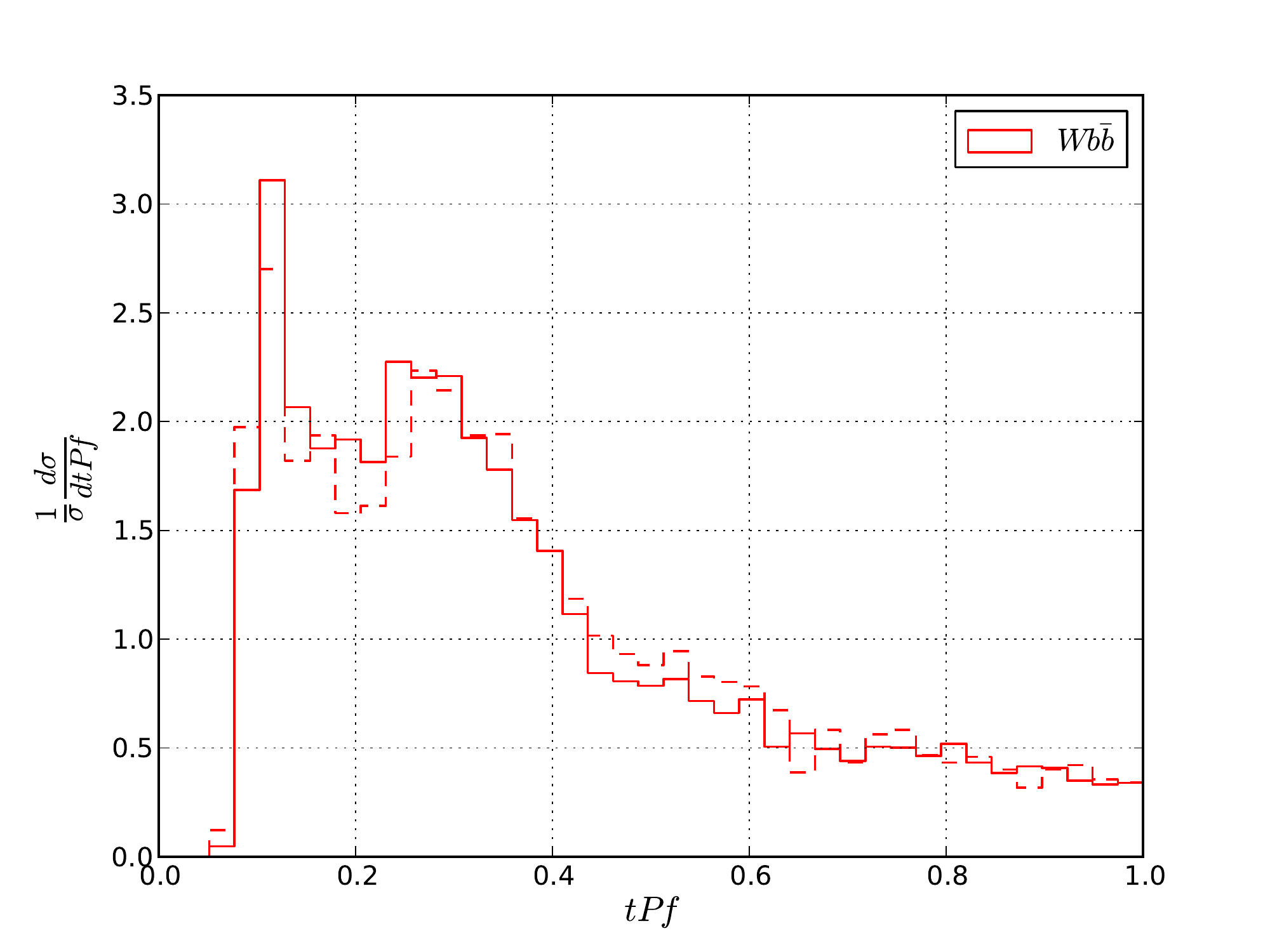} & \includegraphics[width=2in]{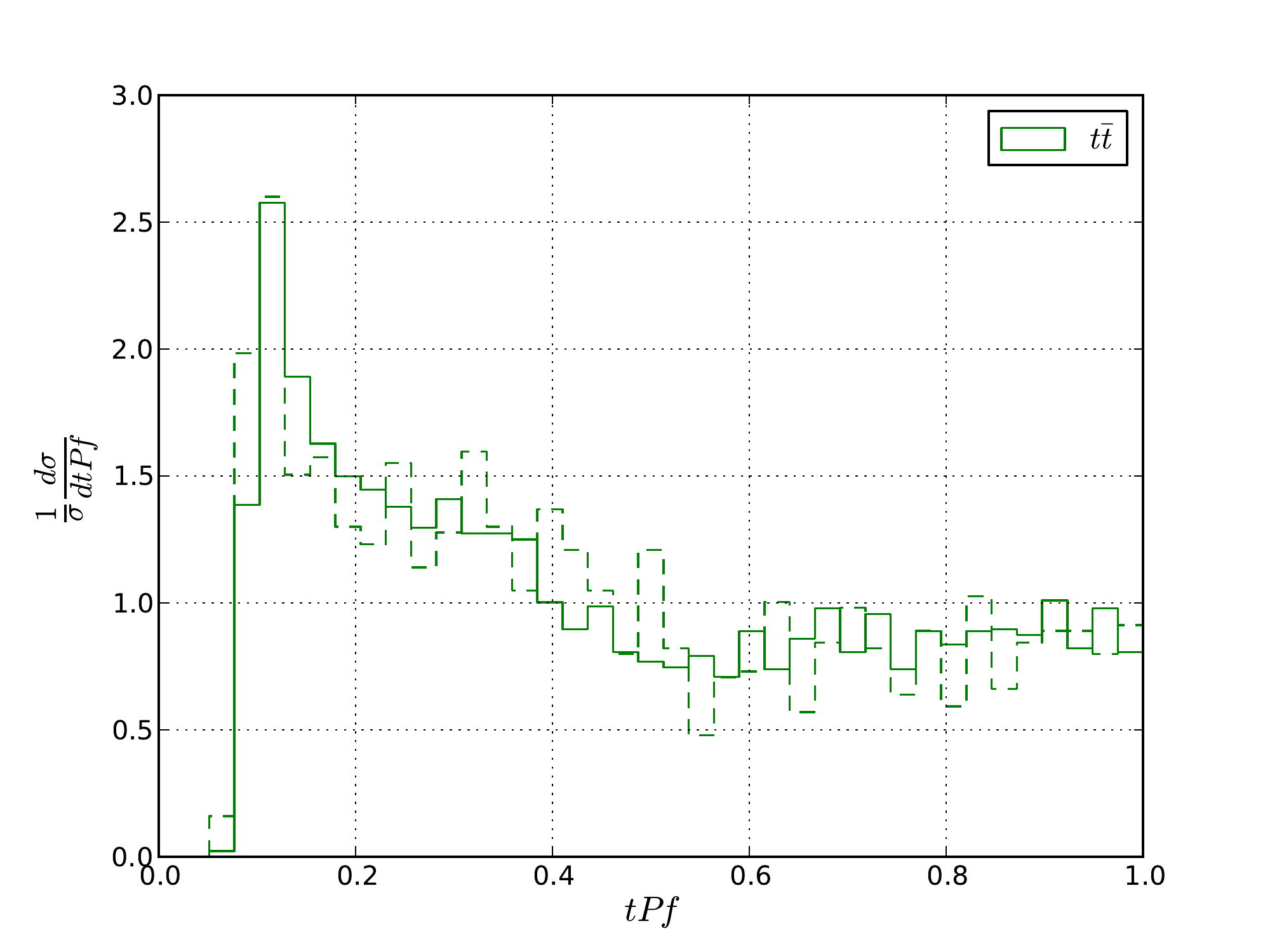}\\
\includegraphics[width=2in]{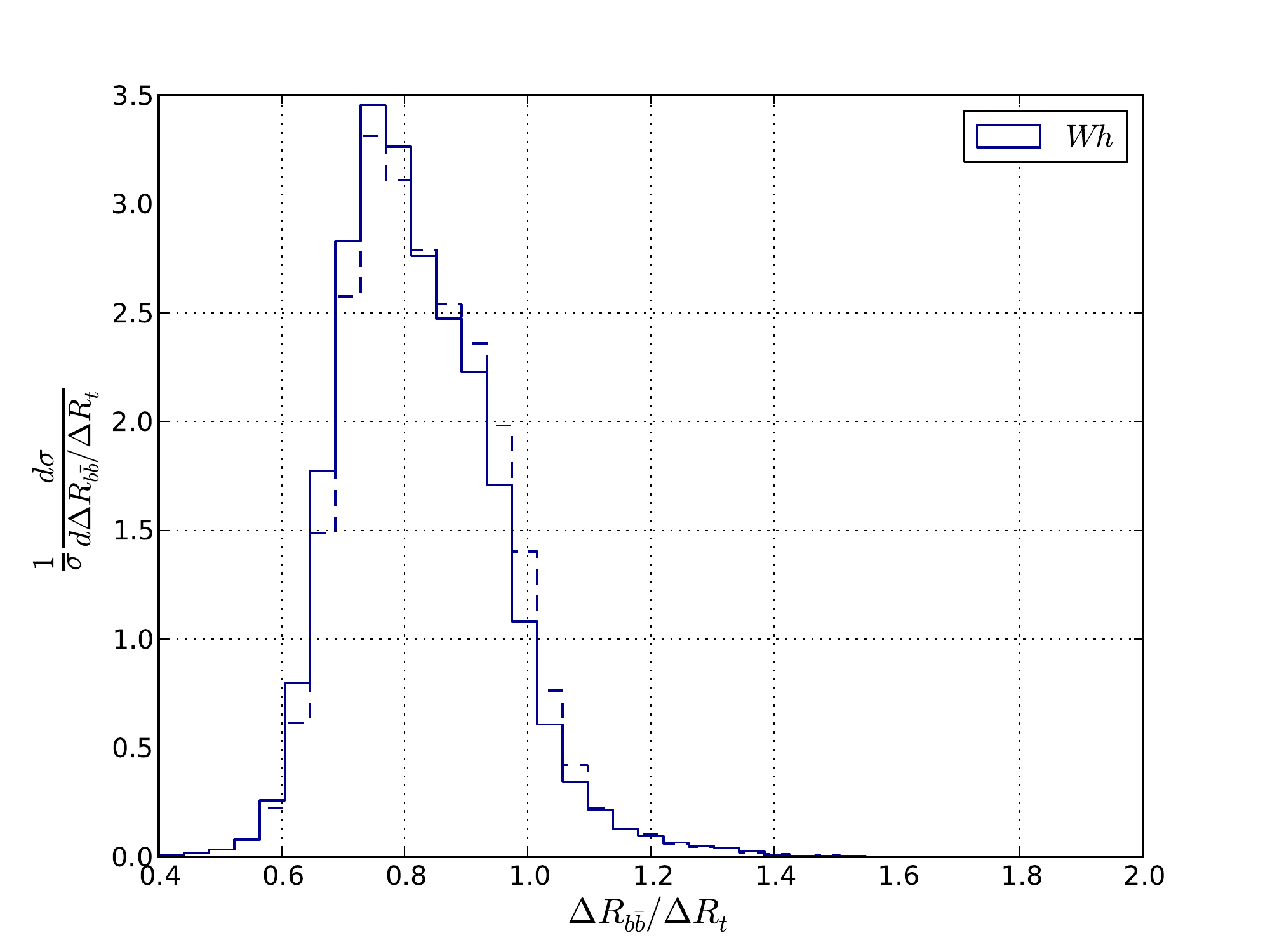} & \includegraphics[width=2in]{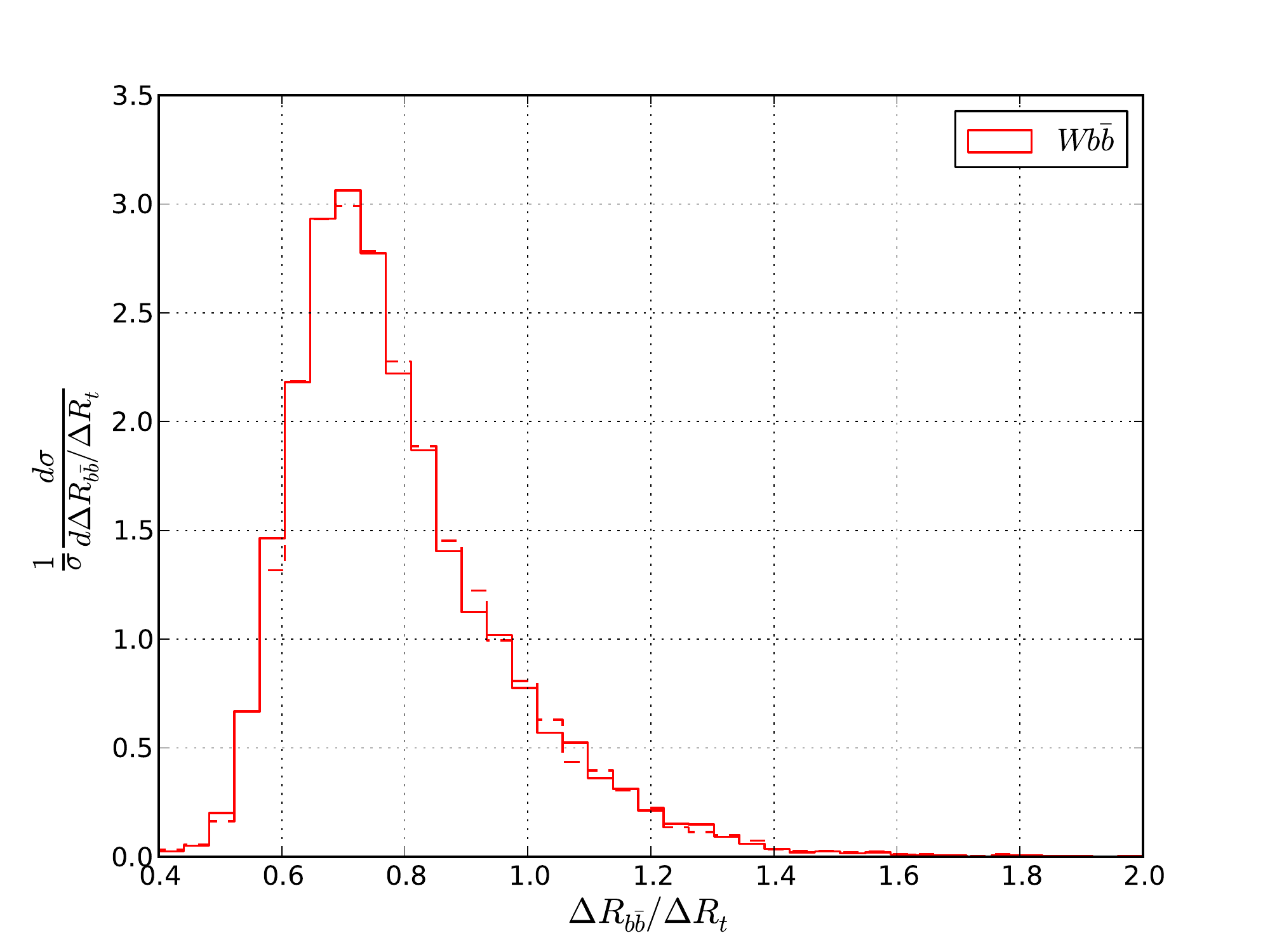} & \includegraphics[width=2in]{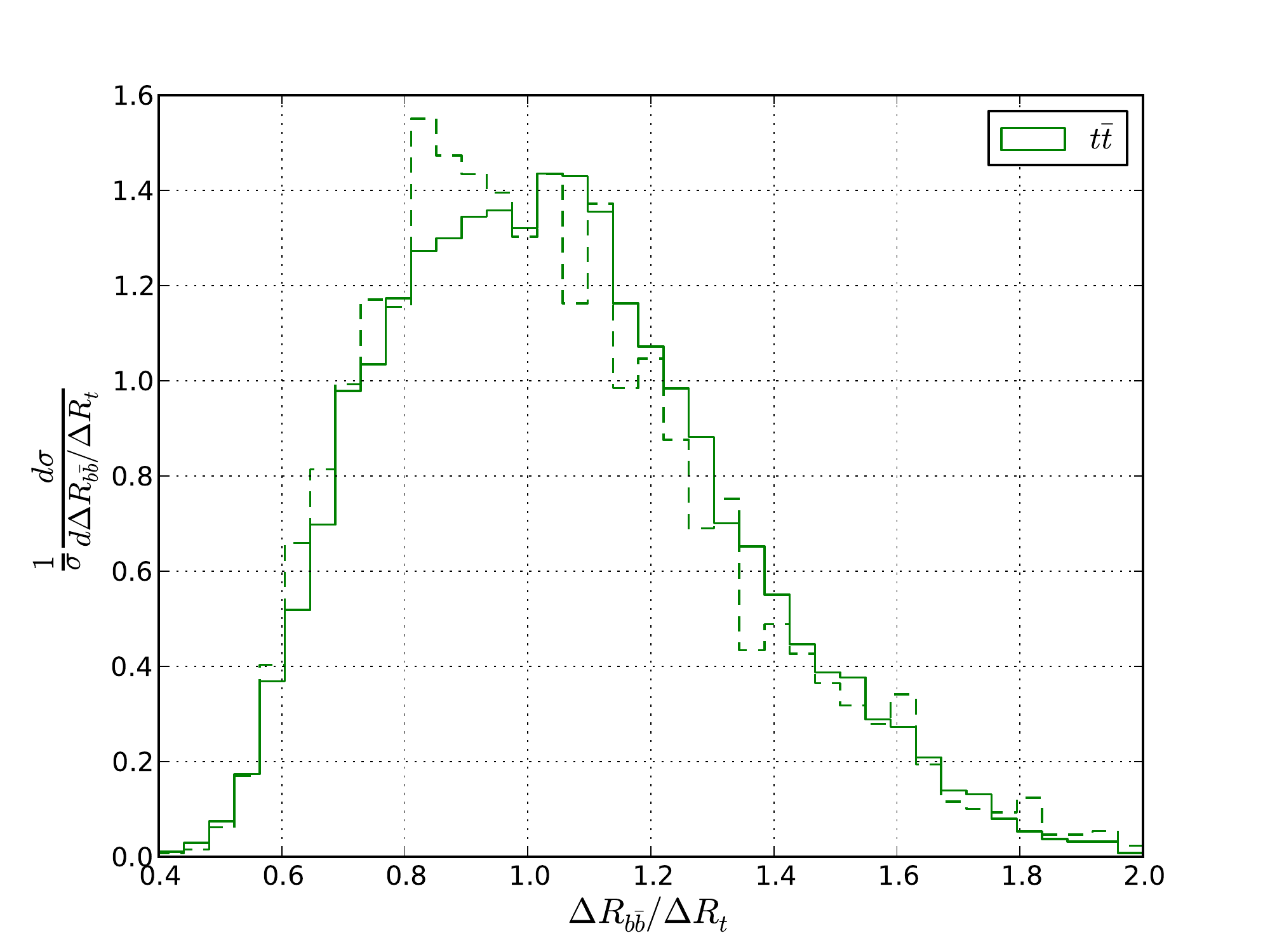}
\end{tabular}
\caption{Effects of pileup on various jet-substructure distributions. Solid distributions were obtained with no pileup, while dashed distributions contain $20$ average pileup events. All plots have a cut of $Ov_2 > 0.9$ for $r_2 = 0.3$. }
\label{fig:TempDistsPileup}
\end{center}
\end{figure}

Matching $p_T$ of a template state to the jet $p_T$ with pileup is problematic. Jet $p_T$ is shifted to higher values by pileup and as such is inappropriate as a criterion for template selection. Furthermore, a lower cut on the jet $p_T$ with pileup will include jets which would not pass the cut if pileup was not present. 
We instead use the $p_T$ of the leptonically decaying $W$ boson, as an infra-red safe, pileup independent observable (recall that since the Higgs recoils agains a $W$ boson, $p_W \approx p_H,$ as we showed in Section \ref{sec:data}).

We simulate the effects of pileup by adding a random number of minimum bias events (MBE) to every event we analyze. The number of added MBEs is distributed according to a Poisson distribution with the mean $N_{\rm vtx} = 20$, consistent with the LHC conditions at $\sqrt{s} = 8 \TeV$. Fig. \ref{fig:EventView1} shows an example of the pileup effects on the overlap analysis of a single event. The analysis of a jet with no pileup (left panel) yields nearly identical peak template state as the jet with pileup (right panel). Notice that the overlap scores remained within $\sim 10 \%$ of each other. On distribution level, the situation is similar. Fig. \ref{fig:TempDistsPileup} shows examples of several template based observables with and without pileup. Even at $20$ interactions per bunch crossing, we find no significant effects of pileup on the distribution shapes. In fact, the difference between susceptibility of $\tPf$ and $\Pf$ to a pileup environment is striking. While jet Planar Flow is significantly shifted to higher values by pileup, template Planar Flow remains mostly unaffected. 

\subsubsection{Rejection Power}

 The overlap method used in this study effectively reduces the area of fat jets and is therefore less sensitivity to pileup. As a simple application of this idea, in Fig. \ref{fig:rejrate3} we show fake rate~{\it vs.}~efficiency with the cuts of Eq. \eqref{eq:cuts}, obtained from {\sc Pythia} data at $\sqrt{s} = 8\TeV$ and $N_{\rm vtx}=20$. For a given cut, the efficiency is controlled by the lower cut on $Ov_3$. The results depend on the choice of cuts, but it is clear that the overall performance of overlap approach remains largely unchanged when compared to the case of events without pileup. 
 Our final results can be summarized in  Table~\ref{tab:Rej_pileup} for a few benchmark efficiency points. We choose to omit the result of adding a mass cut since it requires an additional mechanism for pile up subtraction, and is thus beyond the scope of this paper.
 In each case, we find background rejections  comparable to our results for events wihout pileup.
\begin{table}[htb]
\begin{center}
\begin{tabular}{cc}

{\sc Pythia} &
\begin{tabular}{|c|c|c|c|c|c|}
\hline
Cut Set & $Ov_3^{min}$ & $Wh$ efficiency (\%) &  $Wb\bar{b}$ fake rate (\%) & $t\bar{t}$ fake rate (\%) &  overall rejection power \\
\hline
Cuts 3 & 0.3 & 37.9 & 18.6 & 15.8 & 2.1\\ 
Cuts 3 & 0.8 & 25.1 & 8.7 & 10.1 & 2.9\\ 
\hline
Cuts 4 & 0.3 & 24.5 & 9.7 &  9.9 &  2.5\\ 
Cuts 4 & 0.8 & 16.1  & 4.6 & 5.1 & 3.4\\ 
\hline
\end{tabular}
\end{tabular}
 \caption{Background Rejection Rates at $\sqrt{s} = 8 \TeV$ with $N_{\rm xvt}=20$. The values in the table show the signal efficiencies and fake rates relative to the cross sections with Basic Cuts. The overall rejection power includes both the $Wb\bar{b}$ and $t \bar{t}$. 
 \label{tab:Rej_pileup}}
\end{center}
\end{table}

\begin{figure}[htb]
\begin{center}
\includegraphics[width=3.5in]{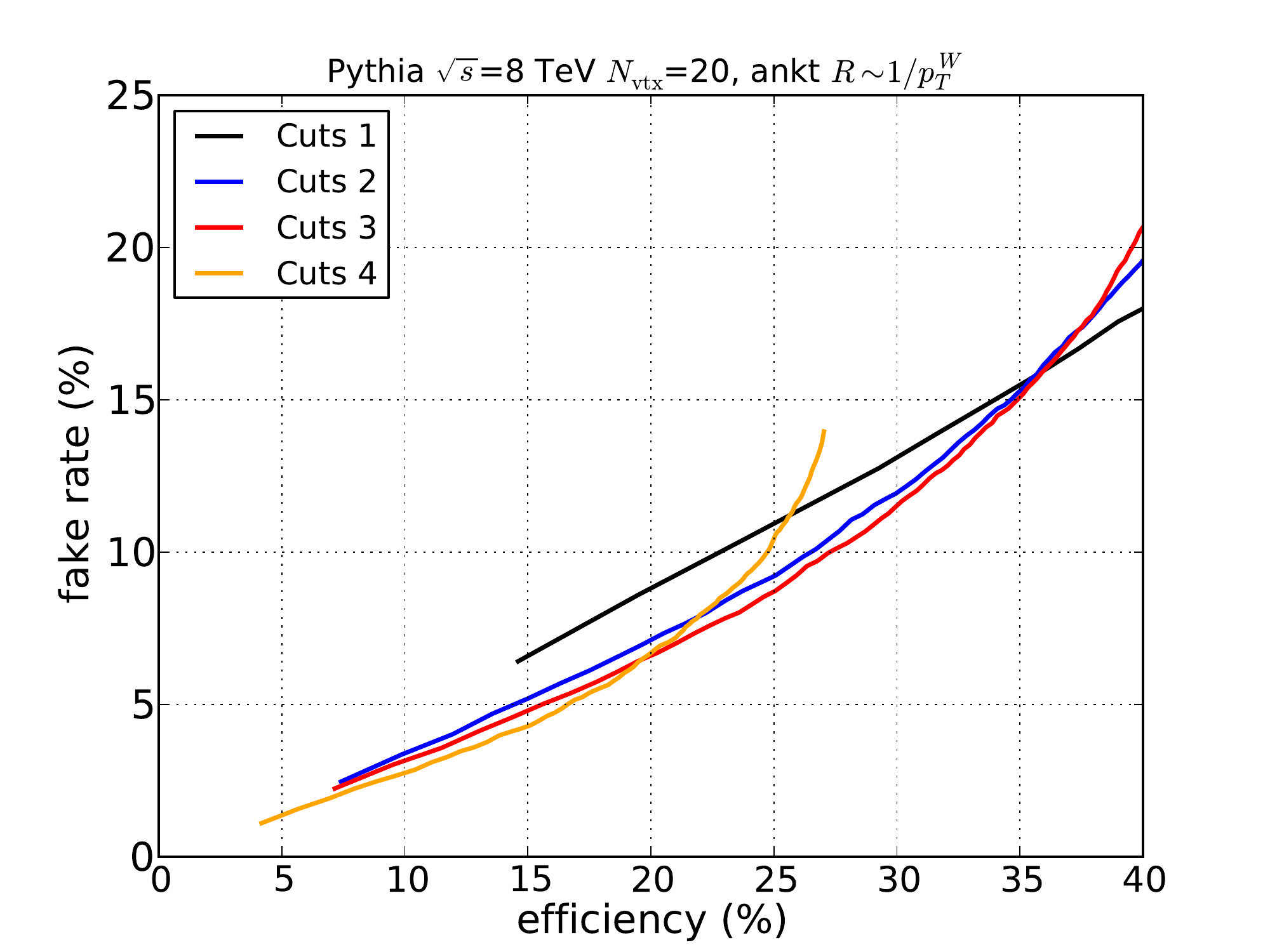}
\caption{ Background rejection power of the Template Overlap Method at $\sqrt{s} = 8 \TeV$ with $N_{\rm vtx}=20$. The figure shows the overall efficiency and fake rate with fixed cuts of Eq. \eqref{eq:cuts}. A cut on $Ov_3^{min}$ runs along the curves. All efficiencies are relative to Basic Cuts of Eq. \eqref{eq:BasicCuts}.}
\label{fig:rejrate3}
\end{center}
\end{figure}

\section{Conclusions}

Hadronic decay channel of the Standard Model Higgs boson is one of the most challenging measurements in Higgs physics at the LHC. Traditional jet observables such as jet mass and $p_T$ are inadequate to combat the large QCD background as well as high luminosity environments characteristic of the LHC. Jet substructure techniques can be used to overcome the large backgrounds in the boosted Higgs regime. We have demonstrated that Template Overlap Method is able to deliver sufficient background rejection power for a viable boosted Higgs search at $\sqrt{s} = 13 \TeV$. The method 
is relatively insensitive to contamination from pileup at an average of 20  interactions per bunch crossing.
 We introduced several improvements into the template overlap framework, which include varying the three-body subcones, sequential template generation, integration of $b$-tagging identification into the peak templates and a new useful substructure variable denoted as template stretch.  Future studies of the boosted Higgs in the framework of template overlap would benefit from a detailed detector simulation and more realistic $b$-tagging.

\section{Acknowledgments}
We thank Raz Alon, Ehud Duchovni, Ohad Silbert and Pekka Sinervo  for useful comments and encouragement and especially Jan Winter for an early collaboration on this project.
We also thank Gavin Salam for useful comments on the {\sc TemplateTagger} code, and Ciaran Williams for discussions about MCFM. MB and JJ would like to thank the CERN theory group for their hospitality, where a portion of this research was conducted. The authors also express their gratitude to 	Pierre Choukroun and Lorne Levinson for their enormous help regarding the computing needs of this work.
GP holds the Shlomo
and Michla Tomarin development chair and is supported by the
Gruber foundation, IRG, ISF and Minerva.

\appendix

\section{More on Template Planar Flow}
\label{app:PF}

To illustrate how energy smearing affects the \textit{parton} level $\tPf$ consider a peak three-body template state with an inertial tensor $I_{t}$. Energy smearing affects only the diagonal elements of $I_{t}$ while off-diagonals remain unchanged due to symmetry, leading to
\be
	I_t  \rightarrow I_{s} = I_t + \frac{P_T}{M} \,  \alpha \,   \mathbb{I} , \label{eq:InertShift}
\ee
where $P_T = \sum_{i=1}^3 p_T^i$ and $p_T^i$ is the transverse momentum of the $i^{th}$ template momentum. The effects of energy smearing are summarized in the symbol 
\be
	\alpha = \int d\Omega \, g(\hat{p}_i),
\ee
where $g(\hat{p}_i)$ is the energy smearing distribution. 

For simplicity, we consider a uniform distribution over a disc of radius $r_3$, giving 
\be
	\alpha_{cone} = \frac{r_3^2}{2}.
\ee
Alternatively, one could also consider a Gaussian distribution centered around the template momentum, with a width $\sigma$, resulting in
\be
	\alpha_{Gauss} = \sigma^2.
\ee

Continuing, the determinant of $I_s$ becomes
\ba
	{\rm det}(I_s) &=& {\rm det}(I_t +\frac{P_T}{M} \, {\alpha}  \mathbb{I}) \nn\\
	 	&=& {\rm det}(I_t (\mathbb{I} + I_t^{-1} \frac{P_T}{M} \alpha \mathbb{I} )) \nn\\ 
		&=& {\rm det}(I_t){\rm det}(\mathbb{I} + I_t^{-1} \frac{P_T}{M} \alpha \mathbb{I}).
\ea
Similarly, the trace becomes
\ba
	{\rm tr}(I_s) &=& {\rm tr}(I_t + \frac{P_T}{M}  \alpha \mathbb{I} )\nn \\
	&=& Tr(I_t) + 2 \frac{P_T}{M} \alpha
\ea

For a jet cone radius $R \sim 1$ it is reasonable to assume that $\alpha  \ll 1.$ Furthermore, $P_T / M$ is typically of $O(1),$ allowing us to expand ${\rm det}(I_s)$ in $\alpha$. Keeping only the leading term in both the trace and the determinant we get
\ba
	\tPf_t \rightarrow  \tPf_s  = \tPf \times \frac{ 1 + {\rm tr}(I_t^{-1}) \frac{P_T  \alpha }{M}}{1 + 4  \frac{P_T \alpha}{M {\rm tr}(I_t)}}. \label{eq:tPfs}
\ea 
Since $I_t$ is by definition a $2 \times 2$ matrix, we can write 
\be {\rm tr}(I_t^{-1}) = \frac{{\rm tr}(I_t)}{{\rm det}(I_t)}  =  \frac{4}{\tPf \times {\rm tr}(I_t)} \label{eq:inverse}. \ee 
Combining Eqns.~\eqref{eq:tPfs} and  \eqref{eq:inverse}  we finally obtain

\be
	 \tPf_s  = \tPf \times \frac{ 1 + \gamma / \tPf}{1 + \gamma}  \label{eq:tPfs3},
\ee
where
\be
	\gamma \equiv \frac{4}{{\rm tr}(I_t) }\frac{P_T \alpha}{M} .
\ee
As $r_3 \rightarrow 0 ,$ Eq. \eqref{eq:tPfs2} correctly reduces to the expression for $\tPf.$ In the limit of $\tPf \rightarrow 1$, the showering effects become irrelevant as well, keeping Eq. \eqref{eq:tPfs} bounded in accordance with the definition of $\tPf$. Notice that Eq. \eqref{eq:tPfs2} is not valid in the $\tPf \rightarrow 0$ limit, as we made an assumption that ${\rm det}(I_t) \neq 0 $ during its derivation. 

\begin{figure}[!]
\begin{center}
 \includegraphics[width=4in]{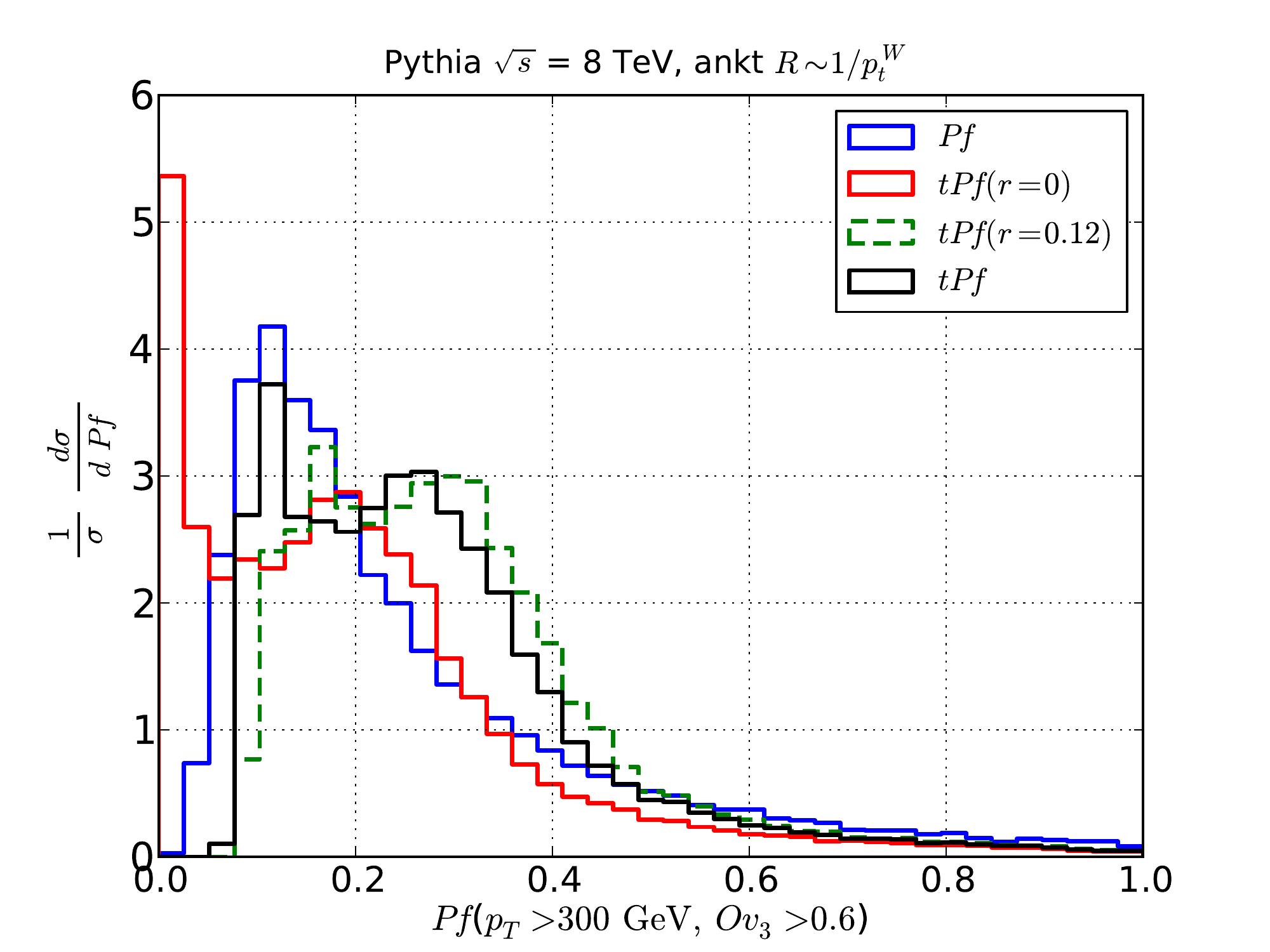}
\caption{ Effects of showering on Planar Flow of the template momenta. The blue solid curve shows the jet $\Pf$ distribution. The red solid curve shows Template Planar Flow ($\tPf$) calculated using only the three momenta from the peak template. The green, dashed curve shows Planar Flow distribution of peak templates with a sub cone of $r_3 = 0.12$ taken into account as in Eq. \eqref{eq:InertShift}. The black, solid line is the result of varying subcone radii according to the scaling rule in Fig. \ref{fig:ConeScaling}
}
\label{fig:tempPf}
\end{center}
\end{figure}

Continuing, in the narrow angle approximation ${\rm tr}(I_t) = M / p^t_T,$ where $p^t_T$ is the total template transverse momentum. We have verified numerically that this result is satisfactory even at R=1.4 
We can then rewrite Eq. \eqref{eq:tPfs} as
\be
	 \tPf_s  = \tPf \times \frac{ 1 + \gamma' / \tPf}{1 + \gamma'}  \label{eq:tPfs2},
\ee
where
\be
	\gamma' \equiv  \left( \frac{4 \,P_T \, p_T^t}{M^2}\right)\alpha.
\ee
Eq. \eqref{eq:tPfs2} provides a useful qualitative insight into the effects of showering on Planar Flow. Template momentum configurations with low Planar Flow are more sensitive to showering effects. This is important when considering Planar Flow of Higgs jets which are expected to peak at low values.   

Fig. \ref{fig:tempPf} shows the results of a data analysis with par tonic $\tPf,$ and template Planar Flow calculated a in Eq. \eqref{eq:InertShift}, for both varying $(\tPf)$ and fixed $r_3$. Addition of subcones to the Planar Flow calculation produces results which match the physical jet distribution much better than the parton level $\tPf(r = 0)$. Notice that the match between $\tPf$ and jet Planar Flow is excellent in the region of $\Pf > 0.4$, where perturbative expansions are expected to hold.

\section{Efficient Generation of Template Libraries }\label{sec:Templates}

Template Overlap Method is a systematic framework 
aimed to identify kinematic characteristics of an boosted jet. A typical template configuration consists of a model template, $f$, calculated in perturbation theory, which describes a ``prong-like'' shape of the underlying hard subprocess of a jet. Template construction typically employs prior,  theoretical knowledge of the signal kinematics and dynamics, as well as possible experimental input. 

The simplest template configurations are the ones describing the kinematics of two-body processes such as the decay of SM Higgs or $W/Z$ bosons into quark-antiquark pairs. 
These are easily dealt with by assuming the rest frame of the parent particle
and producing two decay products with equal and opposite, isotropically-selected momenta and
magnitude, subject to energy conservation. The problem of a $N$-body decay subtracts four constraints from the decay products' $3 N$ degrees of freedom: three for overall conservation of momentum and one for energy \footnote{ For our purpose, a template is an object with no other properties other than its four-momenta.}. The final states can therefore be found on a $(3N-4)$-dimensional
manifold in the multi-particle phase space. Note that the dimensionality of the template space increases rapidly with additional patrons. For instance, the two-body templates require only  two degrees of freedom, while a corresponding four-body template space is already eight dimensional. 

The question of which kinematic frame the templates should be generated in requires careful consideration. 
Authors of Ref. \cite{Almeida:2011aa} argued that a search for the global maximum of $Ov_N$
could be too computationally intensive. To improve the computation time, the template states were generated in the Higgs rest frame using 
a Monte Carlo routine, and then boosted into the lab frame. While this method worked sufficiently well for tagging
a highly-boosted object(\textit{i.e.} a $1\TeV$ Higgs jet), it introduced residual algorithmic dependence and a certain sense
of arbitrariness in the jet shape. At lower $p_T$ the Monte Carlo approach samples mainly the templates within the soft-collinear region, leaving other regions of phase space unpopulated. An enormous number of templates is required to adequately cover the phase space at $p_T \sim O(100 \GeV)$, thus fully diminishing the motivation for a Monte-Carlo approach. The simplest and most robust choice is then to generate templates directly in the lab frame and then rotate them into the frame of the jet axis. The result is a well covered template phase space in all relevant boosted frames. In addition, the lab frame templates result in a  significant decrease in  computation time as a much smaller number of templates are needed.

 We proceed to show how to generate the phase space 
for 2- and 3-parton final states as well as how to generalize the results to arbitrary $N$.

\subsection{The case of 2-body templates}

First, we summarize our notation and conventions. The model template consists of a set of four vectors, $p_1, \cdots, p_N$, on the hyperplane determined by the energy-momentum conservation,
\begin{equation}
 \sum_i p_i = P, \,\,P^2 =M^2, 
\end{equation}
where $M, P$ are the mass and four momentum of a heavy boosted particle,\textit{i.e.} the Higgs.
For simplicity, we treat all template particles to be massless.
We work in an $(\eta,\, \phi,\, p_{T})$ space, where $\eta$ is pseudorapidity, $\phi$ azimuthal angle and $p_T$ transverse momentum. 
Without loss of generality, we can assume that the template points in the $x$ direction ($\eta=\phi=0$).  The templates are distributed according to 
\begin{equation}
p_i = p_{T,i}(\cos\phi_i,\sin\phi_i,\sinh\eta_i,\cosh\eta_i),\,\, i=1,2,3
\end{equation}
subject to the constraint
\begin{equation}
\sum_{i=1}^N p_i = P = (p_T,0,0,E_J) \label{momentum_cons}
\end{equation}
with $E_J=\sqrt{M^2+p_T^2}$. We find it useful to define unit vectors by
\begin{equation}
\hat p_i = (\cos\phi_i,\sin\phi_i,\sinh\eta_i,\cosh\eta_i),\,\, i=1,2, \label{unit_vector}
\end{equation}
so that $p_i = p_{T,i}\, \hat p_i$. 

Phase space for the 2-body decay processes is characterized by particularly simple kinematic parameters. To illustrate, 
first note that the 2-particle templates are uniquely determined by one single four momentum, $p_1$ subject to the condition
\begin{equation}
(P-p_1)^2=0.
\end{equation}
Writing $p_1 = p_{T,1} \hat p_1$, we can solve for $p_{T,1}$ in terms of the angles of the first parton
\begin{equation}
p_{T,1} = \frac{M^2}{2(P\cdot \hat p_1) }.
\end{equation}

We see that a 2-particle template is therefore completely determined in terms of the unit vector $\hat p_1$ as follows:
\begin{eqnarray}
 p_1 &= &\frac{M^2}{2(P \cdot \hat p_1)} \hat p_1 \\
 p_2 & = & P - p_1 .
\end{eqnarray}

Note that we can represent such a template as a point $(\hat \eta,\, \hat \phi)$  in $\eta-\phi$ plane. These are the two degrees of freedom, in accordance with the general result that the dimensionality of the $N$ template space is $3N-4$.

\subsection{The case of 3-body templates}

A space of five degrees of freedom allows for 3-particle templates to differ from one another in more than one way.
The 3-particle templates are determined by  two four momenta, $p_1$ and $p_2$, subject to the constraint, 
\begin{equation}
(P-p_1-p_2)^2=0.
\end{equation}

Using $p_1 = p_{T,1} \hat p_1$ and $p_2 = p_{T,2} \hat p_2$, we can solve for $p_{T,2}$ in terms of the angles of first two partons and $p_{T,1}$,

\begin{equation}
p_{T,2}= \frac{M^2-2 P\cdot p_1}{2( P\cdot \hat p_2 - p_1 \cdot \hat p_2 )}.
\end{equation}
A general 3-particle template is then completely specified by $p_{T,1}$ and two unit vectors (or, equivalently, four angles) $\hat p_1$ and $\hat p_2$.

\subsection{Extension to arbitrary $N$}

A generalization to an arbitrary number of particles is straight-forward. Proceeding as above, 
the $N$-particle templates are determined by $p_1,\cdots ,p_{N-1}$ subject to the constraint, 
\begin{equation}
(P-\sum_{i=1}^{N-1} p_i)^2=0.
\end{equation}
Using $p_i = p_{T,i} \hat p_i$, we can now solve for  $p_{T, N-1}$ in terms of the $p_1,\cdots ,p_{N-2}$ and $\hat p_{N-1}$,

\begin{equation}
p_{T,N-1} = \frac{M^2 + 2 \sum_{i<j}^{N-2} p_i\cdot p_j  - 2\, P\cdot \sum_{i}^{N-2} p_i }{2\, (\hat p_{N-1} \cdot P) - 2\,\hat p_{N-1}\cdot \sum_i^{N-2}p_i }
\end{equation}
For the special cases of $N=2$ and $N=3$, this formula reduces to the above results .
\subsection{Numerical simulations}

We choose to cover the phase space uniformly in the ($3N-4$) variables $\hat p_1, \cdots, \hat p_{N-1}$ and  $ p_{T,1},\cdots p_{T,N-2}$.
For a finite number of points this method covers regions of phase space with large Planar Flow much more uniformly than a Monte Carlo based approach.

We survey the kinematically-allowed templates by fixing the total four momentum of each of the
analyzed configurations and scanning over the possible values of ($p_{T,i}$) and the angles ($\hat p_i$) within the bounded interval. The number of variables depends on the number of degrees of freedom of the template states. The value of $P = (p_T,0,0,E_J)$ can be imposed by an additional equation:
\begin{equation}
 p_N = P -\sum_{i=1}^{N-1} p_i.
\end{equation}

We generate template libraries in
the $\eta-\phi$ plane with a sequential scan in steps of $\Delta \eta = 
\Delta \phi = 0.05$.   Three-particle templates we rehire and additional scan over the transverse momentum with which we perform in steps of  $\Delta p_T = 5 \GeV$. The resulting four momenta are a requirement that they ``fit'' into an anti-kT jet of fixed radius $R$. Templates with particles outside the jet cone are discarded. Our choice of step sizes leads to $O(10^4)$ 2-particle templates and $O(10^6)$ 3-particle templates.

\bibliography{HiggsTemp}

\end{document}